\documentclass[useAMS]{mn2e}
\usepackage{graphicx}
\usepackage{txfonts}
\usepackage{longtable,lscape}
\usepackage{color}
\usepackage{moresize}
\usepackage{pdflscape}
\usepackage{blindtext, rotating}
\usepackage{ulem}

\newcommand\aj{AJ}
\newcommand\apj{ApJ}
\newcommand\apjs{ApJS}

\newcommand\aap{A\&A}
\newcommand\mnras{MNRAS}
\newcommand\apjl{ApJ}
\newcommand\pasp{PASP}
\newcommand\nat{Nature}
\newcommand\aaps{A\&AS}
\newcommand\araa{ARA\&A}

\newcommand\jqsrt{JQSRT}

\def\teff{\mbox{T$_{\rm eff}$}}
\def\logg{\mbox{log~{\it g}}}
\def\vmicro{\mbox{$\xi_{\rm t}$}}
\def\kmsec{\mbox{km~s$^{\rm -1}$}}

\title[]{Iron and $s$-elements abundance variations in NGC\,5286: comparison with ``anomalous'' globular clusters and Milky Way satellites.
}
\author[A.\, F.\, Marino et al.]
{A.\, F.\, Marino$^{1}$,
A.\,P.\, Milone$^{1}$,
A.\,I.\, Karakas$^{1}$,
L. Casagrande$^{1}$,
D. Yong$^{1}$,
L. Shingles$^{1}$,
\newauthor
G. Da Costa$^{1}$,
J.\,E.\, Norris$^{1}$,
P.\,B.\, Stetson$^{2}$,
K. Lind$^{3}$,
M. Asplund$^{1}$,
R. Collet$^{1}$,
H. Jerjen$^{1}$,
\newauthor
L. Sbordone$^{4,5}$,
A.\, Aparicio$^{6}$,
S. Cassisi$^{7}$
\\
$^{1}$Research School of Astronomy \& Astrophysics, Australian National University, Mt Stromlo Observatory, via Cotter Rd, Weston, ACT 2611, Australia \\
$^{2}$Herzberg Institute of Astrophysics, National Research Council Canada, 5071 West Saanich Road, Victoria, BC V9E 2E7, Canada\\
$^{3}$Department of Physics and Astronomy, Division of Astronomy and Space Physics, Uppsala University, Box 515, 75120 Uppsala, Sweden\\
$^{4}$Department of Electrical Engineering, Center for Astro-Engineering, Pontificia Universidad Cat\'{o}lica de Chile, Av. Vicu\~{n}a Mackenna 4860, 782-0436 Macul, Santiago, Chile\\
$^{5}$The Milky Way Millennium Nucleus, Av. Vicu\~{n}a Mackenna 4860, 782-0436 Macul, Santiago, Chile\\
$^{6}$Instituto de Astrof\'isica de Canarias, La Laguna, Tenerife, Spain\\
$^{7}$INAF-Osservatorio Astronomico di Teramo, via M.\, Maggini, 64100 Teramo, Italy \\
}

\begin{document}

\date{Accepted 24 February 2015. Received 24 February 2015; in original form 6 November 2014}

\pagerange{\pageref{firstpage}--\pageref{lastpage}} \pubyear{2013}

\maketitle

\label{firstpage}

\begin{abstract}  
We present a high resolution spectroscopic analysis of 62 red giants in the Milky Way globular cluster NGC\,5286.
We have determined abundances of representative light proton-capture, $\alpha$, Fe-peak and neutron-capture element groups, and combined them with photometry of multiple sequences observed along the colour-magnitude diagram.
Our principal results are: 
(i) a broad, bimodal distribution in $s$-process element abundance ratios, 
with two main groups, the $s$-poor and $s$-rich groups;
(ii) substantial star-to-star Fe variations, with the $s$-rich stars having higher Fe, e.g. $<$[Fe/H]$>_{s{\rm -rich}}$~$- <$[Fe/H]$>_{s{\rm -poor}}$~$\sim$0.2~dex; and (iii) the presence of O-Na-Al (anti-)correlations in both stellar groups.
We have defined a new photometric index, $c_{BVI}$=$(B-V)-(V-I)$, to maximise the separation in the colour-magnitude diagram between the two stellar groups with different Fe and $s$-element content, and this index is not significantly affected by variations in light elements (such as the O-Na anticorrelation). 
The variations in the overall metallicity present in NGC\,5286 add this object to the class of {\it anomalous} GCs. 
Furthermore, the chemical abundance pattern of NGC\,5286 resembles that observed in some of the {\it anomalous} GCs, e.g. M\,22, NGC\,1851, M\,2, and the more extreme $\omega$~Centauri, that also show internal variations in $s$-elements, and in light elements within stars with different Fe and $s$-elements content. In view of the common variations in $s$-elements, we propose the term $s$-Fe-{\it anomalous} GCs to describe this sub-class of objects. 
The similarities in chemical abundance ratios between these objects strongly suggest similar formation and evolution histories, possibly associated with an origin in tidally disrupted dwarf satellites.
\end{abstract}


\section{Introduction}\label{sec:intro}

In recent years, an increasing number of observations have shattered the paradigm of globular clusters (GCs) as simple stellar population systems. 
When the first high resolution spectroscopic data became available,
it was immediately recognised that GCs host stars with different composition in proton ($p$) capture elements (see reviews by Kraft 1994; Gratton et al.\,2004).
Later on, the advent of 8m-class telescopes with multi-object spectrographs enabled a substantial increase in statistics, from which it has become clear that stars with {\it non field-like} abundances constitute a significant fraction of their parent clusters and they were observed along the whole colour-magnitude diagram (e.g.\, Cannon et al.\,1998; Gratton et al.\,2001; Marino et al.\,2008; Carretta et al.\,2009).

The most acknowledged scenario is that the chemical variations in $p$-capture elements are due to the presence of multiple stellar generations in GCs, with some being enriched in the products of high temperature H-burning, such as Na and N.
However, internal star-to-star variations in $p$-capture elements may be due to early disk accretion, rather than to the presence of multiple stellar generations (Bastian et al.\,2013).
In any case, this phenomenon to date, seems to be a typical feature of Galactic GCs. These features are also visible in the form of multiple sequences, mostly in the ultraviolet bands, on the GCs' colour-magnitude diagrams (CMDs), from the main sequence (MS) up to the red-giant branch (RGB; e.g.\,Marino et al.\,2008; Milone et al.\,2012).

For a long time, variations in the overall metallicity, here considered primarily as Fe,  and/or in neutron ($n$) capture elements were considered a strict peculiarity of the most massive cluster $\omega$~Centauri (e.g.\, Norris \& Da Costa\,1995; Johnson \& Pilachowski 2010; Marino et al.\,2011). For this reason, this object was associated with the nuclear remnant of a dwarf galaxy rather than with a real GC (e.g.\,Bekki \& Freeman 2003; Bekki \& Norris 2006).
More recently, other objects with internal variations in metallicity have been identified, including M\,22 (Marino et al.\,2009; Da Costa et al.\,2009). 
Interestingly, a clear bimodality in $slow$ ($s$) $n$-capture elements was found in this GC, with $s$-CNO-enriched stars having a higher overall metallicity (Marino et al.\,2009,\,2011; Roederer et al.\,2012; Alves Brito et al.\,2012). 
Following the discovery of metallicity variations in M\,22, the number of clusters known to have variations in metallicity has increased substantially, and now includes M\,54 (Carretta et al.\,2010), 
Terzan\,5 (Ferraro et al.\,2009), M\,2 (Yong et al.\,2014), for which high-resolution spectroscopy is available, and NGC\,5824 (Da Costa et al.\,2014), whose metallicities have been inferred from the Ca triplet analysis in low-resolution data.
Among these GCs, chemical variations in $s$-process elements have been studied and confirmed, besides $\omega$~Centauri and M\,22, in M\,2 (Yong et al.\,2014); while we are aware of C+N+O variations only in $\omega$~Centauri and M\,22.
A possible small metallicity spread, at the level of a few hundredths of dex, has been proposed also for NGC\,1851 (Carretta et al.\,2010; Gratton et al.\,2013; Marino et al.\,2014), which also shows internal variations in the $s$-elements (Yong \& Grundahl\,2008; Villanova et al.\,2010) and in the overall C+N+O content (Yong et al.\,2014). Although the presence of Fe variations in NGC\,1851 needs further confirmation, enough evidence exists for it to be classified as a GC with chemical {\it anomalies} with respect to the bulk of Galactic GCs, such as the $s$-elements and C+N+O variations, that have been seen also in $\omega$~Centauri and M\,22, but are not typically observed in Milky Way clusters. We therefore include NGC\,1851 in the list of clusters with chemical {\it anomalies} with respect to the bulk of Galactic GCs.\footnote{We note that iron variations have been found in NGC\,3201 by Simmerer et al.\,(2013), but such evidence is not present in Carretta et al.\,(2010) nor in Mu{\~n}oz et al.\,(2013). Specifically, Mu{\~n}oz et al.\,(2013) did not detect any spread in the $s$-elements, as that found in other GCs with Fe variations. At the moment we do not include this object in our list of {\it anomalous} GCs.}

All these findings show that metallicity variations, which were thought to be an exclusive feature of $\omega$~Centauri, is actually a more widespread phenomenon in GCs. 
The degree of the observed metallicity variations varies from cluster to cluster, with $\omega$~Centauri and NGC\,1851 being the extremes with the widest and lowest Fe spreads, respectively. It is tempting to speculate that these objects able to retain fast Supernovae ejecta, were much more massive at their birth, and possibly nuclei of disrupted dwarf galaxies, as suggested for $\omega$~Centauri. 

On the photometric side, these GCs show some {\it peculiarities}, not observed in the other clusters. 
The CMD of $\omega$~Centauri is the most complex ever observed for a GC, with multiple sequences along all the evolutionary stages, from the MS to a well-extended multimodal horizontal branch (HB), through a complex multiple sub-giant branch (SGB, e.g.\ Bellini et al. 2010).
However, while the presence of multiple MSs and RGBs in UV bands, as well as in some cases extended HBs, are in general good proxies for variations in light elements (including He for the HB, see Milone et al.\,2014), multiple SGBs are 
observed in many of the clusters with Fe variations, in all photometric bands (e.g.\,Milone et al.\,2008; Piotto et al.\,2012).
Theoretically, multiple SGBs may reflect differences in the overall metallicity and/or C+N+O and/or age (Cassisi et al. 2008; Marino et al.\ 2011, 2012).
Two notable examples in this respect are the double SGBs of NGC\,1851 (Milone et al.\ 2008) and M22 (e.g. Marino et al.\ 2009, 2012; Piotto et al.\ 2012).   
Like M\,22, NGC\,1851 has a bimodal distribution in the $s$-process elements (Yong \& Grundahl\,2008; Carretta et al.\,2010; Villanova et al.\,2010), and some evidence of variations in the total CNO have been provided by Yong et al. (2009, 2014).
Complex SGB morphologies are present also in M\,54 and M\,2, with the latter exhibiting a triple SGB (Piotto et al.\,2012; Milone et al. 2015).

To date metallicity variations are observed in 8 Galactic GCs, over the $\sim$30 GCs where Fe abundances are available for relatively large samples of stars. Note that the true fraction of these objects in the Milky Way is likely lower 
as many recent spectroscopic observations are biased because they were aimed at the study of these objects previously identified from photometry.     
Despite the number of the GCs with variations in Fe is expected to increase, these objects still constitute a minor component with respect to monometallic GCs. The chemical properties of these objects can be regarded as {\it anomalies} with respect to the bulk of GCs in the Milky Way, indeed we refer to these objects as {\it anomalous} GCs. The term {\it anomalous} will be primarily used to indicate the objects with internal metallicity variation, which on different levels is shared by all these objects.

In this study we further explore the properties of {\it anomalous} GCs, the plausibility to identify a further class of objects where Fe variations are accompanied by variations in $s$-elements, and the possibility that a split SGB in a GC constitutes a proxy for its chemical {\it anomaly}, e.g. internal variations in overall metallicity, heavy elements including slow $n$-capture elements ($s$-elements), and C+N+O. 
Our aims are to trace how frequent these {\it anomalous} objects occur in the Milky Way, and to try to disclose their possible formation and early evolution. A fundamental step to this goal is to understand if they constitute a separate class of objects from typical Galactic GCs, originated in a different way; or they simply form as typical {\it normal} GCs and their chemical anomalies are due to more advanced stages of evolution. The $s$-process enrichment due to low-mass AGB needs some hundreds Myrs to occur; and at the time this enrichment starts to be effective, the Fe-enriched material from supernovae, previously expelled from the cluster, may fall-back into the GC potential well and contribute to the formation of a new stellar generation (e.g.\, D'Antona et al. 2011 for $\omega$~Centauri).  

In this paper we focus on the chemical abundances for a poorly studied GC: NGC\,5286.
$HST$ photometry has demonstrated that this GC shows a split SGB, similar to those observed in NGC\,1851 and M\,22.
In this case, however, the stellar component on the fainter SGB constitutes only $\sim$14\% of the total mass of the cluster, which is significantly lower than in M\,22 and NGC\,1851 (fainter SGB stars in these GCs account for the $\sim$38\% and $\sim$35\% respectively; Piotto et al.\,2012), but larger than in M\,2, where its two faint SGB components account for $\sim$3\% and $\sim$1\% of all the SGB stars (Milone et al.\,2015).

\section{Data}\label{sec:data}

With a mass of $M=10^{5.65}M_{\odot}$ (McLaughlin \& van der Marel 2005), and an absolute visual magnitude of ${\rm M_{V}}=-8.74$, as listed in the Harris catalog (Harris\,1996, updated as in 2010), NGC\,5286 is a relatively massive GC (as a comparison, M\,22 has $\rm {M_{V}}= -8.50$, NGC\,1851 $\rm {M_{V}}= -8.33$ and M\,2 $\rm {M_{V}}= -9.03$).
This GC lies at a distance of 8.9~kpc from the Galactic centre and 11.7~kpc from the Sun, and it is affected by relatively high foreground reddening, with a mean value of $E(B-V)=0.24$ (Harris\,2010).
NGC\,5286 shows a blue horizontal branch, more than a dozen RR~Lyrae variables (e.g., Clement et al.\,2001), whose periods are consistent with an Oosterhoff~II type (Zorotovic et al.\,2010).
In this section we consider in turn the photometric and spectroscopic
data that we have employed in this study.

\subsection{The photometric dataset: multiple populations along the SGB/RGB of NGC\,5286 \label{sec:phot_data}}

\begin{figure*}
   \centering
   \includegraphics[width=15.5cm]{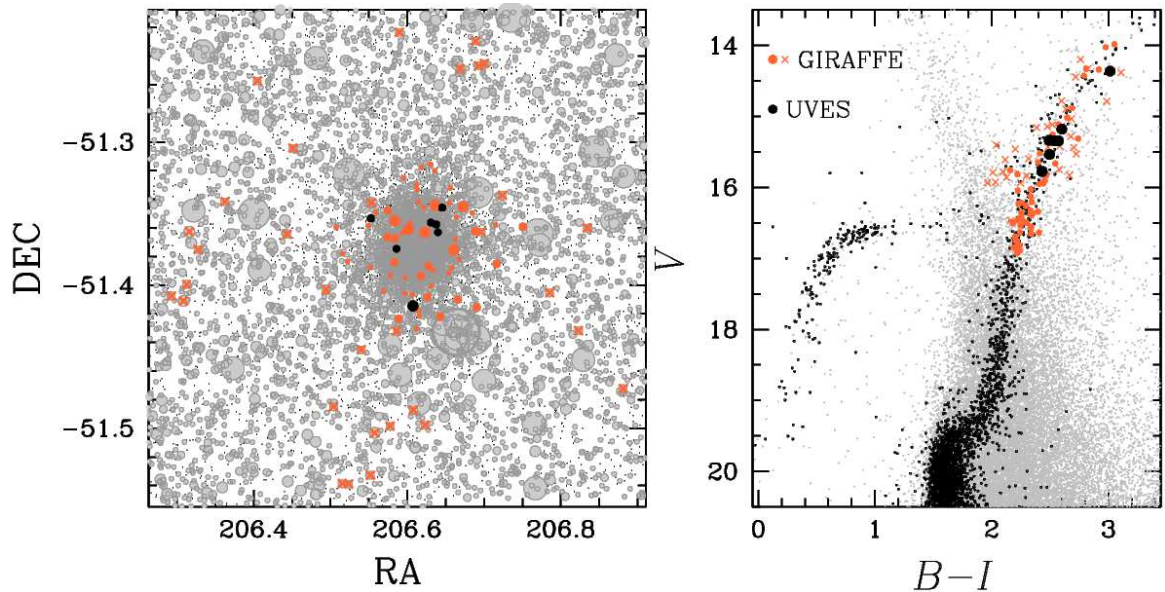}
      \caption{
       Left panel: Location of the spectroscopic targets in RA and DEC. The NGC\,5286 spectroscopic targets have been plotted in orange (GIRAFFE) and black (UVES). For GIRAFFE stars we used different symbols for {\it bona-fide} cluster stars (dots) and field stars (crosses). Right panel: $V$-$(B-I)$ CMD of NGC\,5286  proper-motions members (black) and field stars (grey). The position of the spectroscopic GIRAFFE and UVES targets on the CMD is shown by using the same symbols as on the left panel.
        }
        \label{fig:cmdtarget}
   \end{figure*}

We used photometric data from the Wide Field Imager (WFI) of the Max Planck 2.2m telescope at La Silla collected through the $U$ filter under the SUrvey of Multiple pOpulations in GCs (SUMO; program 088.A-9012-A, PI.\,A.\,F.\,Marino). 
These $U$ images consist of 14$\times$850s$+$3$\times$300s collected on February 2012. 
Additionally, we have used $B$, $V$, and $I$ photometry from the archive maintained by P.\,B.\,Stetson (Stetson\,2000). 
A journal of all the observations is shown in Tab.~\ref{tab:journal}. 
In Fig.~\ref{fig:cmdtarget} we plot a $V$ vs. $(B-I)$ CMD and the location of stars in a $\sim$21$\arcmin \times$21$\arcmin$ field of view around NGC\,5286; 
our spectroscopic UVES and GIRAFFE targets have been marked in black and orange, respectively.

$BVI$ photometry has been used to determine atmospheric parameters, as discussed in Sect.~\ref{sec:atm}.
As the $U$ filter is very efficient to identify multiple stellar populations along the RGB (Marino et al.\,2008), we used $U$ data from our SUMO program to investigate the connection between multiple sequences in the CMD and the chemical composition. 
The photometric and astrometric reduction of WFI data has been carried out by using the software and the procedure described by Anderson et al.\,(2006).
To calibrate the magnitudes in the $U$ Johnson we have matched our photometry with the catalogue of photometric secondary standards by Stetson\,(2000) and derived calibration equation by using least-squares fitting of straight lines of stellar magnitudes and colours. 

Very-accurate photometry is crucial to identify different sequences along the CMD for the analysis of multiple stellar populations.
To this aim, we have followed the recipe by Milone et al.\,(2009, Sect.~2.1) and selected a sample of stars with small astrometric and photometric errors, which are well fitted by the PSF, and relatively isolated. 
Our photometry has been also corrected for differential reddening as in Milone et al.\,(2012).

\begin{figure}
   \centering
   \includegraphics[width=8.3cm]{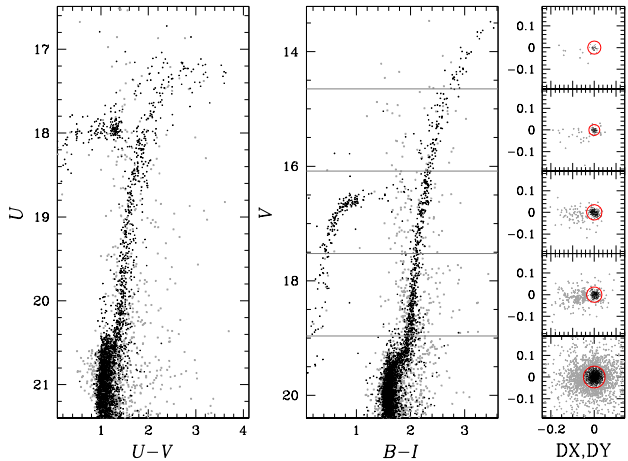}
      \caption{
$U$ vs. $(U-V)$ (left panel) and $V$ vs. $(B-I)$ CMD (middle panel) of stars in the field of view of NGC\,5286 corrected for differential reddening.
Right panels show the vector point diagrams of stellar displacement in the 5 intervals of V magnitude indicated by the horizontal lines in the middle-panel CMD.
The red circles separate probable cluster members and field stars, which have been represented with black dots and grey crosses, respectively in all the panels of this figure.
}
        \label{fig:pm}
   \end{figure}

For NGC\,5286, the analysis of the CMD is affected by strong contamination from background/foreground field stars clearly visible in Fig.~\ref{fig:cmdtarget}. 
Milone et al.\,(2012, see their Fig.~12) have shown that the average proper motion of this GC differs from the motion of most of the field stars. Therefore, we have used proper motions to separate most of the field stars from cluster members. 
Briefly, we have estimated the displacement between the stellar positions measured from WFI data and those in the catalogue by Stetson\,(2000) by using the method described in Anderson \& King (2003; see also Bedin et al.\,2006; Anderson \& van der Marel\,2010). 
Results of our proper motions analysis are illustrated in Fig.~\ref{fig:pm}:
the left panel shows the $V$ vs.\,$(B-I)$ CMD of all the stars with radial distance from the cluster centre smaller than 4.3~\arcmin\ that pass our photometric criteria of selection; the right panels display the vector-point diagrams (VPDs) of the stellar displacement for stars in five luminosity intervals.
Since we have calculated relative proper motions with respect to a sample of cluster members, the bulk of stars around the origin of the VPD is mostly made of NGC\,5286 stars, while field objects have clearly different motion. The red circles have been drawn by eye and are used to separate probable cluster members (black points) from the most-evident field stars (grey crosses).

The probable cluster members, selected by proper motions, have been plotted in the $V$ vs.\,$(B-I)$ and $U$ vs.\,$(U-V)$ CMDs with black dots in Fig.~\ref{fig:pm}, while field stars with grey dots. Our $U$ vs.\,$(U-V)$ CMD shows a complex SGB {\it evolving} into a spread/double RGB. 
For a clearer visualisation of the two SGBs and RGBs in the $U$ vs.\,$(U-V)$ CMD refer to Fig.~\ref{fig:cmd_s}, that provides the first evidence for a double RGB in the $U$-$(U-V)$ CMD for NGC\,5286. 
This intriguing double RGB feature has only been found in a handful of objects observed in the SUMO program. In the next section, we describe the investigation of the chemical composition of these two sequences using FLAMES data. 
We also note that on this diagram the AGB sequence is clearly separated from the RGB.

%
%

%
%

\subsection{The spectroscopic dataset}\label{sec:spec_data}

Our spectroscopic data consist of FLAMES/GIRAFFE and FLAMES/UVES spectra (Pasquini et al.\ 2002)\nocite{pas02} observed under the program 091.D-0578(A) (PI: A.\,F.\,Marino).
The high-resolution HR13 GIRAFFE setup was employed, which covers 
a spectral range of $\sim$300~\AA\ from $\sim$6122~\AA\ to $\sim$6402~\AA, and provides a resolving power $R \equiv \lambda/\Delta\lambda \sim$22,000. The higher resolution fibres available for UVES provided spectra with a
larger wavelength coverage from $\sim$4800~\AA\ to $\sim$6800~\AA, with a resolution of $\sim$45,000. 

In total we gathered spectra for 87 GIRAFFE plus 7 UVES stars,
represented in Fig.~\ref{fig:cmdtarget}.
Our targets have been carefully selected to sample both RGBs of NGC\,5286 that we have found from the analysis of the CMD, as discussed in Sect.~\ref{sec:phot_data}.
Most of the targets are RGB stars of NGC\,5286 with 14$\lesssim V \lesssim$16.5, with some AGB and field stars, and were observed in the same FLAMES plate in 11 different exposures of 46 minutes. 
The UVES targets span a smaller range in magnitude, around $V\sim$15~mag.
The typical S/N of the fully reduced and combined GIRAFFE spectra ranges from $\sim$80 to $\sim$200 at the central wavelength, depending on the brightness of the stars; the UVES final spectra have a S/N around $\sim$70 per pixel at the Na doublet at $\sim$6160~\AA; the most luminous UVES star (\#859U) has S/N$\sim$150 at the same wavelength. 

Data reduction involving bias-subtraction, flat-field correction, 
wavelength-calibration, sky-subtraction, has been done by using the dedicated pipelines\footnote{See {\sf http://girbld-rs.sourceforge.net}}. 
Radial velocities (RVs) were derived using the IRAF@FXCOR task, which cross-correlates the object spectrum with a template. 
For the template we used a synthetic spectrum obtained through the March, 2014 version of MOOG (Sneden 1973). 
This spectrum was computed with a stellar model atmosphere interpolated from the Castelli \& Kurucz (2004) grid, adopting parameters (\teff, \logg, \vmicro, [Fe/H]) = (4500~K, 2.5, 2.0~\kmsec, $-$1.80).
Observed RVs were then corrected to the heliocentric system. 
Heliocentric RVs were used as a membership criterion for our GIRAFFE targets, together with the proper motion selection (see Sect.~\ref{sec:phot_data}). 
First, we selected the stars having velocities in the range between 40 and 100~\kmsec, that is where the major peak in the RV distribution appears; then we considered the stars within 2$\times \sigma$ (where $\sigma$ has been estimated as the $68.27^{{\rm th}}$ percentile of the RV distribution) around the median value of this selected sample as probable cluster members. At the end, our GIRAFFE sample of probable NGC\,5286 stars is composed by 55 stars, whose median RV is 61.5$\pm$1.1~\kmsec (rms=7.8~\kmsec), which is in reasonable agreement with the value reported in the Harris catalog, 57.4$\pm$1.5~\kmsec\ (rms=8.1~\kmsec). The seven stars observed with UVES have mean RV 65.6$\pm$1.3~\kmsec\ (rms=3.1~\kmsec) and were all considered members of NGC\,5286.
Among the {\it bona-fide} GC stars, two GIRAFFE targets lie on the AGB sequence visible on the $U$-$(U-V)$ CMD.
Coordinates, basic $BV$ photometry and RVs for the all the stars observed with GIRAFFE and UVES are listed in Tab.~\ref{tab:data}. 
Only cluster members, selected on the basis of proper motions and RVs have been included in the following analysis.

\section{Model atmospheres}\label{sec:atm}

Given the different resolution and spectral coverage of our GIRAFFE and UVES data we decided to estimate stellar atmospheric parameters using different techniques. For the GIRAFFE spectra, given the relatively low number of Fe lines, we rely on the photometric information to derive effective temperatures (\teff), surface gravities (\logg), and microturbolent velocities (\vmicro). On the other hand, for UVES data we derive atmospheric parameters by using a standard fully spectroscopic approach, independent of the photometry. Details on the estimate of stellar parameters for both sets of spectra are presented below.

\subsection{GIRAFFE spectra}\label{sec:giratm}

We couple our $BVI$ photometry (see Sect.~\ref{sec:phot_data}) with $JHK_S$ from 2MASS and run the Infrared Flux Method (IRFM) described in Casagrande et al.\,(2010). This implementation of the IRFM has been recently validated also for giants, by direct comparison with interferometric angular diameters (Casagrande et al.\,2014). For all stars we assume the cluster metallicity [Fe/H]=$-$1.69 (Harris 2010), while adopting preliminary estimates for \logg\ from isochrones taken from the Dartmouth Stellar Evolution Database (Dotter et al\,2008), which is appropriate since the IRFM depends very mildly on those parameters. 
To investigate the impact of the metallicity and gravity on the temperature values we run the IRFM on all the stars by assuming [Fe/H]=$-$1.49 and [Fe/H]=$-$1.89, which corresponds to a metallicity of $\pm$0.2~dex around the adopted value of [Fe/H]=$-$1.69. The two sets of temperatures are almost identical, and differ on average by $\sim$2~K. We emphasise that a variation of 0.4~dex in [Fe/H] is more than twice larger than the mean metallicity difference measured in NGC\,5286, and conclude that the effect of the adopted metallicity is expected to be negligible on the results of this paper (see Sect.~4). Similarly, a difference in \logg\ by 0.5~dex only marginally affects the derived \teff\ values, as corresponds to a mean variation for this parameter of $\sim$3~K.

As discussed in Sect.~\ref{sec:phot_data}, all our photometry is corrected for differential reddening. This correction is important for GCs like NGC\,5286 that have a high mean reddening; indeed our correction suggests relatively high deviations from the absolute $E(B-V)$ value, with maximum variations being $\sim$0.1~mag. For the sake of deriving the correct effective temperatures we must also account for the absolute value of reddening, which we assume to be $E(B-V)$=0.24 from Harris (2010). In our implementation of the IRFM, the effective temperature of each star is obtained by averaging the values obtained from each infrared band; their standard deviation also provides an estimate of the internal accuracy of our results, which for this dataset is of order 40~K, indirectly confirming the quality of our differential reddening corrections.

However, this estimate for the photometric \teff\ could be derived only for a sample of our stars (39/55), that have good 2MASS photometry. For the remaining stars, we determined \teff\ values from the \teff-$(V-I)$ relation obtained by using the sample for which the photometric \teff\ values could be derived.
The use of the $(V-I)$ colour for this purpose is justified by the fact that it is insensitive to variations in light elements, and we have verified that stars of the two different RGBs of NGC\,5286 overlap in the \teff-$(V-I)$ relation. To ensure homogeneity, we used the \teff\ derived by the \teff-$(V-I)$ relation for all our stars. The spread around this colour-\teff\ relation is 44~K, similar to the internal error associated with the photometric \teff.
We assumed that the internal uncertainty affecting our temperatures is $\sim$50~K. 
As a test to our scale of temperatures, we compared our adopted values with those derived from the projection of the targets on to the best-fitting $\alpha$-enhanced isochrone (Dotter et al.\,2008). The mean difference between the two sets of temperature is $\Delta$\teff$_{{\rm iso - adopted}}$=$-$40$\pm$12~K.

Surface gravities were obtained from the apparent $V$ magnitudes, corrected for differential reddening, the \teff\ from above, bolometric corrections from Alonso et al. (1999), and an apparent distance modulus of $(m-M)_{V}$=16.08 (Harris 2010). We assume masses taken from isochrones of 0.81~$M_{\odot}$. Internal uncertainties associated with these \logg\ determinations are formally small: internal errors in \teff\ values of $\pm$50~K and of $\pm$0.05 unit in mass, affect the \logg\ values by $\pm$0.02 and $\pm$0.03~dex, respectively. The internal photometric uncertainty associated with our $V$ mag modifies our surface gravities by $\sim$0.01~dex. All these effects, added in quadrature, contribute to an internal error in \logg\ $\lesssim$0.05~dex. 

For microturbolent velocities, we adopted the latest version of the relation used in the Gaia-ESO survey (GES, Gilmore et al. 2012; Bergemann et al., in preparation), which depends on \teff, \logg\ and metallicity{\footnote{{\sf {http://great.ast.cam.ac.uk/GESwiki/GesWg/GesWg11/Microturbulence}}}}.
Temperatures and gravities were already set from above, while for metallicity we adopted [A/H]=$-$1.75 as first guess, and then the [Fe/H] abundance derived from Fe lines (as explained below).
The dispersion of the recommended \vmicro\ values for the GES UVES spectra around the adopted relation is about 0.20~\kmsec, which is a reasonable internal uncertainty to be associated with our adopted values.
The dispersion in [Fe/H] obtained for various RGB groups (see Sect.~\ref{sec:abb}) has been considered as an estimate for the internal error in the metallicity used in the stellar atmosphere model.

\subsection{UVES spectra}

The high resolution and the large spectral coverage of UVES spectra allowed us to derive \teff, \logg\, and \vmicro\  solely from spectroscopy.
We determine \teff\ by imposing the excitation potential (E.P.) equilibrium of the Fe\,{\sc i} lines and gravity with the ionisation equilibrium between Fe\,{\sc i} and Fe\,{\sc ii} lines. For \logg\ we account for non-local thermodynamic equilibrium effects (NLTE) by imposing Fe\,{\sc ii} abundances slightly higher (by 0.05-0.07 dex) than the Fe\,{\sc i} ones (Bergemann et al. 2012; Lind, Bergemann \& Asplund 2012). For this analysis, microturbolent velocities, \vmicro\ were set to minimise any dependence on Fe\,{\sc i} abundances as a function of EWs.

In order to have an estimate of the internal errors associated with our spectroscopic atmospheric parameters we have compared our \teff/\logg\ values with those derived from the projection of the UVES targets on the best-fitting isochrone (as in Sect.~\ref{sec:giratm}). We obtain: 
$\Delta$\teff=\teff$_{\rm {(Fe~lines)}} -$\teff$_{\rm {(isochrone)}}=-83\pm15$~K (rms=36~K), and $\Delta$\logg=\logg$_{\rm {(Fe~lines)}} -$\logg$_{\rm {(isochrone)}}=-0.29\pm0.07$ (rms=0.16).
Comparing with the \teff\ values derived from the IRFM we obtain a larger systematic, that is $\Delta$\teff=\teff$_{\rm {(Fe~lines)}} -$\teff$_{\rm {(IRFM)}}=-132\pm12$~K (rms=28~K), reflecting the fact that \teff\ from the IRFM are $\sim$40~K higher than those derived from the best-fit isochrone.
Regarding the most reliable \teff-scale, both spectroscopic and photometric scales are likely affected by systematics. These systematics are due to the used Fe lines, adopted log$gf$, residual NLTE effects in the case of the spectroscopic \teff\ scale, and mostly due to the adopted absolute reddening in the case of the photometric \teff\ values. 
A systematic difference in \teff\ by $\sim$100~K can be easily obtained by varying the mean reddening by $\sim$0.03 mag. 
In any case, our comparisons suggest that even if the spectroscopic \teff/\logg\ scales are systematically lower, the internal errors in these parameters are expected to be relatively small, comparable with the rms of the average differences, e.g. about 30-40~K and 0.16~dex, in temperature and gravity, respectively.

As a further check on internal errors associated to our spectroscopic \teff\ we calculated, for each star, the errors on the slopes of the best least squares fit in the relations between abundance vs.\,E.P. The average of the errors corresponds to the typical error on the slope. Then, we fixed the other parameters and varied the temperature until the slope of the line that best fits the relation between abundances and E.P. became equal to the respective mean error. This difference in temperature can be considered a rough estimate of the error in temperature itself. The value we found is 50~K.

The same procedure applied for \teff\ was also applied for \vmicro, but using the relation between abundance and the reduced EWs. We obtained a mean error of 0.11~\kmsec.

As explained above, surface gravities for the UVES data have been obtained by imposing the ionisation equilibrium between Fe\,{\sc i} and Fe\,{\sc ii} lines (accounting for NLTE effects). The measures of Fe\,{\sc i} and Fe\,{\sc ii} have averaged uncertainties of $<\sigma{\rm {(Fe\,{\textsc i})}}>$ and $<\sigma{\rm {(Fe\,{\textsc {ii}})}}>$ (where $\sigma$(Fe\,{\sc i,ii}) is the dispersion of the iron abundances derived by the various spectral lines in each spectrum given by MOOG, divided by  $\sqrt{ {N_{\rm lines}-1}}$). Hence, in order to have an estimate of the error associated with the adopted \logg\ values we have varied the gravity of our stars such that the ionisation equilibrium is satisfied between Fe\,{\sc i}$- <\sigma{\rm {(Fe\,{\textsc i})}}>$ and Fe\,{\sc ii}$+ <\sigma{\rm {(Fe\,{\textsc {ii}})}}>$, including the additional difference due to NLTE effects. The obtained mean error is  $\Delta$\logg=0.14$\pm$0.02. This error agrees with that estimated from the comparison with photometric values (0.16), hence we adopted an error of 0.16 for our adopted \logg\ values.

\section{Chemical abundances analysis}\label{sec:abundances}

Chemical abundances were derived from a local thermodynamic
equilibrium (LTE) analysis by using the March, 2014 version of the
spectral analysis code MOOG (Sneden 1973), 
and the alpha-enhanced Kurucz model atmospheres of
Castelli \& Kurucz (2004), whose parameters have been obtained as
described in Sect.~\ref{sec:atm}.
We used the abundances by Asplund et al. (2009) as reference solar abundances. 

A list of our analysed spectral lines, with excitational potentials (EPs) and the adopted total oscillator strengths (log~$gf$) is provided in Tab.~\ref{tab:linelist}.
At the higher resolution of UVES we computed an EW-based analysis, with EWs estimated from gaussian fitting of well isolated lines (Tab.~\ref{tab:linelist}), computed by using a home-made routine (see Marino et al.\,2008). The exceptions from the EW analysis are discussed below. For GIRAFFE, given the lower resolution, we synthesised all spectral features.
When required and atomic data is available from the literature, we considered hyperfine and/or isotopic splitting in our analysis (last column of Tab.~\ref{tab:linelist}).
We comment in the following on the transitions that we used for UVES and GIRAFFE, depending on the spectral coverage, resolution and S/N of the two different datasets.

{\it Iron:} Iron abundances were derived from the EWs of a number of isolated spectral lines for UVES data. Typically, we used a number of $\sim$30-35 lines for Fe\,{\sc i}, and of $\sim$10 for Fe\,{\sc ii}. From GIRAFFE data we synthesise a typical number of $\sim$20 Fe\,{\sc i} lines. 

{\it Proton-capture elements:}
For UVES data we determined Na abundances from spectral synthesis of the two Na\,{\sc i} doublets at $\sim$5680~\AA\ and $\sim$6150~\AA; while in the smaller spectral range available for GIRAFFE we used only the doublet at $\sim$6150~\AA. 
NLTE corrections from Lind et al.\,(2011) have been applied to all our Na spectral lines. 
Oxygen abundances were inferred from the synthesis of the forbidden [O\,{\sc i}] line at 6300~\AA\ both for UVES and GIRAFFE data. 
Telluric O$_{2}$ and H$_{2}$O spectral absorptions often affect the O line at 6300~\AA. Indeed, for our NGC\,5286 targets  the analysed O transition is contaminated by O$_{2}$ lines. We have removed tellurics by using the software MOLECFIT\footnote{{\sf http://www.eso.org/sci/software/pipelines/skytools/molecfit}} provided by ESO (Smette et al.\,2014; Kausch et al.\,2014). But, even with such a subtraction procedure, we caution that residual telluric feature contamination might be of concern for the analysis of the 6300.3 [O\,{\sc i}] line. Magnesium and aluminium abundances were possible only for the UVES data. Aluminium was determined from the synthesis of the doublet at 6696~\AA. Spectral synthesis of the analysed Al transitions allow us to account for possible blending caused by CN molecules, that are substantial in the case of the star \#859U, which is the coolest in our UVES sample.
Magnesium has been inferred from EWs of the transitions at $\sim$5528, 5711, and 6318~\AA.

{\it $\alpha$ elements:}
For UVES spectra we determined abundances from EWs of Si, Mg (see above), Ca, and Ti (I and II). All these $\alpha$ elements, except Mg, could be inferred also for the smaller GIRAFFE spectral range, where we measured abundances for a subsample of lines using spectral synthesis.

{\it Iron-peak elements:}
From UVES spectra we determined abundances for Sc, V, Cr, Ni, Zn using EWs.
Abundances for Cu were inferred by synthesising the Cu\,{\sc i} lines at 5105, 5218~\AA. Both hyperfine and isotopic splitting were included in the Cu analysis, with well-studied spectral line component structure from the Kurucz (2009) compendium\footnote{Available at: {\sf http://kurucz.harvard.edu/}}. Solar-system isotopic fractions were assumed in the computations: f(${\phantom{}}^{63}$Cu)=0.69 and f(${\phantom{}}^{65}$Cu)=0.31. 
For Zn we analysed the Zn\,{\sc i} line at 4810~\AA, for which we determined EWs. This line has no significant hyperfine or isotopic substructures, and was treated as a single absorber. 
From GIRAFFE spectra we inferred only Ni and Sc using spectral synthesis. 

{\it Neutron-capture elements:}
We derived Y, Zr, La, Ce, Pr, Nd, Eu and Ba from the UVES spectra, and Ba and La from the GIRAFFE spectra.
An EWs-analysis was performed for Y, Ce, Nd, and Ba, and spectral synthesis for the other elements for which hyperfine and/or isotopic splitting and/or blending features needed to be taken into account.
Specifically, we have employed spectrum syntheses to derive the La and Eu abundances, because the spectral features of both La\,{\sc ii} and Eu\,{\sc ii} have significant hyperfine substructure, and the Eu\,{\sc ii} lines also have isotopic splitting. 
Because barium lines suffer from both hyperfine and isotopic substructures, and in the case of the 6141~\AA\ line blending by Fe, we used the blended-line EW analysis option available in MOOG.
Zirconium abundances are available for all the seven stars observed with UVES from the Zr\,{\sc ii} line at 5112~\AA. 

\begin{figure*}
    \includegraphics[width=8.1cm]{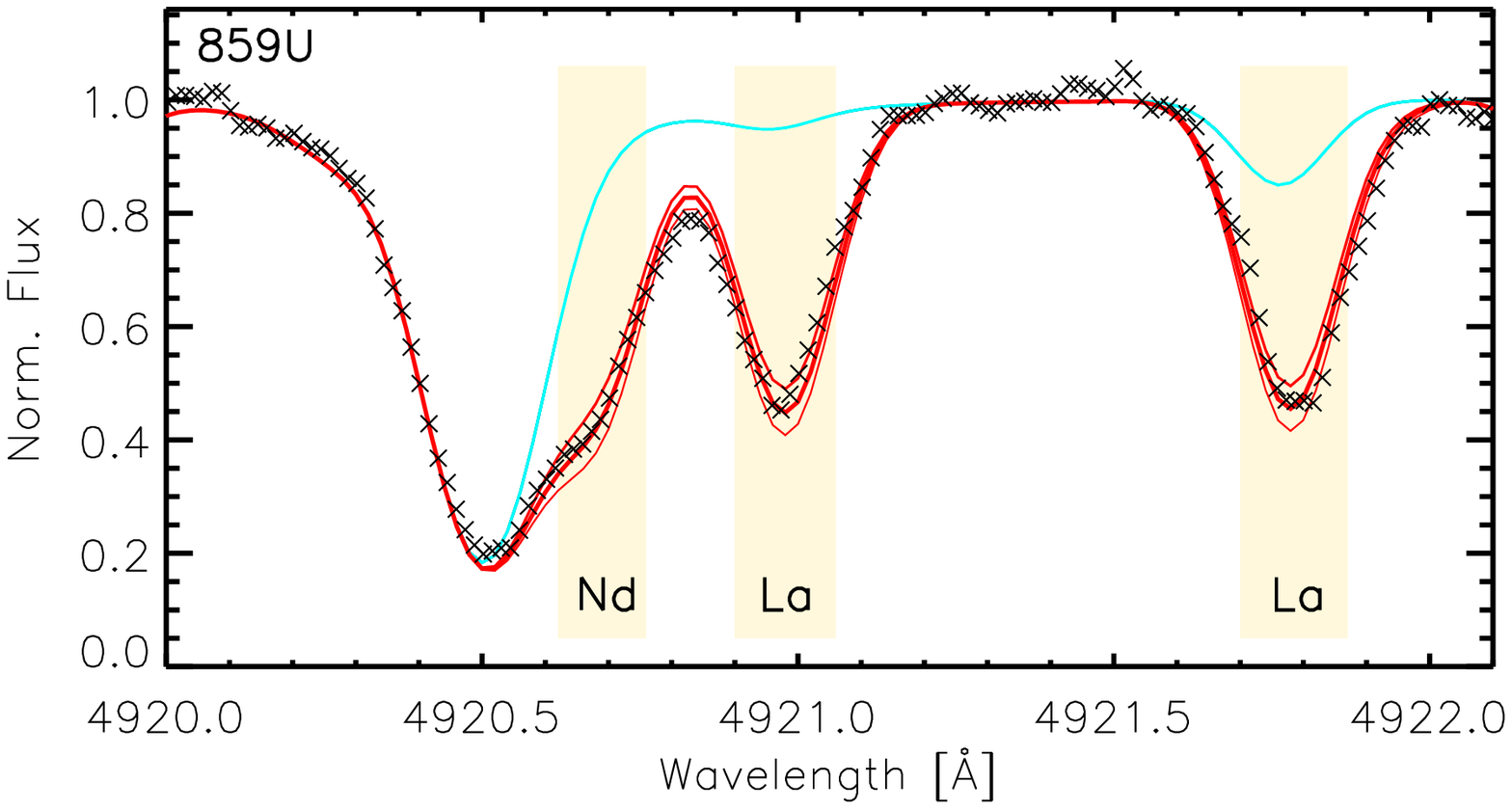}
    \includegraphics[width=8.1cm]{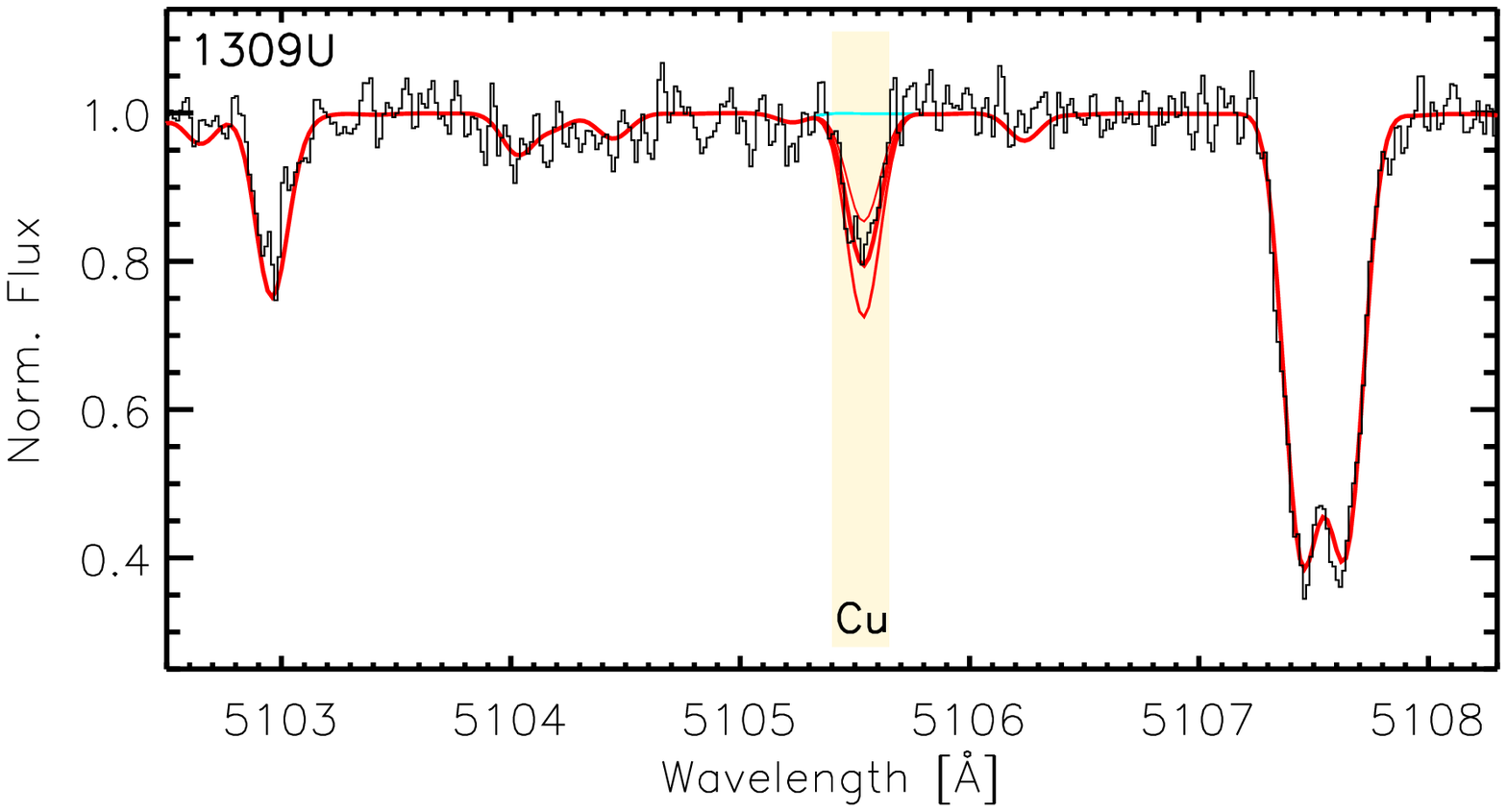}
    \includegraphics[width=8.1cm]{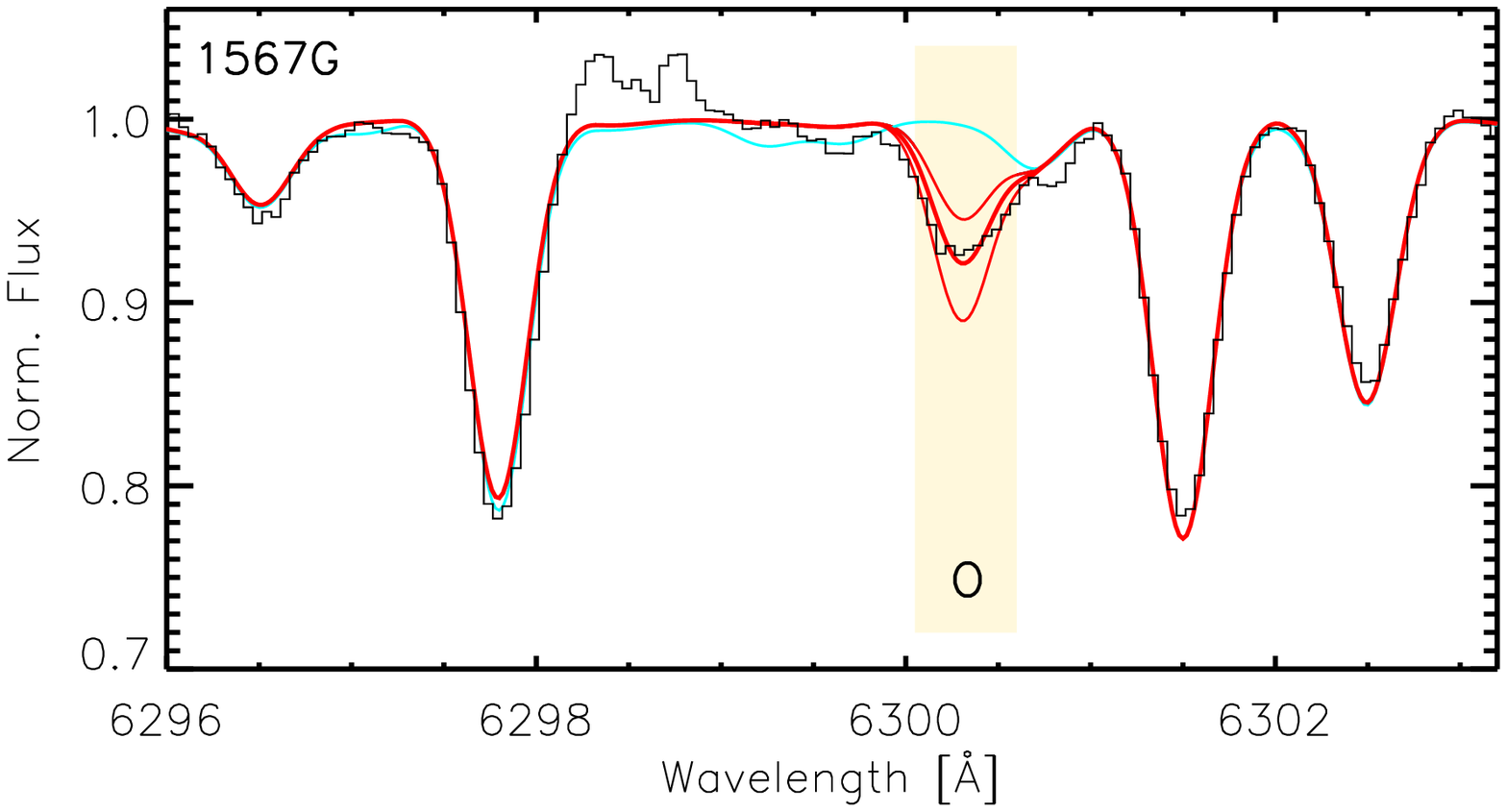}
    \includegraphics[width=8.1cm]{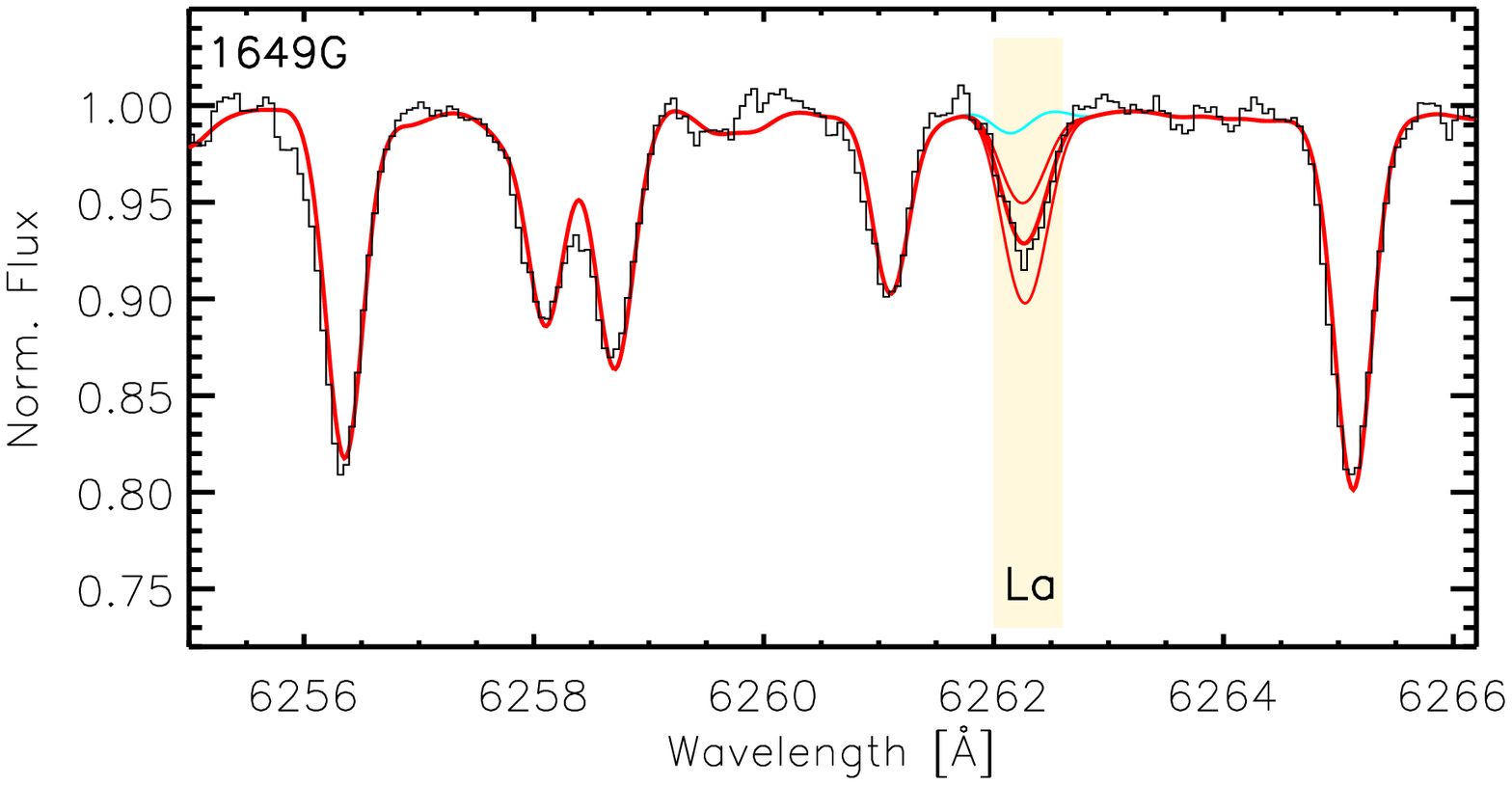}
    \caption{Observed and synthetic spectra around some analysed transitions for two stars observed with UVES (upper panels), and two stars observed with GIRAFFE (lower panels). In each panel the observed spectrum has been represented in black. The cyan spectra have been computed with no contribution from La and Nd, and Cu for stars 859U and 1309U, respectively, and no contribution from O and La for stars 1567G and 1649G, respectively. The thick red line is the best-fitting synthesis; while the red thin lines are the syntheses computed with abundances altered by $\pm$0.2~dex from the best value.}
   \label{fig:synthUVES}
   \end{figure*}

Some examples of our spectral synthesis are plotted in Fig.~\ref{fig:synthUVES} for two stars observed with UVES (859U and 1309U), and two stars observed with GIRAFFE (1567G and 1649G). The represented spectra are shown around some $n$-capture features (La and Nd) for 859U and 1649G, the Cu line at $\sim$5105~\AA\ for the star 1309U and around the forbidden O line for 1567G.
A list of all the derived chemical abundances and adopted atmospheric parameters is provided in Tab.~\ref{tab:gir_abundances} and Tab~\ref{tab:abbUVES} for GIRAFFE and UVES, respectively.

Internal uncertainties in chemical abundances due to the adopted model atmospheres were estimated by varying the stellar parameters, one at a time, by the amounts derived in Sect.~\ref{sec:atm}. 
The internal uncertainty associated with the photometric surface gravities are formally small, so we conservatively adopt an error of 0.2~dex. Thus, we vary
\teff/\logg/[Fe/H]/\vmicro=$\pm$50~K/$\pm$0.16~dex/$\pm$0.05~dex/$\pm$0.11~\kmsec\ for UVES, and 
$\pm$50~K/$\pm$0.20~dex/$\pm$0.05~dex/$\pm$0.20~\kmsec\ for GIRAFFE.
Variations in chemical abundances due to variations in atmospheric parameters are listed in Tab.~\ref{tab:errUVE} and Tab.~\ref{tab:errGIR}. 

In addition to the contribution introduced by internal errors in atmospheric parameters, we estimated the contribution due to the limits of our spectra, e.g. due to the finite S/N, fit quality, which affect the measurements of EWs and the spectral synthesis.
The contribution due to EWs, used in the case of UVES data, has been calculated by varying the EWs of spectral lines by $\pm$4.5~m\AA, that is the typical error associated to our EWs measurements as we have verified by comparing Fe lines for stars with similar atmospheric parameters (e.g. \#1219U and \#579U).
The variations in the abundances obtained by varying the EWs have been then 
divided by the square root of the number of available spectral lines minus one. 
Since the EWs measurement errors are random, the error associated to those elements with a larger number of lines is lower.
For the other elements we have a lower number of lines, so the error contribution introduced by EWs uncertainties is higher.

For GIRAFFE data we derived EWs only for the Ba line at 6141~\AA. To evaluate the error affecting this measurement we re-derive EWs from single (not combined) exposures for three stars observed with GIRAFFE, and derived the error associated with the mean EW obtained for each star. The derived error in EW for this Ba line is $\sim$5~m\AA. The impact of this uncertainty to the Ba abundance has been derived in the same manner as for the UVES data, e.g. by changing the EWs of the Ba line by this quantity and re-determining the abundances. The mean difference in [Ba/Fe] due to a change by 5~m\AA\ in the EWs of the Ba transition is 0.04~dex (see Tab~\ref{tab:errGIR}).  

To estimate the uncertainties introduced by the limited S/N in the fitting procedure used in the spectral synthesis we computed a set of 100 synthetic spectra for two stars representative of the UVES sample (1439U and 859U) and two stars representative of the GIRAFFE sample (527G and 1077G). These set of synthetic spectra were calculated by using the best-fit inferred abundances, and were then degraded to the S/N of the observed spectra. We then analysed the chemical abundances of all these synthetic spectra in the same manner as the observed spectra. The scatter that we obtain from the abundances from each spectral line for a set of synthetic spectra corresponding to a given star, represents a fair estimate of the uncertainty introduced by the fitting procedure, due to the S/N, the pixel size and the continuum estimate. 
These uncertainties strongly depend on the S/N, and are higher for less luminous stars.
Indeed, our set of synthetic spectra has been computed  at two different S/N values, e.g. S/N=80 and S/N=200, representing the lower and higher S/N of our spectra.
These errors are listed as $\sigma_{\rm fit}$ in Tab.~\ref{tab:errGIR}, and Tab.~\ref{tab:errUVE}, for GIRAFFE and UVES, respectively.
Double entries in these errors correspond to the different values obtained at the two different S/N values. Similarly to the discussion for EWs, these errors are random, and the corresponding uncertainty in chemical abundances is lower for those elements with a large number of lines (e.g., Fe). 

All the contributions both from atmospheric parameters and S/N are included in the total uncertainty values $\sigma_{\rm total}$ listed in the last columns of Tab.~\ref{tab:errGIR}, and Tab.~\ref{tab:errUVE}. These total uncertainties have been obtained following the formalism given in Johnson (2002), 
and also account for correlations in the atmospheric parameters determination. For GIRAFFE, correlations between \teff\ and \logg\ are small, and we considered only covariance terms including \vmicro. 

We remark here that we are interested in star-to-star abundance variations. For this reason, we are only marginally interested in external sources of error which are systematic and much more difficult to evaluate. Later in the paper, we will discuss mostly internal uncertainties affecting our abundances, while systematic effects will be discussed only when relevant, e.g. when comparing abundances inferred from two the different analysed data sets (GIRAFFE and UVES).

%
%

\section{The chemical composition of NGC\,5286}\label{sec:abb}

From our abundance analysis NGC\,5286 is a metal-poor GC, with the typical enhancements in $\alpha$-elements (Si, Ca, Ti). 
The mean metallicity obtained from our sample of GIRAFFE probable NGC\,5286 members, composed of 55 stars, is [Fe/H]=$-$1.72$\pm$0.01~dex, with a dispersion $\sigma$=0.11~dex. From the UVES sample, composed of only seven stars, we obtain a mean [Fe/H]=$-$1.80$\pm$0.05~dex, with a similar dispersion, e.g. $\sigma$=0.12~dex.
The systematically lower Fe abundances inferred from UVES can be easily explained by systematic differences in atmospheric parameters, which have been derived in a different manner for the two sets of data. Indeed, as discussed in Sect.~\ref{sec:atm}, spectroscopic \teff\ and \logg\ are lower than the photometric ones. Neutral iron abundances are more sensitive to temperature variations (see Tab.~\ref{tab:errUVE} and Tab.~\ref{tab:errGIR}); so a systematic difference of $\sim$100~K, such as that estimated in Sect.~\ref{sec:atm}, can explain the difference found in the mean Fe abundances from GIRAFFE and UVES data.
We remark here that we are mostly interested in the internal variations in chemical abundances present in the cluster. We are aware of systematics in the abundances derived from GIRAFFE and UVES, which are due in part to the systematics in atmospheric parameters, but also to the different transitions used for the two data-sets, as UVES spectra span a significantly wider range in wavelength.

\begin{figure*}
    \includegraphics[width=16cm]{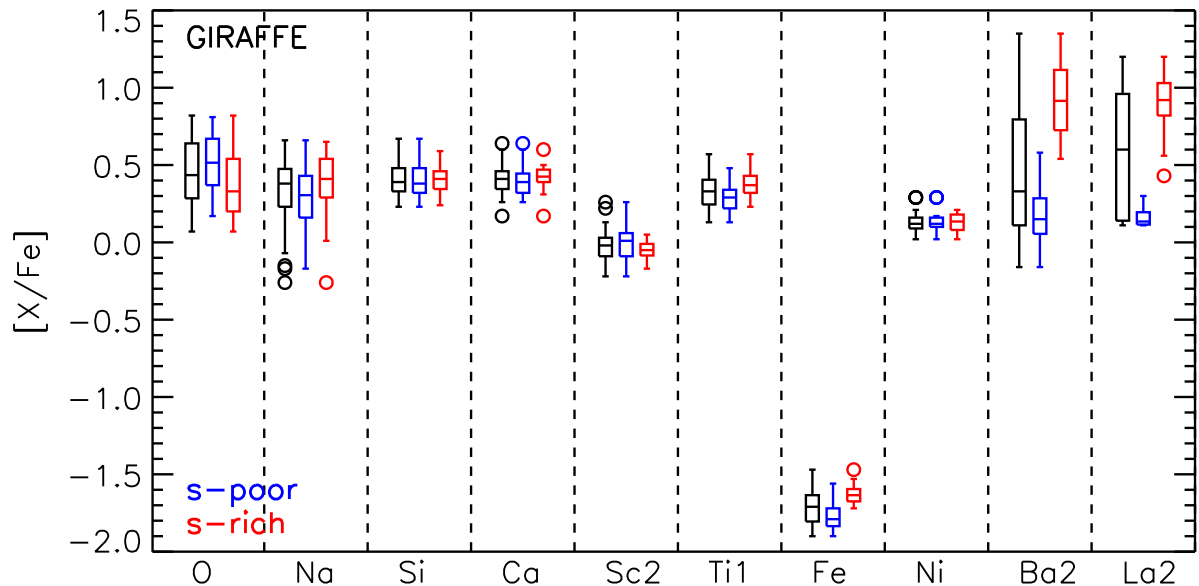}
    \caption{Box and whisker plot of the Fe-poor/$s$-poor (blue) and the Fe-rich/$s$-rich (red) abundances observed with GIRAFFE. The black box includes all the stars of NGC\,5286 stars analysed with GIRAFFE. For the non-Fe species, their [X/Fe] relative abundances are shown, for Fe we plotted [Fe/H]. For a given elements, the box represents the interquartile range (middle 50\% of the data) and the median is indicated by the horizontal line. The vertical tails extending from the boxes indicate the total range of abundances determined for each element, excluding outliers. Outliers (those 1.5 times the interquartile range) are denoted by open circles.}
   \label{fig:boxGIR}
   \end{figure*}

\begin{figure*}
    \includegraphics[width=16cm]{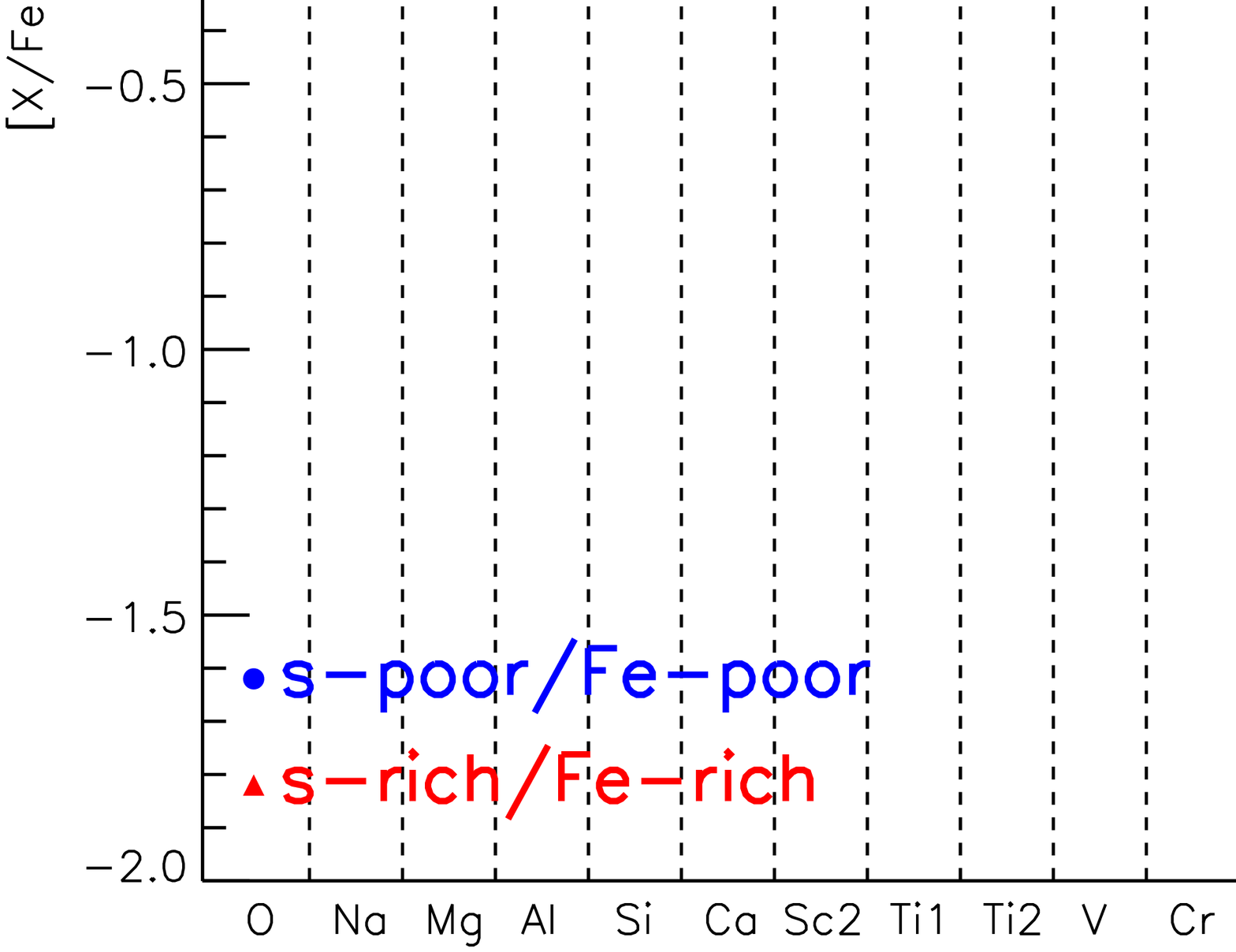}
    \caption{Summary of the abundance results obtained from UVES spectra. For the non-Fe species, their [X/Fe] relative abundances are plotted, for Fe we plotted [Fe/H]. Filled red triangles are used for stars with $s$-process and Fe enhancements, filled blue circles are for stars without such enhancements, while the green cross shows the abundances for the star 1309U, with Fe-rich/$s$-poor composition (see Sect.~\ref{sec:abb} for definitions of these stellar groups).}
   \label{fig:boxUVES}
   \end{figure*}

A summary of our elemental abundance results is shown in Fig.~\ref{fig:boxGIR} and Fig.~\ref{fig:boxUVES} for GIRAFFE and UVES, respectively. 
We anticipate here that in these plots, our abundance results are best represented by dividing the two samples of stars into different groups, having different abundance patterns in $n$-capture elements and overall metallicity.
In the next few subsections we consider and discuss all the abundance trends we observe in NGC\,5286, starting with the $n$-capture process elements.

%
%

%
%

\subsection{The overall metallicity and neutron-capture elements}\label{sec:ncapture}

Figure~\ref{fig:boxGIR} clearly suggests that the chemical elements with the largest internal variations in NGC\,5286 are the two $n$-capture elements Ba and La (see the box representative of the entire GIRAFFE sample represented in black).
Expected observational errors, listed in Tab.~\ref{tab:errGIR}, cannot account for the large internal variations observed in both these elements.
Although both Ba and La are expected to be produced mostly by $s$-process in the solar system, their production can be influenced by the $r$-process at low metallicity (e.g.\,Sneden et al.\,2008).
However, we will refer to them as $s$-process elements because results from UVES suggest that the enrichment in this object has been due primarily to material that has undergone $s$-process nucleosynthesis (see the discussion below in this section). 

\begin{figure*}
   \centering
   \includegraphics[width=11cm]{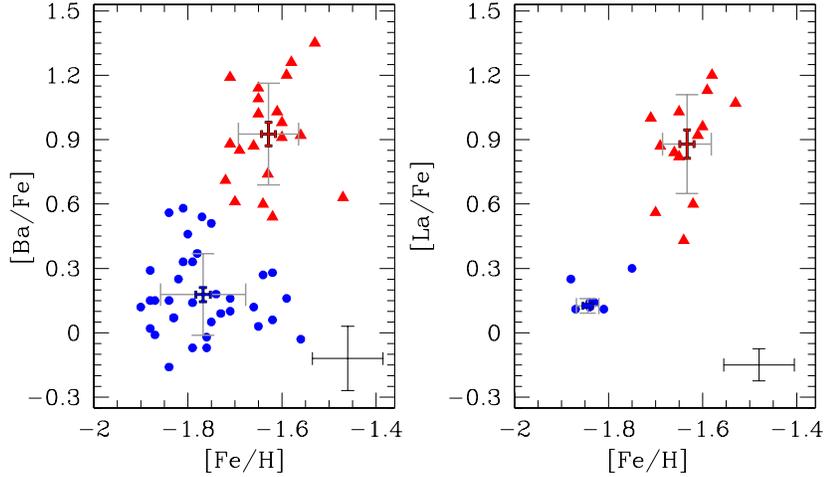}
      \caption{
Abundance ratios [Ba/Fe] and [La/Fe] as functions of [Fe/H] derived from the GIRAFFE sample. In each panel, blue circles represent $s$-poor/Fe-poor stars, red triangles represent $s$-rich/Fe-rich stars.
For each stellar group we show the average values, with associated dispersions (grey bars) and errors (blue and red bars). The typical uncertainty (from Tab.~\ref{tab:errGIR}) associated with single measurements is plotted on the right-bottom corner.
        }
        \label{fig:sgir}
   \end{figure*}

The abundance patterns of Ba and La are clearer when we consider how they vary with the overall metallicity. Indeed, a visual inspection of [Ba/Fe] and [La/Fe] as a function of [Fe/H], represented in Fig.~\ref{fig:sgir}, immediately suggests a complex chemical pattern: overall there is an increase in [Ba/Fe] and [La/Fe] as a function of [Fe/H], with one group of stars showing higher Ba and Fe; stars with lower [Ba/Fe] appear to span a larger range in metallicity.

On the basis of the position of our stars analysed with GIRAFFE in the [Ba/Fe] vs.\,[Fe/H] plane, we selected two different groups of stars (as plotted in Fig.~\ref{fig:sgir}).
Because of the observed variations in Ba (and other $s$-elements as discussed below) and Fe we will refer to our populations as: 
{\it (i)} $s$-rich/Fe-rich for the stars with both higher $s$ and Fe, 
selected as the stars with [Fe/H]$> -$1.73 and [Ba/Fe]$>$0.50 (red triangles);
{\it (ii)} $s$-poor/Fe-poor are all the other stars, having lower $s$ and, on average, lower Fe content (blue circles).
The difference in the chemical composition of these identified stellar groups can be evaluated by their mean abundances listed in Tab.~\ref{tab:meanGIRAFFE}. 
The mean [Ba/Fe] of the $s$-rich/Fe-rich group is a factor of five higher.
Lanthanum abundances confirm the presence of both a $s$-poor/Fe-poor and a $s$-rich/Fe-rich group of stars in NGC\,5286, with the second having an over-abundance in [La/Fe], similar to that present in [Ba/Fe]. 

The $s$-rich/Fe-rich stars show larger dispersions in both Ba and La. 
As listed in Tab.~\ref{tab:meanGIRAFFE}, the rms values for the $s$-poor/Fe-poor stars are $\sigma_{\rm [Ba/Fe]}$=0.19~dex and $\sigma_{\rm [La/Fe]}$=0.03~dex, both significantly lower than those of the $s$-rich/Fe-rich stars, that are $\sigma_{\rm [Ba/Fe]}$=0.24~dex and $\sigma_{\rm [La/Fe]}$=0.23~dex. 
We note that the chemical content of La has been inferred only for a subsample of stars (8 $s$-poor/Fe-poor and 13 $s$-rich/Fe-rich as selected in the [Ba/Fe]-[Fe/H] plane) due to the fact that La lines are much weaker than the Ba line, and has been possible only for the higher-S/N spectra. In particular, the mean and rms values of the La content for the 8 GIRAFFE $s$-poor/Fe-poor stars may not be representative, as we cannot exclude stars with lower contents difficult to be inferred from our limited-S/N data.

The mean difference in Fe between the $s$-rich and the $s$-poor stars is 
$\overline{{\rm [Fe/H]}_{s {\rm -rich}}}$~$-$~$\overline{{\rm [Fe/H]}_{s{\rm -poor}}}$=$+$0.14$\pm$0.03~dex, 
i.e. a difference of a factor of $\sim$1.4, with a significance at the $\sim$4.5~$\sigma$ level (Tab.~\ref{tab:meanGIRAFFE}). 
$s$-poor/Fe-poor stars have a larger scatter in [Fe/H], that is $\sigma_{\rm [Fe/H]}$=0.09, to be compared with the value obtained for the $s$-rich/Fe-rich stars $\sigma_{\rm [Fe/H]}$=0.06~dex.  
Figure~\ref{fig:boxGIR} summarises the chemical abundances in the various analysed elements obtained from GIRAFFE data for the total (black), $s$-poor/Fe-poor and $s$-rich/Fe-rich stellar components of NGC\,5286.
We note that the two AGB observed with GIRAFFE both belong to the $s$-poor/Fe-poor group.

Having identified the main stellar groups by means of the large sample available from GIRAFFE data, we were able to better chemically characterise them by using the higher-resolution and larger-spectral range of the UVES sample. 
Indeed, although the UVES sample is composed of only seven RGBs, all these stars are probable cluster members, as suggested both by RVs and proper motions (see Sects.~\ref{sec:phot_data} and \ref{sec:spec_data}).
The UVES sample was carefully chosen to ensure that stars on both RGBs were selected to allow us to conduct a more detailed chemical characterisation. 
   
From UVES, we infer chemical abundances of many $n$-capture elements, including Y, Zr, Ba, La, Ce, Pr, Nd, Eu. 
Figure~\ref{fig:boxUVES} suggests that these elements, excluding Eu, are those displaying the higher dispersions.
From the UVES results the separation between the $s$-poor stars and the $s$-rich stars is clearer, making the identification of the two $s$-groups straightforward. 

\begin{figure*}
   \centering
   \includegraphics[width=14cm]{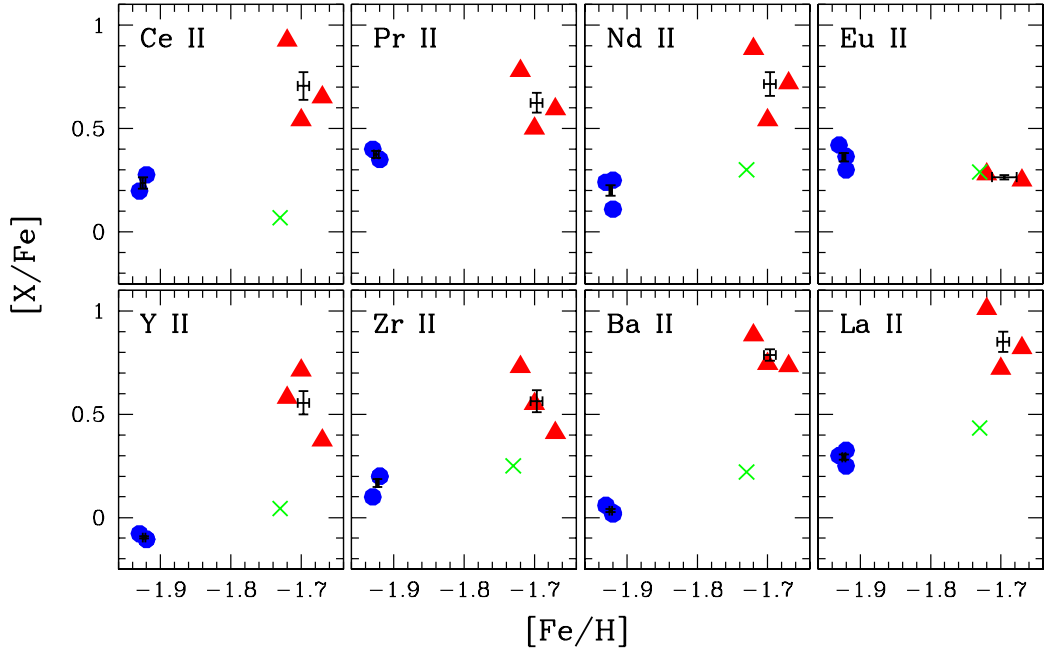}
      \caption{
Summary of the abundance results for $n$-capture process elements obtained from the UVES sample. Abundance ratios of $n$-capture elements from Y to Eu relative to Fe are shown as a function of their [Fe/H] metallicities. The horizontal and vertical ranges are identical in all panels. Symbols are as in Fig.~\ref{fig:sgir}. For the $s$-poor/Fe-poor and $s$-rich/Fe-rich stellar groups we show the means and associated error bars.
        }
        \label{fig:suve}
   \end{figure*}

In Fig.~\ref{fig:suve} we show the abundances of all $n$-capture elements relative to iron as a function of [Fe/H] for the UVES sample. The UVES stars appear to cluster around two different values in all these plots: at a lower and at a higher level of Fe and $n$-capture element contents. One star in our UVES sample has relatively high Fe, but lower content in $n$-capture elements.
Summarising, the UVES sample comprises of three $s$-poor/Fe-poor stars (blue in Fig.~\ref{fig:suve}), three $s$-rich/Fe-rich stars (red in Fig.~\ref{fig:suve}) and one star that apparently stands away from the two main component, being $s$-poor, but higher Fe relatively to the $s$-poor/Fe-poor group. 
The presence of one UVES star with low $s$-elements, and relatively Fe-rich may suggest that a minor stellar component, that is Fe-rich and $s$-poor may be present in NGC\,5286. We note that this would confirm the higher dispersion in [Fe/H] that we found for the $s$-poor/Fe-poor group in the GIRAFFE sample. 
The possible presence of this minor stellar component will be discussed in more details in Sect.~\ref{sec:thirdgroup}. 

The average UVES abundances obtained for the $s$-poor/Fe-poor and the $s$-rich/Fe-rich group, excluding the $s$-poor star with relatively high Fe,  are listed in Tab.~\ref{tab:meanUVES}.
The differences between the mean content in $n$-capture elements between the $s$-poor/Fe-poor and the $s$-rich/Fe-rich exceed a 3~$\sigma$ level for the abundance ratios of Y ($\gtrsim$4.5~$\sigma$), Ba ($\gtrsim$9~$\sigma$), La ($\gtrsim$4~$\sigma$).  
The mean abundance of Zr, Ce and Nd over Fe are also higher for the $s$-rich/Fe-rich stars at a level of $\sim$2.5~$\sigma$. 
Praseodymium is mildly enhanced in the $s$-rich/Fe-rich stars too, but the difference with the $s$-poor/Fe-poor group is at a $\sim$1.7~$\sigma$ level. 
[Eu/Fe] reverses the general trend displayed by the other $n$-capture elements, as it is slightly lower in the $s$-rich/Fe-rich stars. However, the difference in [Eu/Fe] is only $<$0.10~dex, and it is significant at a $\sim$1.5~$\sigma$ level. To detect a possible small difference in this element (if any) higher quality data (in terms of resolution and S/N) is required. We note here that a similar (low significance) small difference in Eu in the same sense has been observed in M\,22 (Marino et al.\,2011). 
For the moment, we assume that, at odds with the other $n$-capture elements, Eu does not show any strong evidence for internal increase in NGC\,5286. 

From UVES data, the [Ba/Fe] abundance is almost a factor of 6 higher in the $s$-rich/Fe-rich stars, similarly to that inferred from GIRAFFE data. Lanthanum abundance is a factor of $\sim$4 higher in the $s$-rich/Fe-rich group.
UVES data also confirm the difference in metallicity among different stellar groups, with the $s$-rich/Fe-rich stars again being enhanced in Fe by a factor of $\sim$1.5.

\begin{figure}
   \centering
   \includegraphics[width=8.5cm]{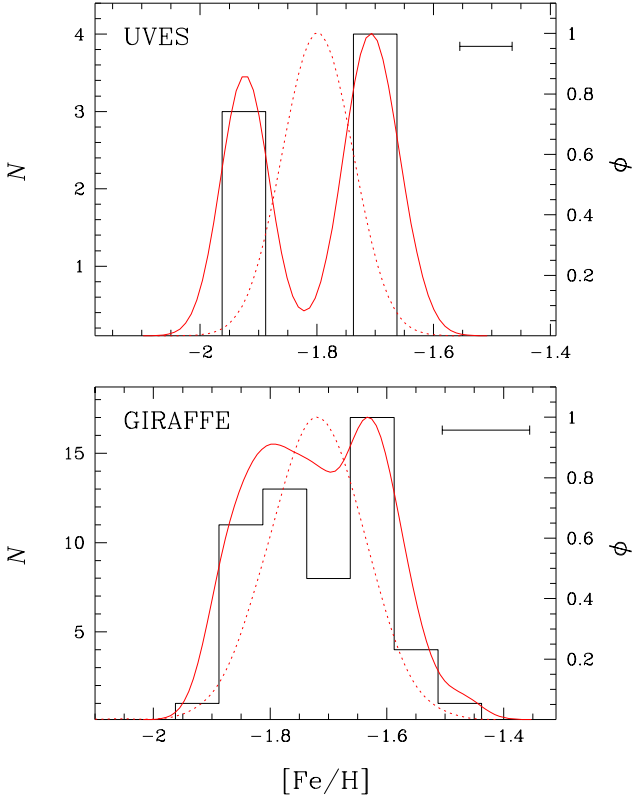}
      \caption{
Histograms of [Fe/H] distribution for the GIRAFFE (lower panel) and UVES (upper panel). The measurement error in [Fe/H] is plotted in the upper-left corner. The red continuous lines are the normalised kernel density distributions of the observed metallicities, while the red dotted lines are the normalised kernel-density distributions corresponding to measurement errors only.
 To derive each kernel density distribution we used a Gaussian Kernel and  a dispersion equal to the measurement error.
        }
        \label{fig:histoFe5286}
   \end{figure}

The histogram distribution of the [Fe/H] values obtained from GIRAFFE and UVES data is represented in Fig.~\ref{fig:histoFe5286}. These Fe distributions alone suggest the presence of a genuine Fe spread in NGC\,5286. The kernel-density distributions corresponding to the observed data strongly differ from the distribution for a mono-metallic GCs expected from our observational errors. The probability that the $s$-poor and $s$-rich stars in the GIRAFFE sample come from the same parent distribution is $\sim 10^{-7}$, as verified by computing a Kolmogorov-Smirnov test.

\begin{figure}
   \centering
   \includegraphics[width=8cm]{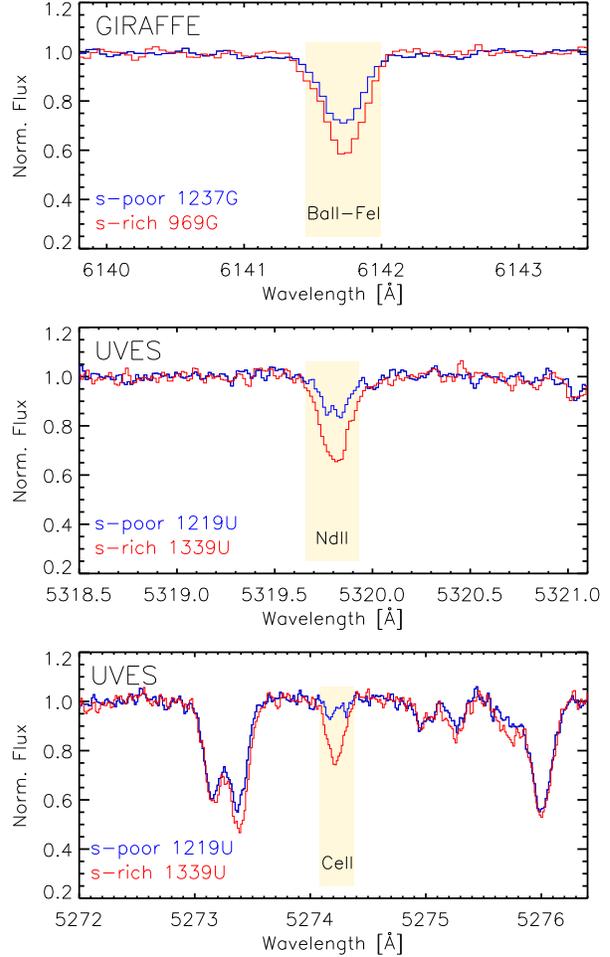}
      \caption{Observed spectra around some analysed $n$-capture transitions for two stars observed with GIRAFFE (upper panel), and two stars observed with UVES (lower panels). In each panel we represent pairs of stars with similar atmospheric parameters, so that the difference in the represented lines (Ba, Nd, Ce) are due almost entirely to a different chemical content in these elements. The blue spectrum represents a star belonging to the $s$-poor group, the red spectrum to a star belonging to the $s$-rich one.                }
        \label{fig:spectra_s}
   \end{figure}

In Fig.~\ref{fig:spectra_s} we show some Ba, Ce and Nd transitions in stars with similar atmospheric parameters but very different derived $s$-elements chemical contents. The $s$-rich GIRAFFE star 969G clearly has a much stronger Ba line $\lambda$6141~\AA\ than does the $s$-poor GIRAFFE star 1237G, and the same is observed in the pair of UVES stars 1339U and 1219U. Inspection of other contrasting pairs of stars and other spectral lines yields the same conclusion. The average chemical abundances for the $s$-poor and the $s$-rich stellar groups of NGC\,5286 are listed in Tab.~\ref{tab:meanGIRAFFE} and Tab.~\ref{tab:meanUVES}, for GIRAFFE and UVES, respectively.

%
%

%
%

%
%

%
%

\subsection{The $p$-capture elements and light-elements (anti-)correlations}\label{sec:pcap}

The elements we have inferred that can be affected by proton-capture ($p$-capture) reactions include O and Na analysed from both UVES and GIRAFFE, and Mg and Al, available only for UVES data. 
All of these elements show internal dispersions larger than those expected from observational errors, suggesting that NGC\,5286 shares with the typical Milky Way GCs the presence of light-elements variations (see Tab.~\ref{tab:meanGIRAFFE} and Tab.~\ref{tab:meanUVES}).
For most elements (including O, Na and Al), internal dispersions remain high even when the sample is separated into $s$-poor and $s$-rich groups.

\begin{figure}
   \centering
   \includegraphics[width=8.1cm]{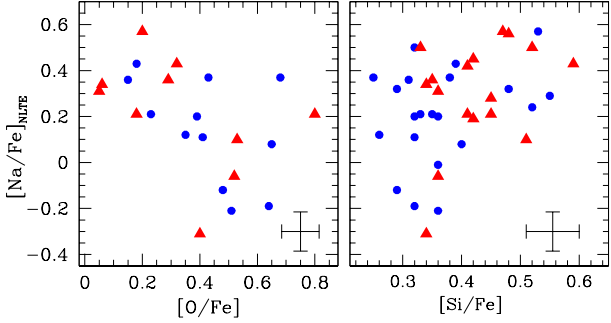}
      \caption{[Na/Fe] abundances corrected for NLTE as a function of [O/Fe] (left panel) and [Si/Fe] (right panel). Symbols and colours are a in Fig.~\ref{fig:sgir}.}
        \label{fig:pgir}
   \end{figure}

In Fig.~\ref{fig:pgir}, we show [Na/Fe] vs.\,[O/Fe] (left) and vs.\,[Si/Fe] (right) for the GIRAFFE sample.
These data suggest that both the $s$-rich/Fe-rich and the $s$-poor/Fe-poor groups independently exhibit an O-Na anticorrelation. Additionally, the $s$-rich/Na-rich stars typically have higher Si content, and a subsample of the $s$-poor/Na-rich stars are also Si-richer.
As shown in Fig.~\ref{fig:sgir}, no obvious correlations are present between O and Na with $s$-elements Ba and La, although the mean [Na/Fe] and [O/Fe] are respectively higher and lower in the $s$-rich stars, but these differences have only a 1~$\sigma$ significance.

The smaller sample of UVES stars confirms the presence of a O-Na anticorrelation, showing also a well-defined Na-Al correlation (Fig.~\ref{fig:puves}), with variations in these elements internally present in both the $s$-groups (see upper-right panel in Fig.~\ref{fig:puves}).
Despite clear variations in Al, which correlates with Na, also implying a O-Al anticorrelation, there is no strong evidence for a Mg-Al anticorrelation, at least from our relatively small sample of UVES targets. 
We note, however, that the star with the highest Al also has the lowest Mg and low O, and possibly a larger sample of stars with Al and Mg abundances can reveal the presence of clear Mg-Al anticorrelation.
For now we can say that the possible lack of a clear Mg-Al anticorrelation does not necessarily mean that $p$-captures on Mg are ruled out. 
Our average abundances, as listed in Tab.~\ref{tab:meanUVES}, suggests that Mg is slightly depleted in $s$-rich stars; however we notice that, given the associated errors, this difference is marginal.
If we suppose that the higher observed Mg abundances are representative of ``primordial'' NGC\,5286 material (that is, prior to any $p$-capture synthesis events), and if primordial Al is indicated by the lower observed Al abundances ($\sim$0.2~dex), then for this material [Mg/Al]$\sim +$0.4. 
Then, if (for example) 10\% of this Mg were to be converted to Al by $p$-capture in the primordial material, the resulting Al would go up by a factor of four, nearly the range covered by our data. The 10\% decrease in Mg would be difficult to detect. Additionally, if the ab initio abundance of Mg contains substantial amounts of ${\phantom{}^{25}}$Mg and/or ${\phantom{}^{26}}$Mg, then the final Al abundance would be even larger after $p$-captures.
It is worth noticing that similar weak Mg dependence on $p$-capture elements, such as Na and Al, are present also in other ``anomalous'' GCs, such as $\omega$~Centauri (Norris \& Da Costa\,1995), M\,22 (Marino et al.\,2009, 2011) and M\,2 (Yong et al.\,2014). 

%
%

%
   \begin{figure*}
   \centering
   \includegraphics[width=11cm]{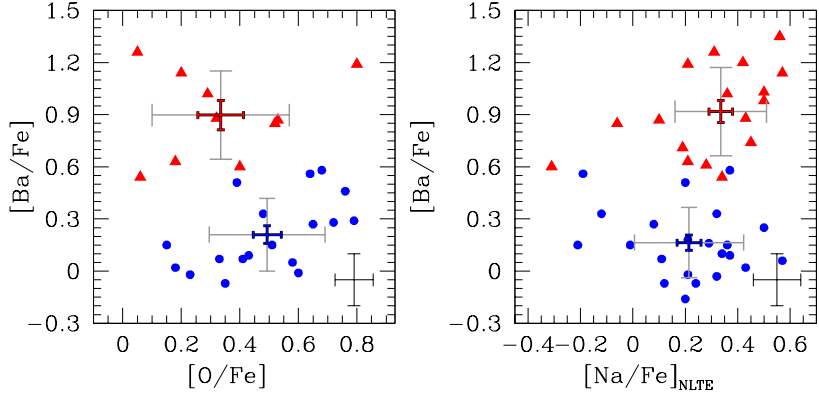}
      \caption{Abundance ratios [Ba/Fe] as functions of [O/Fe] and [Na/Fe]$_{\rm NLTE}$ derived from the GIRAFFE sample. In each panel, blue circles represent $s$-poor/Fe-poor stars, red triangles represent $s$-rich/Fe-rich stars.
For each stellar group we show the average values, with associated dispersions (grey bars) and errors (blue and red bars). The typical uncertainty (from Tab.~\ref{tab:errGIR}) associated with single measurements is plotted on the right-bottom corner.}
        \label{fig:psgir}
   \end{figure*}
%

%
   \begin{figure}
   \centering
   \includegraphics[width=8cm]{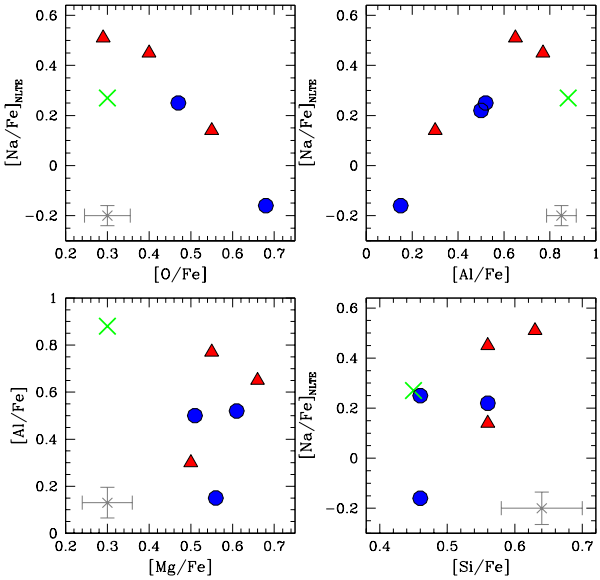}
      \caption{Abundance patterns for elements involved in hot H-burning for stars analysed with UVES spectra. The upper panels show the typical O-Na anticorrelation and a clear Na-Al correlation.
        Mg and Al do not show any strong evidence for anticorrelation, within observational errors (bottom-left panel); while Si is mildly correlated with Na (bottom-right panel). Error bars in the bottom corners of each panel represent estimated errors for single measurements. Symbols and colours are as in Fig.~\ref{fig:sgir}.
        }
        \label{fig:puves}
   \end{figure}
%

\subsection{The Fe-peak elements}

Chemical abundances for Fe-peak elements V, Cr, Mn, Co, Ni and Zn relative to Fe do not show any evidence for internal variations between the two groups, exceeding a 1$\sigma$ level (see Fig.~\ref{fig:boxGIR} and Fig.~\ref{fig:boxUVES}, and values listed in Tab.~\ref{tab:meanGIRAFFE} and Tab.~\ref{tab:meanUVES}).
As is the case with low-metallicity field stars (Sneden et al. 1991; Mishenina et al. 2002) and GCs (Simmerer et al. (2003), copper is under-abundant in NGC\,5286.
However, [Cu/Fe] may vary slightly in concert with the $s$-process elements, being higher in the $s$-rich than the $s$-poor group by [Cu/Fe]=$+$0.18$\pm$0.09 (Tab.~\ref{tab:meanUVES}). 
The difference among the two groups is only at a 2~$\sigma$ level, and, given the uncertainties associated with individual Cu abundance measurements and the low statistics available (three $s$-poor and three $s$-rich stars observed with UVES), interpretation of this difference should be viewed with caution. 

\subsection{The CMD/chemistry connection}\label{sec:cmd}

The analysis of our ground-based photometric data shows that multiple branches (both on the RGB and the SGB) are present in the CMD of NGC\,5286.
Piotto et al.\,(2012) have used multi-wavelength {\it HST} photometry to demonstrate that NGC\,5286 hosts a broad SGB with at least two components that correspond with two main stellar populations. A bright SGB, which hosts about 86\% of SGB stars, and a faint-SGB component made of $\sim$ 14\% of stars. As discussed in Sect.~\ref{sec:phot_data}, by using the $U$ filter, we have identified a split in the RGB in the $(U-V)$ colour. The double RGB seems to merge in the broad SGB of NGC\,5286, with the red-RGB connected with the faint part of the SGB, and the blue-RGB associated with the bright SGB, in close analogy with what observed in NGC\,1851 and M\,22 (Lee et al.\,2009; Marino et al.\,2011).

Multiple SGBs and RGBs can be detected in the CMD and two-colour diagrams of most GCs only when appropriate combination of ultraviolet colours and magnitudes are used. In {\it anomalous} GCs, multiple SGBs are clearly visible in CMDs made with visual filters only. Furthermore, in the $U$ vs.\,($U-V$) CMD of anomalous GCs, the faint and the bright SGB evolve into the red and blue RGB and are made of metal/$s$-rich and metal/$s$-poor stars, respectively (e.g. Marino et al.\,2012).

\begin{figure*}
    \includegraphics[width=15cm]{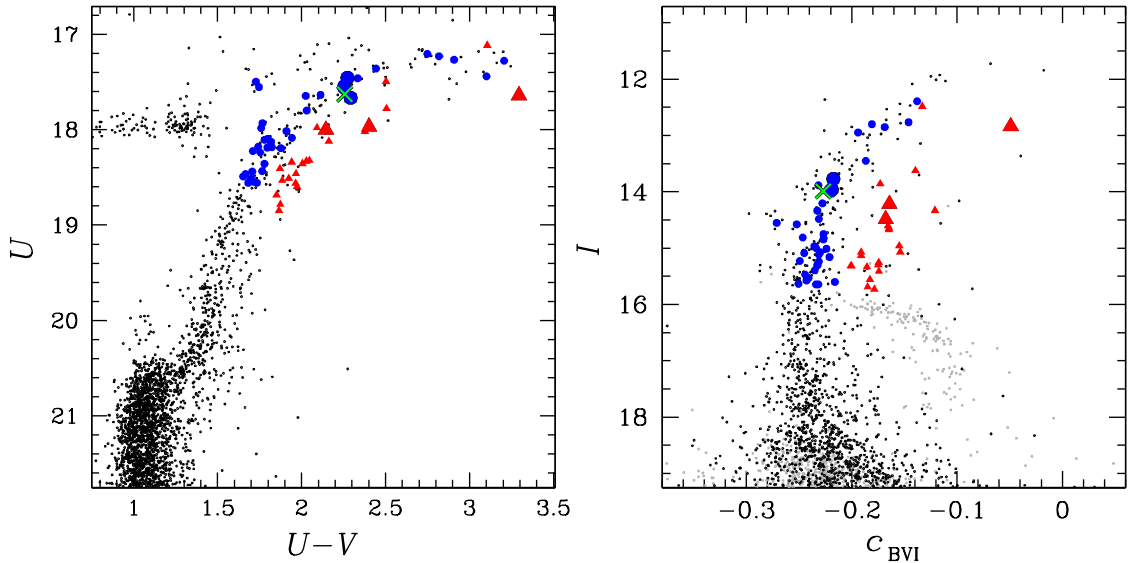}
    \caption{$U$-$(U-V)$ CMD (left panel) and $I$-$c_{BVI}$ diagram (right panel)
      for NGC\,5286, corrected for differential reddening. The grey
      symbols on the right panel represent HB stars. Spectroscopic
      data are superimposed with $s$-rich/Fe-rich stars represented by
      red triangles, $s$-poor/Fe-poor stars by blue circles, and
      the $s$-poor/Fe-rich UVES star by a green cross, according
      with the other figures. 
    }
   \label{fig:cmd_s}
   \end{figure*}

The location of our spectroscopic sample on the $U$-$(U-V)$ CMD, suggests that the two identified RGBs are populated by stars belonging to the two groups with different Fe and $s$-elements (left panel of Fig.~\ref{fig:cmd_s}).
Among GCs, the same behaviour observed in Fig.~\ref{fig:cmd_s} for NGC\,5286, is seen on the CMD of M\,22.
Similar to NGC\,5286, M\,22 also has a similar RGB-SGB split in the $U$-$(U-V)$ CMD and spectroscopy on the SGB has demonstrated that the faint SGB stars are more enriched in $s$-elements compared to bright SGB stars (Marino et al.\,2012).
We emphasis that the difference in the $s$-element content does not directly affect the separation of the two RGBs along the $U$-$(U-V)$ CMD. Additionally, the observed variation in metallicity cannot account for the relatively large separation in colour among the Fe-$s$-rich and the Fe-$s$-poor RGBs. It is tempting to speculate that internal variations in the overall C+N+O abundance, together with iron variations, are responsible for the SGB split as shown by Marino et al.\,(2012) for the case of M\,22. The fact that the faint and the bright SGB of NGC\,5286 are consistent with two stellar populations with different C+N+O has been already shown by Piotto et al.\,(2012) on the basis of their comparison of isochrone and {\it HST} photometry. Spectroscopic measurement of C, N, O in NGC\,5286, together with the iron measurements provided in this paper, are mandatory to understand if the SGB and RGB morphology of this cluster can be entirely explained in terms of metallicty and C+N+O.
Here we can conclude that, as the two RGBs {\it evolve} from a spread SGB in the $U$-$(U-V)$ CMD, the SGB morphology of NGC\,5286 is indirectly connected with the presence of the two stellar groups with different chemical composition: the bright SGB is composed of $s$-poor stars, and the faint SGB of $s$-rich stars.

Previous papers have shown that a number of colours or photometric indices based on ultraviolet and far-ultraviolet photometry are very effective for detecting  multiple sequences along the RGB. These include the $(U-B)$ colour (Marino et al.\,2008), the Str{\" o}mgren index $c_{y}=c_{1}-(b-y)$ (Grundahl\,1999; Yong et al.\,2008), and the $c_{\rm F275W, F336W, F438W}=(m_{\rm F275W}-m_{\rm F336W})-(m_{\rm F336W}-m_{\rm F438W})$ and $c_{\rm F336W, F438W, F814W}$ `($c_{\rm U, B, I}$)'=$(m_{\rm F336W}-m_{\rm F438W})-(m_{\rm F438W}-m_{\rm F814W})$ indices introduced by Milone et al.\,(2013) by using {\it HST} filters F275W, F336W$\sim U$, F438W$\sim B$, and F814W$\sim I$. 
An appropriate combinations of $U$, $B$, $I$ ground-based photometry can efficiently separate multiple populations with different content of nitrogen (through the CN line at $\sim$3300~\AA) and helium (Marino et al.\,2008; Milone et al.\,2012, 2013; Monelli et al.\,2013).

However, since O-Na-Al (anti-)correlations are present among both $s$-rich and $s$-poor stars of NGC\,5286, an index that is sensitive to the light-element patterns, like the $c_{\rm U, B, I}$ index, is not able to provide a clear separation between the two $s$ and Fe groups of stars hosted in the cluster. 
We have defined a new photometric index based on a combination of $B$, $V$, and $I$ magnitudes, $c_{BVI}$=$(B-V)-(V-I)$ , that maximises the separation between the $s$-poor and the $s$-rich groups of NGC\,5286 (right panel of Fig.~\ref{fig:cmd_s}). 
In contrast with the $c_{\rm U, B, I}$ index, the newly defined  $c_{\rm B, V, I}$ index,  is largely insensitive to nitrogen variations on the atmosphere of the stars and provides a clear separation between $s$-rich and $s$-poor stars of NGC\,5286. This is similar to what observed in M\,22, i.e.\, the $s$-rich and $s$-poor stars define two RGBs in the $V$ versus $m_{1}$ diagram (Marino et al.\,2011). 

While a difference in age may account for the SGB structure of NGC\,5286, it cannot reproduce the large split seen in the $c_{BVI}$ index (Marino et al., in prep.), and also the observed difference in metallicity is too small to cause the wide separation observed on the $I$-$c_{BVI}$ diagram.
As the $s$-process abundances are not directly affecting broad band colours,
this is another indication that the major cause of the observed split in the $c_{BVI}$ is likely to be the presence of metallicity and C+N+O variations among the $s$-poor/Fe-poor and the Fe-rich/$s$-rich. 
Future spectroscopic investigations may be enlightening in this regard, and should prove if C+N+O variations exist in NGC\,5286, further constraining the nature of the polluters.
We note here that the observed variation in the overall metallicity is not able to reproduce such a large split in $(U-V)$ and $c_{BVI}$.
For M\,22 we have demonstrated that the observed difference in C+N+O (plus the variation in metallicity) can account for the entire split SGB/RGB without any significant variation in age among the $s$-poor (bright SGB) and the $s$-rich stars (faint SGB).
Even if we could not measure abundances of C and N for our stars in NGC\,5286, given the similar chemical abundance and photometric patterns shared with M\,22, it is tempting to speculate that internal variations in C+N+O are present also in this GC and can account (in part or totally) for its SGB split. 
We note that, in order to determine relative ages among the two SGB populations in NGC\,5286, future studies that measure the total C+N+O of the two $s$-groups is essential.

%
%

\subsection{On the possible presence of a metallicity spread in the $s$-poor/Fe-poor group}\label{sec:thirdgroup}

From Fig.~\ref{fig:sgir} we note that although on average, the $s$-poor group is more Fe-poor than the $s$-rich group, there are some $s$-poor stars with [Fe/H] similar to the $s$-rich group. 
Also in the UVES sample, one star shows a similar behaviour and we preferred not to include this star in either of the two main $s$-groups. 
The possible presence of a minor third stellar component or a metallicity spread in the $s$-poor group warrants a detailed discussion, and we have performed additional tests to investigate whether this group is present in NGC\,5286, or if the spread in [Fe/H] is merely due to observational errors.

\begin{figure*}
    \includegraphics[width=15cm]{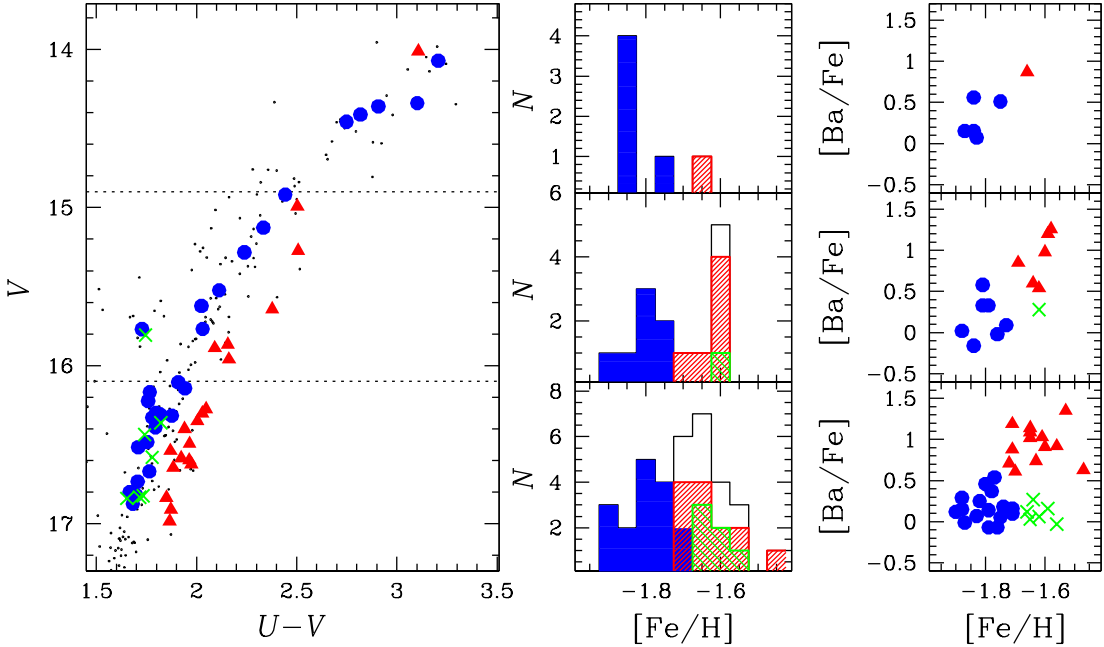}
    \caption{Left panel: $V$-$(U-V)$ CMD (left) of NGC\,5286, corrected for differential reddening. Spectroscopic GIRAFFE targets are superimposed with $s$-rich/Fe-rich stars represented by red triangles, $s$-poor/Fe-poor stars by blue circles, and possible $s$-poor/Fe-rich stars by green crosses. Middle panels: histograms of [Fe/H] for GIRAFFE stars in the three different bins in $V$, represented on the $V$-$(U-V)$ CMD. Right panels: [Ba/Fe] vs.\,[Fe/H] for stars in the three defined bins in $V$ mag. Stars belonging to different Fe/$s$-groups have been represented with different colours. The sources for the photometric data are given in Sect.~\ref{sec:phot_data}.}
   \label{fig:3bin}
   \end{figure*}

First we have inspected the position of these stars in the CMD.
As shown in Fig.~\ref{fig:3bin}, we have selected a group of GIRAFFE stars, composed of seven objects, showing lower $s$-content ([Ba/Fe]$<$0.40) and higher Fe ([Fe/H]$> -$1.70; green crosses in the [Ba/Fe] vs.\,[Fe/H] plane).
The position of these stars in the $V$-$(U-V)$ CMD is shown in the left panel of Fig.~\ref{fig:3bin}. Clearly, the main $s$-poor and $s$-rich groups define two different branches on this CMD (as discussed in Sect.~\ref{sec:cmd}), but the seven stars with higher Fe and lower $s$-content do not appear to define distinct branches, and their position is consistent with that of the RGB as defined by the $s$-poor stars.

All seven GIRAFFE stars are at fainter magnitudes, and none of them is observed in the upper RGB, so it is tempting to state that  the higher Fe abundances inferred for these fainter stars could be due to larger observational errors.
On the middle and right panels of Fig.~\ref{fig:3bin}, we note that in the lowest luminosity bin the dispersions in [Fe/H] are higher than those in the middle bin, both for the $s$-rich and the $s$-poor stars, suggesting that the internal errors are higher, as expected.
However, although we expect a larger dispersion due to the lower S/N, it is difficult to explain with just errors the presence  of these stars that reside preferentially on one side of the [Fe/H] distribution. 
On the other hand, as stars with different metallicity populate different RGB sequences, we cannot exclude selection effects in our sample that have selected preferentially stars belonging to the main $s$-rich and $s$-poor RGBs at brighter magnitudes where the separation among them is higher. 

\begin{figure}
   \centering
   \includegraphics[width=7.8cm]{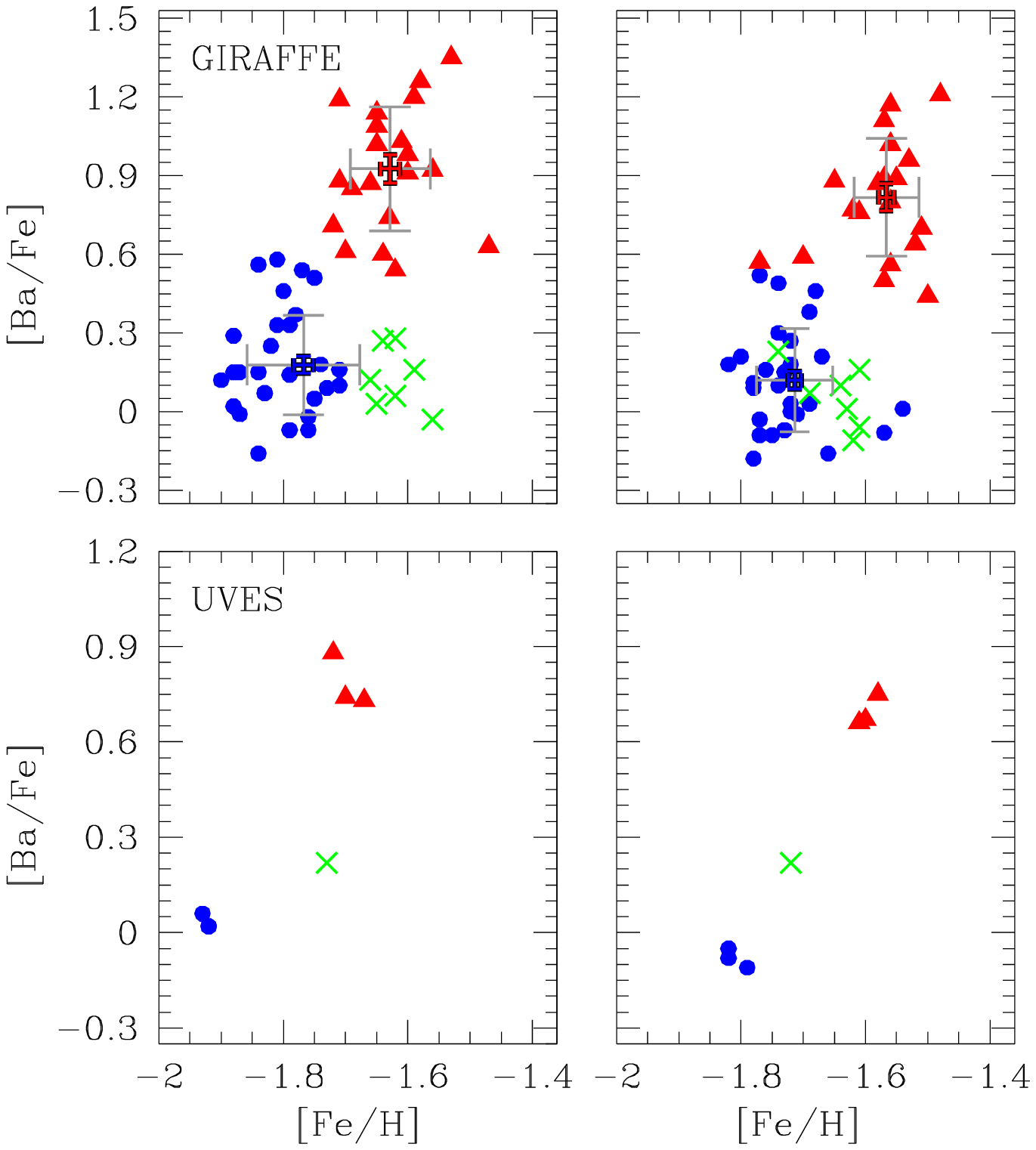}
      \caption{Abundance ratios [Ba/Fe] as functions of [Fe/H] for GIRAFFE (upper panels) and UVES data (lower panels). In each panel, blue circles represent $s$-poor/Fe-poor stars, red triangles represent probable $s$-rich/Fe-rich stars.
The assumed abundances, from atmospheric parameters derived as explained in Sect.~\ref{sec:atm}, are shown on left panels, while abundances derived by assuming parameters from isochrones are displayed on the right panels.
For GIRAFFE we show, for the $s$-poor (including the probable stars with higher Fe) and the $s$-rich groups, the average values, associated dispersions (grey bars) and errors (blue and red bars). 
        }
        \label{fig:test}
   \end{figure}

We recall that internal errors in atmospheric parameters are not able to produce the large variations we observe in $s$-elements, such as barium, but may be more important for [Fe/H] that shows much smaller variations.
To check this possibility, we re-determined Fe and Ba for the entire GIRAFFE and UVES samples of stars by using a different set of atmospheric parameters. Specifically, we used effective temperatures from isochrone fitting, by projecting spectroscopic targets on the isochrone on the $V$-$(V-I)$ CMD. Surface gravities and microturbolent velocities were determined as explained in Sect.~\ref{sec:atm}. Results obtained by using parameters based on isochrones are shown in Fig.~\ref{fig:test} for GIRAFFE (upper panels) and UVES (lower panels). Overall, for the GIRAFFE sample we note that by using atmospheric parameters from isochrones we get higher precision. This is not surprising as the internal photometric errors cancel out by projecting on a fiducial or isochrone. On the other hand, in cases like NGC\,5286 the use of one single isochrone is not appropriate: it can 
affect either, or both, of the two populations, e.g., introducing spurious abundance patterns and decreasing the accuracy. 
A few stars, due to observational errors, migrate from one group to the other, but apparently there is still a tail of $s$-poor stars with slightly higher metallicity.

Due to higher resolution and spectral coverage, UVES data provide more precise results. Adopting photometric stellar parameters based on isochrones, instead of purely spectroscopic parameters, we note that the star 1309U (green cross) approaches the metallicity of the $s$-poor stars, but still it is slightly metal rich.
Our data set suggest that the $s$-poor group may be not homogenous in metallicity, but since the variation in metals is small, and not coupled with any large variation in other elements, such as the $s$-process elements, larger sample of these stars analysed with high-resolution/S/N data are required to definitively assess this issue.

%
%

%
%

\section{The chemical pattern of NGC\,5286 and other anomalous GCs compared to field stars}

\begin{figure*}
   \centering
   \includegraphics[width=13cm]{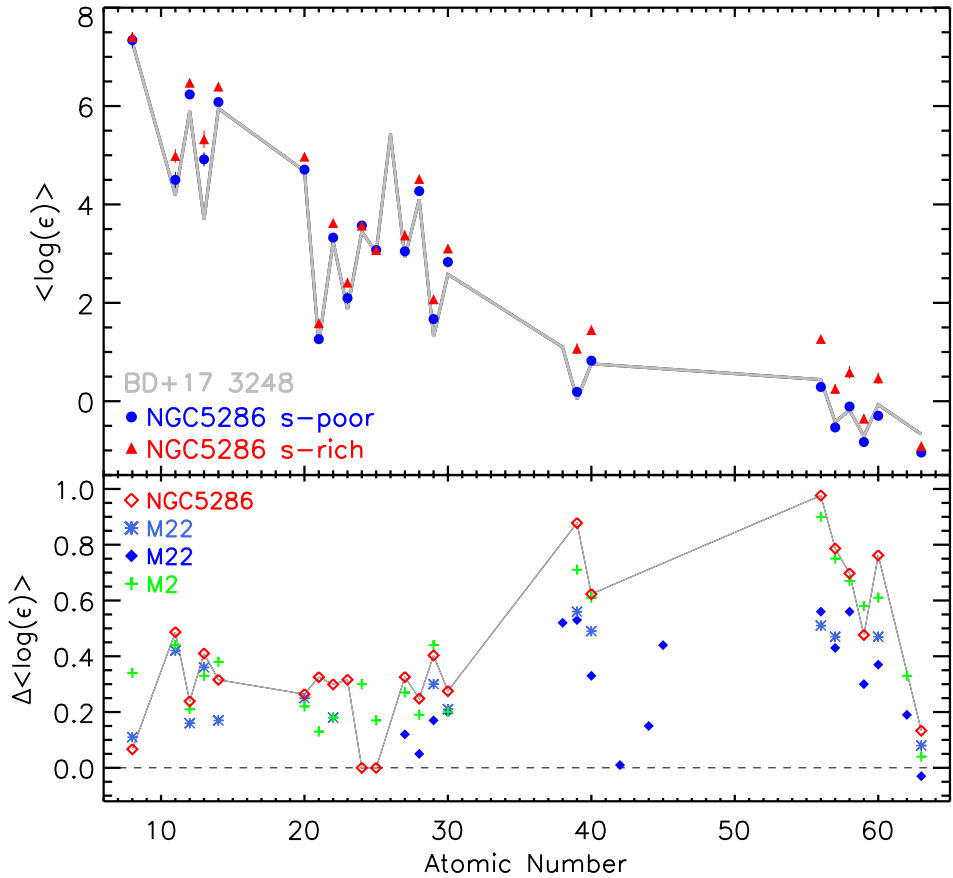}
      \caption{Top panel: Mean logarithmic abundances for the three $s$-poor stars (blue circles) and the three $s$-rich stars in NGC\,5286 (red triangles). The grey line illustrates the abundances in the $r$-process standard star BD+17$^{\circ}$~3248 from Cowan et al.\,(2002). Bottom panel: Differences in these mean abundances for the two $s$-groups ($s$-rich$- s$-poor) in NGC\,5286 (red symbols), M\,22 (blue stars from Marino et al.\,2009, 2011; filled blue symbols from Roederer et al.\,2012), and M\,2  (green crosses from Yong et al.\,2014). The dashed line indicates zero difference.
        }
        \label{fig:rs}
   \end{figure*}

The dominant feature of the chemical pattern of NGC\,5286 is the presence of two main stellar groups with different metallicities and the $n$-capture elements abundances.
In Fig.~\ref{fig:rs} we show the mean logarithmic abundances for the three $s$-rich stars (red) and the three $s$-poor stars (blue) observed with UVES.

If we suppose that the abundances of the $s$-poor group are representative of ``primordial'' NGC\,5286 $n$-capture element material (that is, prior to any internal $n$-capture synthesis event), then we can compare the abundance patterns of this stellar group with those of BD+17$^{\circ}$~3248, that is a $r$-process standard star whose metallicity is only a factor of $\sim$2 lower than stars in NGC\,5286 (Cowan et al. 2002; Roederer et al. 2011).
The $s$-poor stars of NGC\,5286 (blue circles in Fig.~\ref{fig:rs}) have element abundance patterns very similar to those in BD+17$^{\circ}$~3248. The larger differences are in Mg and Al, that appear higher in NGC\,5286, and in Eu which is lower{\footnote{Note that each $s$-group has its own internal variations in $p$-capture elements, so only on a first approximation is the $s$-poor composition representative of the primordial material.}}. Overall, the chemical pattern of the $s$-poor group in NGC\,5286 is well approximated by the $r$-process standard BD+17$^{\circ}$~3248.

Larger differences are observed between the abundance pattern of the $s$-rich group and BD+17$^{\circ}$~3248.
The mean chemical abundances for most of the analysed elements with $Z>39$ are  higher than in the $r$-process standard star, with the most significant differences in the first peak element Y, and the second peak elements Ba and La (see Sect.~\ref{sec:ncapture}), while Pr shows more modest variations. The exception among these heavy elements is Eu, which does not show evidence for any enrichment in the $s$-rich stars.  

\begin{figure}
   \centering
   \includegraphics[width=8.5cm]{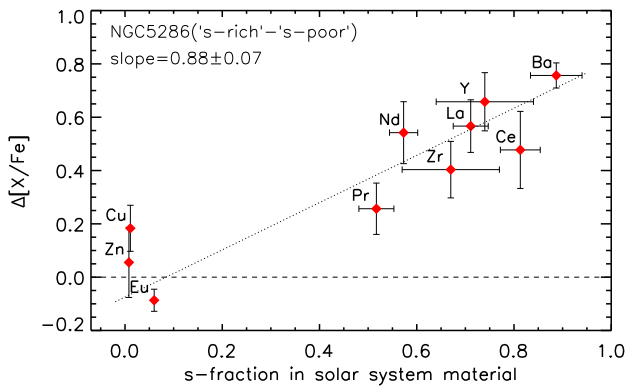}
      \caption{Differences in the mean abundances relative to Fe between the $s$-rich and  the $s$-poor group of NGC\,5286 as a function of the fraction of each element attributed to an $s$-process origin in solar material (Travaglio et al. 2004; Bisterzo et al. 2011). The dashed line indicates zero difference. We overplot the linear fit to the data and write the slope and associated error.         }
        \label{fig:solar}
   \end{figure}

The solar-system abundances of $n$-capture elements showing internal variations in NGC\,5286 are largely produced by the $s$-process: Y (72\%), Zr (81\%), La (75\%), Ce (81\%), and Ba (85\%) (e.g., Simmerer et al.\,2004; Sneden et al.\,2008). 
The origins of Pr and Nd in the solar system are attributed to both the $r$- and $s$-processes in similar proportions: 51\% $r$-process, 49\% $s$-process for Pr, and 42\% $r$-process, 58\% $s$-process for Nd.
We note however that at metallicities significantly lower than solar, as in this case, the relative production of $n$-capture  elements by the $r$- or $s$-processes may be different. 
Among the inferred $n$-capture elements, Eu is the one with the most sharply contrasting solar-system $s$-process/$r$-process origins, as it is only 3\%\ $s$-process. 
The fact that Eu does not show any strong evidence of internal variation suggests that the chemical enrichment in NGC\,5286 has been mostly due to the $s$-process, rather than the $r$-process. 
The excess of heavy elements found in the $s$-rich group relative to the $s$-poor group exhibits a correlation with the fraction of each element attributed to a $s$-process origin in solar system material, as shown in Fig.~\ref{fig:solar}. The element with the highest $s$-process fraction in the solar system is Ba, which is also the most overabundant in the $s$-rich group. 
The more modest excess of Pr is in line with its smaller $s$-process fraction in the solar system.

The enrichment in $s$-process elements in NGC\,5286 has been accompanied by an enrichment in the overall metallicity. The hints of intrinsic internal dispersions in Fe among the $s$-poor stars, if confirmed by larger samples, would imply that stars with a moderate increase in overall metallicity form continuously, or in discrete bursts, prior to the enrichment in $s$-process elements. Additionally, the $s$-rich stars appear to have larger internal scatters in $s$-elements, favouring the idea of a more prolonged star formation for this stellar group.

The chemical enrichment history of NGC\,5286 is made even more complex by the internal variations in $p$-capture elements, including evidence for O-Na anticorrelations and Na-Al correlations present in both groups with different $s$-elements. So, we have to assume either the pollution from high-mass AGB and/or fast-rotating massive stars has occurred before the enrichment in $s$-elements. Alternatively, the variations in light elements are not due to different stellar generations, but to other mechanisms, such as early accretion disks in pre-MS binary systems (Bastian et al.\,2014).

NGC\,5286 shares many similarities with some of the other {\it anomalous} GCs, e.g. those with metallicity variations (see Sect.~\ref{sec:intro}), studied so far. 
First, it shows a genuine internal variation in the overall metallicity and, for this reason, it can be included among the class of {\it anomalous} GCs. Second, it exhibits large internal variations in $s$-process elements, that have been also detected in other {\it anomalous} GCs, e.g. $\omega$~Centauri, NGC\,1851, M\,22, and M\,2.
In these clusters, the $s$-rich population(s) exhibit higher metallicity with respect to the $s$-poor/metal-poor population(s). In the other considered {\it anomalous} GCs, specifically M\,54, Terzan\,5, NGC\,5824, there is no information on the $s$-process element abundances in populations with different Fe, and we cannot prove or disprove, at the moment, $s$-process enrichment in these objects.

A list of the known {\it anomalous} GCs, and their chemical properties, is provided in Tab.~\ref{tab:anomaliclass}. 
The objects exhibiting also $s$-process variations have been classified as $s$-Fe {\it anomalous}. The class of $s$-Fe {\it anomalous} has to be intended as a  sub-class of the {\it anomalous} GCs. 
Of the eight {\it anomalous} GCs, five are s-Fe {\it anomalous}. 
The $s$-process enrichment is observed in a significant fraction of {\it anomalous} GCs, suggesting that, even if the range in metallicity variations is different in these objects, they have experienced some contribution from low-mass AGBs. We emphasise that the Fe and the $s$ enrichments are very likely due to different mass ranges and polluters, e.g. to high mass and low mass first-generation stars, respectively. 
A possibility is that in these more massive proto-clusters, the star-formation proceeded for longer times than in {\it normal} clusters, giving the possibility to low-mass AGBs to contribute to the enrichment of the proto-cluster. At the time these low-mass stars start to pollute the intra-cluster medium, material enriched from fast SNe, and previously expelled, may have had the time to be {\it fall-back} into the cluster. A similar scenario has been proposed by D'Antona et al.\,(2011) for $\omega$~Centauri.

All the known $s$-Fe {\it anomalous} GCs also have internal variations in light elements (such as a Na-O anticorrelation) within stars with different metallicity and $s$-content (see Tab.~\ref{tab:anomaliclass}). For the objects in which a mono-metallic group can be defined (e.g. M\,22 $s$-Fe-poor, M\,22 $s$-Fe-rich), they resemble a {\it normal} GC. 

The comparison of the chemical properties of NGC\,5286 with other {\it anomalous} GCs may shed some light on the origin of these objects.  
Specifically, M\,22 and M\,2 are more suitable for comparison with NGC\,5286 because they have a similar metallicity, and comparable variations in the [Fe/H] between the $s$-poor and $s$-rich stars. Note that M\,2 also shows a third group with much higher metallicity and no $s$-element enrichment. It is tempting to speculate that this third group in M\,2 is the counterpart of the few stars with low $s$-elements and relatively high Fe possibly present in NGC\,5286. 
However, for simplicity we consider just the $s$-poor and the $s$-rich stars of M\,2 to be compared with the stellar groups of NGC\,5286

In Fig.~\ref{fig:rs} we compare the abundance pattern inferred here for NGC\,5286 with those of M~22 and M\,2, considering the mean abundance differences between the $s$-rich and $s$-poor ($s$-rich$- s$-poor) stars observed in each of these clusters.
We note that for elements with $Z>$27 the chemical variations observed in the two stellar groups of NGC\,5286 are very similar to those observed in M\,2. Yttrium ($Z$=39) variations appear to be larger in NGC\,5286. Among the $p$-capture elements the most notable difference is in O, which varies more between the $s$-poor and the $s$-rich groups of M\,2 than in those of NGC\,5286. Note however that since light elements vary in each single $s$-group, the total observed variation in these elements may not be representative of the entire variations in these elements due to small-number statistics.

More distinctive differences are seen with respect to M\,22. For this cluster we plotted both abundances from Marino et al.\, (2009, 2011, blue stars) and from Roederer et al. (2012, blue diamonds). Both data-sets for M\,22 suggest that it has more moderate variations in all the $s$-elements than in M\,2 and NGC\,5286.
A common feature displayed by NGC\,5286, M\,22 and M\,2 is the apparent constancy of [Eu/Fe] suggesting that in all of these GCs the chemical enrichment was due to pollution of material that has undergone $s$-processing, rather than $r$-processing.

As already discussed, the most commonly discussed stellar site where $s$-neutron capture occurs is AGB stars.
Recently, Shingles et al.\,(2014) and Straniero et al.\,(2014), suggested that both AGB stars with a ${\phantom{}^{22}}$Ne source and lower-mass AGB stars with ${\phantom{}^{13}}$C pockets are required to account for the $s$-elements enrichment in M\,22 and to explain the large $s$-process elements abundances in M\,4 (relative to clusters with same metallicity like M\,5). 
The contribution from stars with masses as low as 2.75-4.5~$\rm {M_{\odot}}$ may be required to explain the enrichment, with the precise lower limit depending on which assumptions are made about ${\phantom{}^{13}}$C-pocket formation in AGB models (Shingles et al.\,2014; Straniero et al.\,2014).
We may think that similar mechanisms have worked also in NGC\,5286, but future proper analysis for this specific case would be enlightening to understand the higher $s$-enrichment.

\begin{figure*}
   \centering
   \includegraphics[width=15cm]{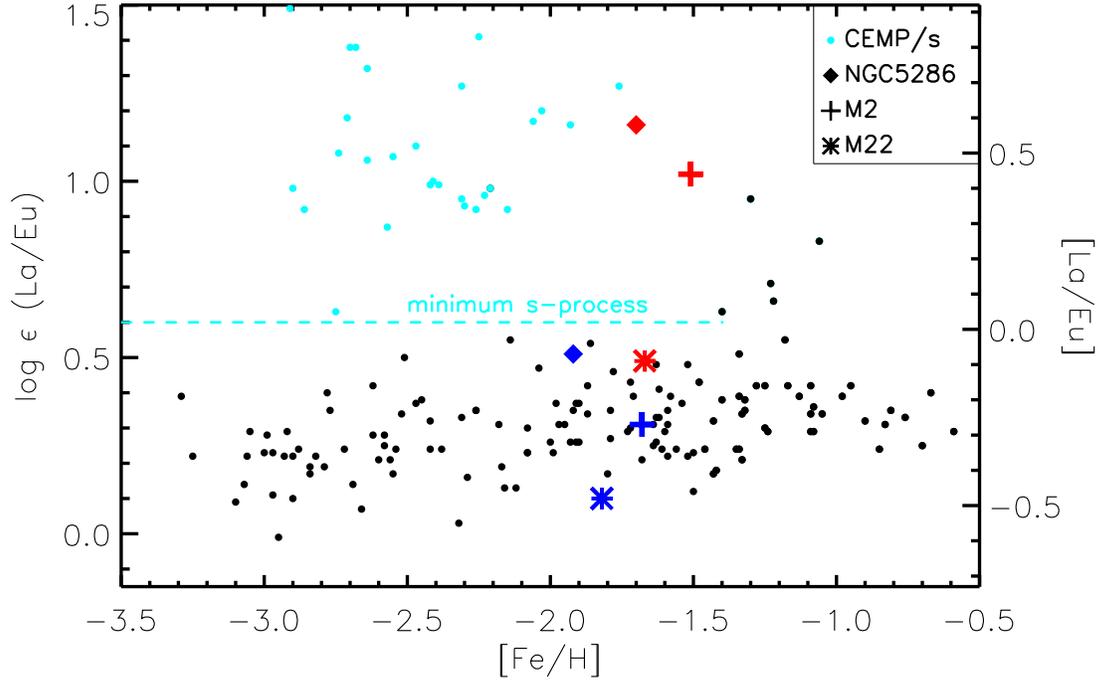}
      \caption{Logarithmic abundance ratios of La/Eu as a function of [Fe/H] for metal-poor halo stars. All measurements for field stars are indicated by small black and cyan circles, with cyan circles denoting stars enriched in $s$-process material.
The literature sources  of these data are Roederer et al.\,(2010a), Aoki et al.\,(2001, 2002), Barbuy et al.\,(2005), Cohen et al.\,(2003, 2006), Goswami et al.\,(2006), Ivans et al.\,(2005), Johnson \& Bolte\,(2004), Jonsell et al.\,(2006), Preston \& Sneden\,(2001), Roederer et al.\,(2010b), Simmerer et al.\,(2004), Thompson et al.\,(2008).
 The dashed line indicates log~(La/Eu)$\leq +$0.6, the approximate minimum ratios expected from AGB pollution (appropriate for [Fe/H]$< -$1.4~dex). Superimposed on to the field stars are the average abundances for the $s$-poor (blue) and $s$-rich stars (red) in NGC\,5286 (diamonds), M\,2 (plus, Yong et al.\,2014), and M\,22 (stars, Marino et al\,2009, 2011).}
        \label{fig:emp}
   \end{figure*}

Figure~\ref{fig:emp} reproduces Fig.~3 from Roederer et al.\,(2010). It shows the logarithmic abundance ratios of La/Eu as a function of the metallicity for metal-poor stars, including C-enriched metal-poor stars with overabundances of $s$-process material (CEMP/$s$).
Many CEMP/$s$ stars are known binary systems, indeed all of them may be in binaries (e.g.\,McClure et al.\,1980; McClure\,1983; Lucatello et al.\,2005). 
The approximate minimum La/Eu ratio expected from AGB pollution is shown in Fig.~\ref{fig:emp} as a dashed cyan line.
Superimposed on to the field stars are the average abundances for the $s$-poor and $s$-rich stars in NGC\,5286 together with those for M\,2 (Yong et al.\,2014), and M\,22 (Marino et al.\, 2009, 2011). 
We note that the two groups of M\,22, besides showing the smaller difference, also lie below the expected minimum $s$-process enrichment. On the other hand, $s$-rich stars in NGC\,5286 and M\,2 are well above this minimum, with La/Eu ratios being higher by $\sim$0.2~dex in both the $s$-groups of NGC\,5286. The differences on the La/Eu variations present in these clusters may indicate different degrees of intra-cluster pollutions.
While starting abundance levels may be affected by systematics between different studies, it is clear that $s$-rich stars of NGC\,5286 and M\,2 lie in a region very close to that occupied by CEMP/$s$ field stars. 
CEMP/$s$ stars are well fit by AGB models of low-mass indicating that ${\phantom{}^{13}}$C neutron source is dominant (Lugaro et al.\,2012; Bisterzo et al\,2012).
The fact that NGC\,5286 and M\,2 stars are closer to the CEMP/$s$ stars in the La/Eu-Fe diagram hints to a similar pollution source.

Assuming that in all these GCs low-mass AGB stars have contributed, possibly at different degrees, to the chemical self-enrichment, we emphasise that the real situation is more complex. Two observables make the situation much more difficult to understand: {\it (i)} the light element variations within each $s$-group; and {\it (ii)} the increase in the overall metallicity in the $s$-rich stars.
We thus
require the operation of at least two different nucleosynthetic processes during the formation and evolution of these GCs,
e.g. higher mass AGB/fast-rotating massive stars for the light element variations and supernovae for the metallicity increase. 

The higher relative variations in the $s$-process elements in NGC\,5286 and M\,2, relative to  M\,22, occur despite the very similar variation in metallicity between the $s$-poor and the $s$-rich stars.
Quantitatively, the variations ($s$-rich$- s$-poor) in [Fe/H] and [Ba/Fe] in the three GCs are listed in Table~\ref{tab:anomali}.
From these values we see that the total range in $s$-process elements in NGC\,5286 is about twice as large as it is in M\,22, and it is similar to M\,2.
This may suggest that there may be some differences in the polluters that enriched the intra-cluster medium in NGC\,5286 and M\,2 relatively to M\,22.
The $s$-process nucleosynthesis per increase in Fe varies between the clusters, and this must eventually tell us something useful about timescales, mass functions, or maybe dilution with pristine gas.

\section{Anomalous GCs and Milky Way satellites}

\begin{figure*}
    \includegraphics[width=17.3cm]{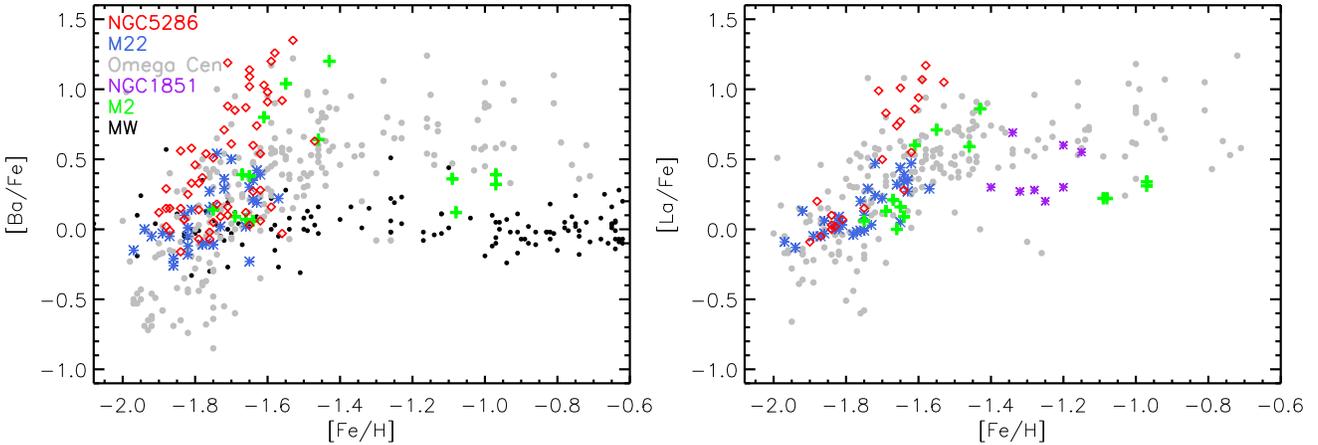}
    \caption{[Ba/Fe] and [La/Fe] as a function of [Fe/H] for anomalous GCs and field stars. Different symbols and colours are used for different GCs, as shown in the legend.
    The literature sources of the plotted abundances are: this paper for NGC\,5286; Marino et al.\,(2009; 2011) for M\,22; Marino et al.\,(2011) for $\omega$~Centauri; Yong et al.\,(2008) for NGC\,1851; Yong et al.\,(2014) for M\,2; and Fulbright\,(2000) for the Milky Way field stars.}
   \label{fig:anomali}
   \end{figure*}

Apart from the chemical/photometric features discussed above, comparing $s$-Fe-{\it anomalous} GCs with {\it anomalous} (non-$s$) GCs or with those clusters not identified as {\it anomalous} there is no evidence (to date) for different global/non-global parameters, except that {\it anomalous} and $s$-Fe {\it anomalous} GCs are among the most massive in the Milky Way.
In Fig.~\ref{fig:anomali} we show the Ba and La abundances relative to Fe as a function of [Fe/H] for $s$-Fe-{\it anomalous} GCs and field stars taken from this study and the literature. For NGC\,5286 we used the larger sample observed with GIRAFFE. In these plots it is clear that there is a common rise of $s$-elements with metallicity in the common metallicity regime. The data for $\omega$~Centauri suggest that in this case there is a significant extension towards higher [Fe/H], but above [Fe/H]$\sim -$1.5 there is a plateau in $s$-elements. In the common metallicity regime there is a similar rise in $s$ in NGC\,5286 and M\,2.

Internal variations in metallicity are intriguing because, assumed that they arise from internal chemical evolution, they suggest that material ejected at fast velocities from supernovae (SNII) could has been retained by these GCs at their early stages of evolution.
This would imply that their initial masses were much higher.
The idea that the very extreme GC $\omega$~Centauri is the nucleus of a dwarf galaxy was first suggested by Bekki \& Freeman (2003). The discovery of more objects with similar properties, although less extreme, may suggest that these objects may also be surviving nuclei of dwarf galaxies. This scenario would provide 8 more candidates to alleviate the missing satellites problem, i.e.\, 
the lack of observed MW satellites compared to the numbers expected from theoretical simulations (e.g.\,Kauffmann et al.\,1993, Klypin et al.\,1999, Moore et al.\,1999).
 
Additional support for the idea that the {\it anomalous} GCs may constitute the central remnants of dwarf galaxies after the outer layers have been stripped away come from other observations. 
The position of the GC M\,54 coincides with the nucleus of the Sagittarius dwarf galaxy (Layden \& Sarajedini 2000) and has metallicity variations (Carretta et al.\,2010). Also, a low-density halo of stars surrounding NGC\,1851 has been discovered by Olsweski et al.\,(2009), whose chemistry is compatible with the $s$-poor group observed in this cluster (Marino et al.\,2014). The absence of $s$-rich stars in this halo may suggest either that this GC is preferentially losing $s$-poor stars into the field, or that this sparse structure is the remnant of a dwarf galaxy, as its composition is compatible with field stars at similar metallicity (Marino et al.\,2014).

\begin{figure}
    \includegraphics[width=8.6cm]{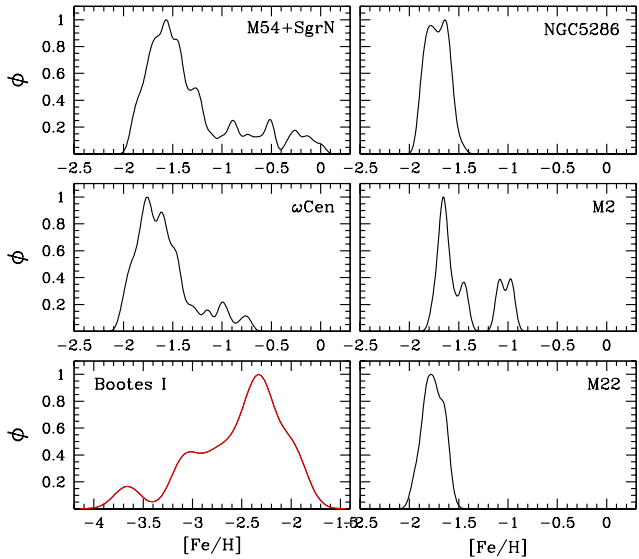}
    \caption{Metallicity histograms for the GCs with detected internal variations in metallicity plus $s$-process elements. The literature sources of these data are: Marino et al\,(2009, 2011) for M\,22, Marino et al.\,(2011) for $\omega$~Centauri and Yong et al.\,(2014) for M\,2. As a comparison we also show the metallicity distribution for the UFD galaxy Bootes~I (data from Norris et al.\,2010) and the GC M\,54+SgrN (data from Carretta et al.\,2010).}
   \label{fig:histo}
   \end{figure}

The lower-left panel of Fig.~\ref{fig:histo} shows the metallicity distribution function (MDF) for 16 stars in the ultra-faint dwarf galaxy (UFD) Bootes~I (from Norris et al.\,2010). 
Norris and collaborators noted that Bootes~I, similarly to that observed in dwarf spheroidals, exhibits a slow increase from lowest abundance to the MDF peak, while, by contrast, $\omega$~Centauri shows a steep rise.
In the other panels of Fig.~\ref{fig:histo} we compare the [Fe/H] kernel-density distributions for the {\it anomalous} GCs studied through high-resolution spectroscopy, including $\omega$~Centauri and the M\,54 plus the Sagittarius nucleus (SgrN) system.
An inspection of these distributions reveals that a sharp rise in metallicity to the metal-poorer peak\footnote{We note that the distributions shown in Fig.~\ref{fig:histo} for the {\it anomalous} GCs are biased in the number of metal-richer stars for NGC\,5286, M\,2 and M\,54+SgrN because metal-richer stars have been preferentially selected.} is a common feature among {\it anomalous} GCs. 

Another difference among {\it anomalous} GCs and dwarf galaxies is the lack, in the latter, of typical (anti-)correlation patterns among light-elements (e.g.\, Norris et al.\,in prep. for Carina). 
We note however that, if the hypothesis of the origin of {\it anomalous} GCs as nuclei of disrupted dwarf galaxies will be confirmed, the chemistry of the dwarfs does not have to necessarily resemble the GCs' one, as the latter would constitute just their nuclear regions. 
In {\it anomalous} GCs, the variations in light elements within each $s$/Fe-group (such as the individual Na-O anticorrelation) is difficult to understand within a self-pollution scenario, even in the hypothesis that the {\it anomalous} GCs are the nuclear remnants of more massive systems.
As individual $s$-groups appear similar to mono-metallic GCs, with their own Fe content and their own Na-O anticorrelation, it has been proposed that they can result from mergers between different clusters (e.g.\, Bekki \& Yong 2012). If this scenario will apply to every anomalous GC, we will need to understand why in almost all the GCs with internal metallicity variations found so far the increase in metals is coupled with a $s$-enrichment. In other words, we should find an explanation for the very similar properties of these objects, while in the hypothesis of a merger of two (or more) GCs one would expect a much more heterogenous observational scenario.

It is worth noticing that nearby dwarf and irregular galaxies host their more massive GCs in their central regions, these GCs being {\it nuclear} star clusters (e.g.\,Georgiev et al.\,2009).  We do not have, at the moment, any evidence for these nuclear GCs to share the same Fe distribution of the parent galaxies, or they show instead Fe distributions and chemical patterns more similar to the Galactic {\it anomalous} GCs.

\begin{figure}
    \includegraphics[width=8.5cm]{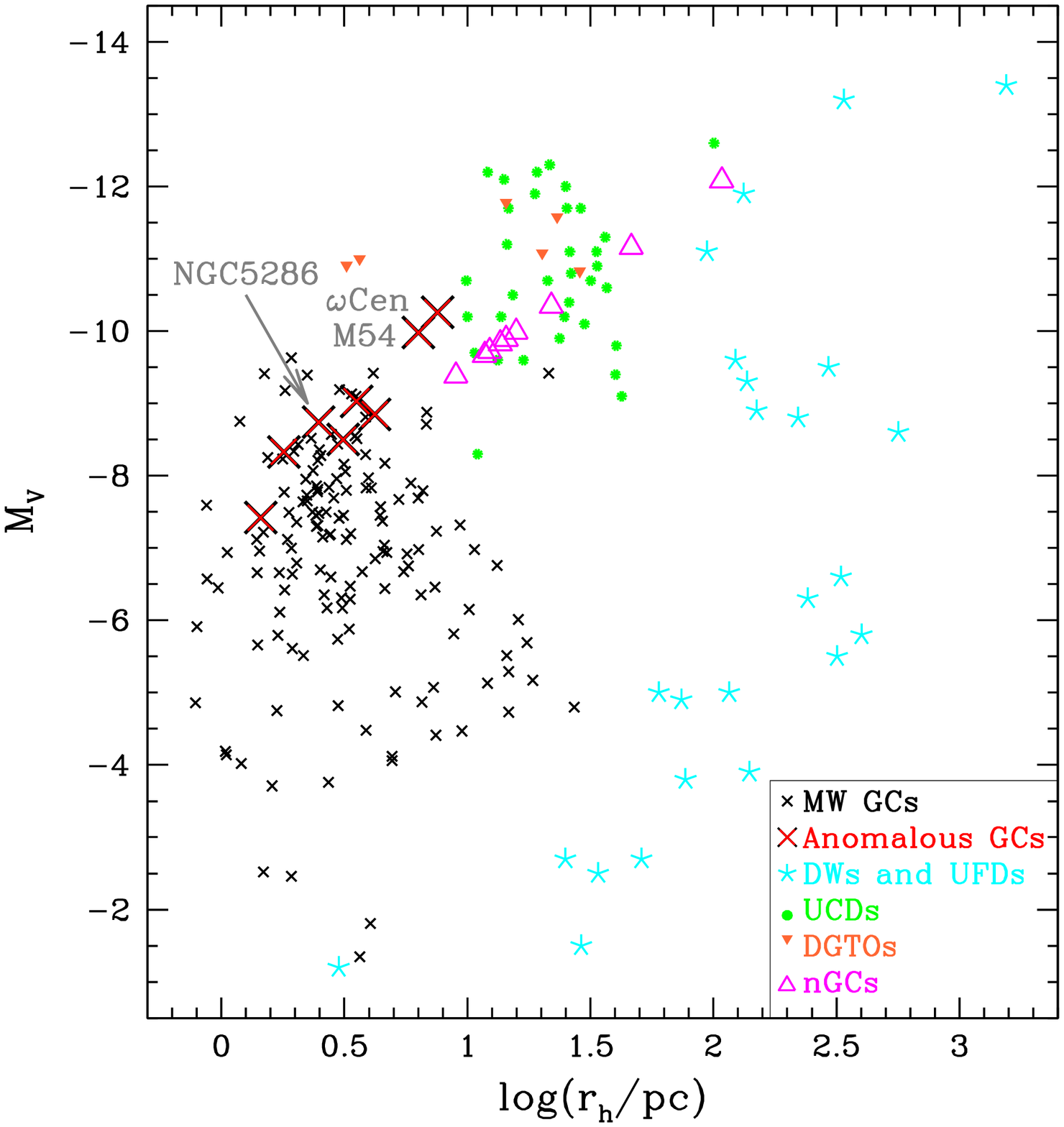}
    \caption{Absolute magnitude as a function of the half-light radius (r$_{\rm h}$). Different colours and symbols represent different class of objects: Milky Way GCs (black crosses, Harris 2010), Milky Way satellites including dwarf (DW, Irwin \& Hatzidimitriou\,1995; Mateo\,1998), ultra-faint dwarf galaxies (UFDs) and all the objects discovered by the SDSS (Willman et al.\,2005, 2006; Belokurov et al.\,2006, 2007; Zucker et al.\,2006; Jerjen\,2010), ultra compact dwarf galaxies (UCDs, Brodie et al.\,2011), nucleated GCs (nGCs; Georgiev et al.\,2009), dwarf-globular transition objects (DGTOs, Ha{\c s}egan et al.\,2005). Anomalous GCs, here considered as all the objects with internal metallicity and/or heavy-elements variations, have been marked with red crosses. }
   \label{fig:gal}
   \end{figure}

To explore further a possible galaxies-{\it anomalous} GCs connection we plotted the position of various stellar systems in the half-light radius (log~(r$_{\rm h}$/pc)) vs.\,absolute-mag ($\rm {M_{V}}$) plane (Fig~\ref{fig:gal}). We include the position of Milky Way GCs, classical dwarfs and UFDs, ultra-compact dwarfs (UCDs), dwarf-globular transition objects (DGTOs), nucleated GCs (nGCs). 
The position of GCs with internal variations in the overall metallicity (including those with not-investigated $s$-element abundances) shows that they are in general among the most massive GCs. Furthermore, various stellar systems appear to clump in different regions of the $\rm {M_{V}}$-log~(r$_{\rm h}$/pc) plane, with GCs with metallicity variations interestingly falling in regions abutting (or {\it evolving}) UCDs, DGTOs, nGCs.

It is worth noticing that the mean [Fe/H] is very similar between NGC\,5286, M\,22, and M\,2. Is this due to some selection effects or could this similarity in metallicity coincide with a particular phase in the MW or galaxies' evolution? For example, if these objects formed as dwarf galaxies' nuclei, we may suppose that the accreted dwarfs had nuclei with similar metallicities. This would imply that these nuclei formed out chemically similar clouds, probably at similar evolutionary phases of their host galaxies. 
On the other hand, if we suppose that these objects are the result of mergers, their similar metallicities may indicate a phase of abundant mergers in early galaxies.

%
%

%
%

%
%

%
%

%
%

%
%

\section{Conclusion}\label{sec:discussion}

We have found genuine metallicity and $s$-process abundance variations in the Galactic GC NGC\,5286.
To date large Fe variations have been confirmed to be present in eight GCs,
including M\,22 and M\,2, which have a similar mean metallicity as NGC\,5286.
As the mono-metallicity is a typical feature of Galactic GCs, we classify all the GCs with internal variations in metals as {\it anomalous} GCs. A sub-class of {\it anomalous} GCs, 5/8, that we define $s$-Fe {\it anomalous} have confirmed variations in $s$-elements.
The class of $s$-Fe {\it anomalous} GCs show common observational features:
\begin{itemize}
\item{Photometrically:
\begin{enumerate} 
\item{On the CMD, the most striking feature of all these GCs is a split in the SGB in visual bands. This split can be considered as an indication of $s$-elements$+$(C+N+O)$+$Fe variations, and can guide future spectroscopic observations aimed at the identification of other ``anomalous'' GCs; }
\item{large separations in the different $s$-populations along the RGBs, evolving from multiple SGBs, in the $U$-$(U-V)$ CMDs, and a maximum separation obtained using our new defined photometric index $c_{{\rm BVI}}$;}
\end{enumerate}
}
\item{Spectroscopically: 

\begin{enumerate}
\item{by definition, internal variations and/or multi-modalities in the $s$-process elements;} 
\item{by definition, different degrees of variations in the main metallicity, but in all cases the Fe content is higher in the $s$-richer stars.}
\item{no variations in the $r$-process elements detectable within observational errors. More specifically, for a given element, the degree of the abundance difference between the two stellar groups is strongly correlated with the fraction attributed to the $s$-process in solar system material;} 
\item{variations in light-elements (e.g. Na, O) often present in each main stellar group with different $s$-elements abundance;}
\end{enumerate}
}
\end{itemize}

NGC\,5286 shows internal variations in metals similar to those present between the two groups of $s$-rich and $s$-poor stars of M\,22 and M\,2, and $s$-abundance variations much higher than in M\,22, but similar to M\,2.
Among field stars, the stars that show largest similarities with the $s$-rich stars of NGC\,5286 and M\,2 are the CEMP/$s$ stars (while M\,22 appears to have undergone lower levels of $s$-enrichments).

We conclude that the observational scenario for the {\it anomalous} GCs is not compatible with an origin of these objects as {\it normal} GCs, with typical initial masses in the range observed today. 
It is intriguing to think, that these objects, much more massive at their birth,
may be nuclei of dwarfs tidally disrupted through interactions with the Milky Way, just as the {\it anomalous} GC M\,54 lies at the central region of Sagittarius. If this hypothesis will be confirmed, the 8 {\it anomalous} GCs should count as Milky Way satellites, with the number of those being substantially enhanced with respect to the number of the 27 confirmed known satellites (McConnachie 2012).  

\section*{acknowledgments}

\small
This research has been supported in part by the Australian Research Council through grants FL110100012, DP120100991 and DP120101237. APM acknowledges support by the Australian Research Council through Discovery Early Career Researcher Award DE150101816.
\normalsize

\bibliographystyle{aa}

\begin{table*}
\caption{Description of the photometric images used in this work.}\label{tab:journal}
\begin{tabular}{llrrrrl}
\hline\hline
   Telescope         &  Detector      & $U$ & $B$ & $V$ & $I$  &  Date            \\  \hline               
   CTIO 0.9m         &  RCA           &   - &   8 &   8 &   -  &  Jan, 21-24, 1987\\
   CTIO 0.9m         &  Tek2K$_3$     &   - &   - &  10 &  10  &  Apr, 16-19, 1996\\
   CTIO 0.9m         &  Tek2K$_3$     &   - &  24 &  27 &   -  &  May, 1998 - Jun, 26, 2004 \\
   CTIO 0.9m         &  Tek2K$_3$     &   - &   2 &   2 &   2  &  Jun, 13, 1999\\
   CTIO 0.9m         &  Tek2K$_3$     &   - &   1 &   1 &   1  &  Mar, 25, 2001\\
   MPI/ESO 2.2m      &  WFI           &   - &   4 &   4 &  10  &  Feb, 20, 2002\\
   ESO NTT 3.6m      &  SUSI          &   - &   - &   4 &   0  &  May, 30, 2003\\
   CTIO 0.9m         &  Tek2K$_3$     &   - &   - &   2 &   -  &  Jun, 22, 2007\\
   SOAR 4.1m         &  SOI           &   - &  65 &  63 &  61  &  Feb, 12-18, 2008\\
   MPI/ESO 2.2m\textsuperscript{a}&WFI&  17 &   - &   - &   -  &  Feb, 26, 2012\\   
\hline
\multicolumn{7}{l}{\textsuperscript{a}\scriptsize{SUMO project: program 088.A9012(A).}}
\end{tabular}
\end{table*}

\begin{table*}
\caption{Coordinates, basic photometric data and radial velocities for the stars in the field of view of NGC\,5286. The status of probable cluster members is listed in the last column. We list both the original magnitudes ($BVI_{\rm ori}$) and those corrected for differential reddening ($BVI_{\rm cor}$). }\label{tab:data}
\begin{tabular}{lccccccccrr}
\hline\hline
ID            &  RA (J2000) & DEC (J2000) & $B_{\rm ori}$ & $V_{\rm ori}$ & $I_{\rm ori}$ & $B_{\rm cor}$ & $V_{\rm cor}$ & $I_{\rm cor}$ & RV [\kmsec] &         \\ \hline
   \multicolumn{11}{c}{UVES}  \\
\\
 N5286-1219U   &  13:46:31.38   & $-$51:21:22.8 &  16.532 &   15.347 &  13.959 & 16.495 & 15.319 & 13.943 & 66.99 &   member\\  
 N5286-1439U   &  13:46:34.90   & $-$51:20:45.6 &  16.917 &   15.781 &  14.485 & 16.977 & 15.827 & 14.512 & 70.18 &   member\\  
 N5286-859U    &  13:46:25.75   & $-$51:24:53.0 &  15.850 &   14.369 &  12.836 & 15.801 & 14.331 & 12.814 & 60.76 &   member\\
\\
   \multicolumn{11}{c}{GIRAFFE}  \\
\\ 
  N5286-667G   &  13:46:22.38  & $-$51:21:48.6  &  17.849 &   16.853 &  15.644 & 17.875 & 16.873 & 15.656 &    66.92 & member \\  
  N5286-527G   &  13:46:20.17  & $-$51:21:18.3  &  15.802 &   14.351 &  12.769 & 15.702 & 14.274 & 12.724 &    66.03 & member \\  
  N5286-1117G  &  13:46:29.51  & $-$51:21:47.3  &  15.739 &   14.429 &  12.948 & 15.767 & 14.450 & 12.961 &    66.33 & member \\  
\hline
\multicolumn{11}{l}{Only a portion of this table is shown here to demonstrate its form and content. A machine-readable version will be available.}
\end{tabular}
\end{table*}

\begin{table*}
\caption{Line list for the program stars. For the UVES targets we list the measured EWs. For GIRAFFE we synthetised the same lines used for UVES, in the common spectral range 6120-6400~\AA; a few lines have been used only for GIRAFFE. }\label{tab:linelist}
\begin{tabular}{c r cc rrrrrrr c r}
\hline\hline
Wavelength & Species & L.E.P. &log~$gf$ & 1219U & 1439U & 859U  & 1309U & 579U  & 1339U & 177U &  & Ref. for log~$gf$ \\

 [\AA]     &         &  [eV]  &         & [m\AA]& [m\AA]& [m\AA]& [m\AA]& [m\AA]& [m\AA]&[m\AA]&  &                 \\ 
\hline                                                                   
6300.304 &  8.0 &  0.000 &$-$9.819 &   syn  &    syn  &    syn   &     syn   &   --    &   syn  &  syn  &  & 1      \\
5682.633 & 11.0 &  2.102 &$-$0.710 &   syn  &    syn  &    syn	 &     syn   &   syn   &   syn  &  syn  &  & 2      \\
5688.205 & 11.0 &  2.104 &$-$0.450 &   syn  &    syn  &    syn   &     syn   &   syn   &   syn  &  syn  &  & 2      \\
6154.226 & 11.0 &  2.102 &$-$1.550 &   --   &     --  &    syn	 &     syn   &   --    &   syn  &  --   &  & 2      \\
6160.747 & 11.0 &  2.104 &$-$1.250 &   syn  &    syn  &    syn	 &     syn   &   syn   &   syn  &  syn  &  & 2      \\
\hline
\multicolumn{13}{l}{\normalsize{References: (1) Kurucz compendium; 
(2) Fuhr \& Wiese (2009); 
(3) Yong et al. (2014); 
(4) NIST; 
(5) Gaia-ESO linelist (Heiter et al., in prep.);}} \\
\multicolumn{13}{l}{\normalsize{
(6) Yong et al. (2012);
(7) Mel{\'e}ndez \& Barbuy (2009);
(8) Johnson \& Pilachowsky (2010);
(9) Bielski (1975), Kurucz \& Bell (1995);}}\\
\multicolumn{13}{l}{\normalsize{
(10) Hannaford et al. (1982);
(11) Biemont et al. (1981);
(12) Gallagher (1967), Sneden et al. (2000);
(13) Gallagher (1967), Burris et al. (2000);}}\\
\multicolumn{13}{l}{\normalsize{
(14) Gallagher (1967), Sneden et al. (1996);
(15) Lawler, Bonvallet \& Sneden (2001a);
(16) Sneden et al. (2009);
(17) Ivarsson et al. (2001);}}\\
\multicolumn{13}{l}{\normalsize{
(18) Li et al. (2007);
(19) Lawler et al. (2001b).}}\\
\multicolumn{13}{l}{Only a portion of this table is shown here to demonstrate its form and content. A machine-readable version will be available.}
\end{tabular}
\end{table*}

\onecolumn

\newpage
\begin{landscape}
\begin{table*}
\caption{Adopted atmospheric parameters and chemical abundances derived for the NGC\,5286 stars observed with GIRAFFE. For sodium we list both the LTE ([Na/Fe]$_{\rm {LTE}}$) and the NLTE ([Na/Fe]$_{\rm {NLTE}}$) abundances. Line-to-line scatter is listed when more than one line has been analysed. The last column lists the status of $s$-rich or $s$-poor assigned to each star.}\label{tab:gir_abundances}
\scriptsize
\begin{tabular}{lcccrcrrcrcrcrcrcrrcrrcc}
\hline\hline 
  ID         &\teff  &\logg &\vmicro & [Fe/H]& $\pm$&[O/Fe] &[Na/Fe] &$\pm$&[Na/Fe]  &$\pm$ &[Si/Fe]&$\pm$&[Ca/Fe]&$\pm$&[Sc/Fe]&$\pm$&[Ti/Fe]&[Ni/Fe]&$\pm$&[Ba/Fe]&[La/Fe]&$\pm$& group\\ 
             &  [K]  & [cgs]&[\kmsec]&       &      &       & NLTE   &     &  LTE    &      &       &     &       &     &       &     &       &       &     &       &       &     &      \\ \hline
  N5286-667G &  5181 & 2.47&  1.60 & $-$1.66 & 0.07 &  --   &    --   & --   &   --   & --   &  0.54 & --   &  0.30 & 0.04 &$-$0.14 & --   &  --   &  --   &  --   &   0.12 &  --   & --    & $s$-poor  \\
  N5286-527G &  4429 & 1.00&  1.95 & $-$1.75 & 0.03 &  0.39 &    0.20 & 0.08 &   0.26 & 0.09 &  0.36 & 0.02 &  0.32 & 0.06 &   0.06 & 0.00 &  0.37 &  0.15 &  0.19 &   0.51 &  0.30 & --    & $s$-poor  \\
 N5286-1117G &  4520 & 1.20&  1.91 & $-$1.87 & 0.03 &  0.51 & $-$0.21 & --   &$-$0.17 & --   &  0.36 & 0.04 &  0.32 & 0.10 &   0.05 & 0.09 &  0.31 &  0.16 &  0.10 &   0.15 &  0.11 & 0.02  & $s$-poor  \\
  N5286-757G &  4432 & 0.99&  1.94 & $-$1.66 & 0.04 &  0.53 &    0.10 & 0.12 &   0.14 & 0.13 &  0.51 & 0.10 &  0.37 & 0.07 &   0.00 & 0.17 &  0.57 &  0.19 &  0.33 &   0.87 &  0.84 & --    & $s$-rich  \\
  N5286-697G &  4958 & 2.15&  1.67 & $-$1.79 & 0.04 &  0.35 &    0.12 & --   &   0.20 & --   &  0.26 & 0.18 &  0.26 & 0.13 &$-$0.08 & --   &  0.20 &  0.02 &  --   &$-$0.07 &  --   & --    & $s$-poor  \\
  N5286-779G &  5102 & 2.31&  1.61 & $-$1.56 & 0.04 &  --   &    --   & --   &   --   & --   &  0.24 & 0.13 &  0.17 & 0.09 &$-$0.17 & --   &  --   &  0.16 &  --   &   0.92 &  --   & --    & $s$-rich  \\
  N5286-399G &  4965 & 2.00&  1.69 & $-$1.60 & 0.03 &  --   &    0.50 & --   &   0.58 & --   &  0.33 & 0.06 &  0.39 & 0.13 &$-$0.03 & --   &  0.32 &  0.03 &  --   &   0.98 &  0.96 & --    & $s$-rich  \\
  N5286-989G &  5147 & 2.48&  1.58 & $-$1.60 & 0.04 &  --   &    --   & --   &   --   & --   &  0.34 & --   &  0.31 & 0.06 &$-$0.12 & --   &  0.40 &  --   &  --   &   0.91 &  --   & --    & $s$-rich  \\
 N5286-1297G &  4350 & 0.90&  1.99 & $-$1.84 & 0.03 &  0.64 & $-$0.19 & --   &$-$0.15 & --   &  0.32 & 0.04 &  0.26 & 0.10 &   0.13 & 0.08 &  0.29 &  0.06 &  0.18 &   0.56 &  0.12 & 0.01  & $s$-poor  \\
 N5286-1767G &  5021 & 2.31&  1.63 & $-$1.71 & 0.03 &  --   &    0.29 & --   &   0.38 & --   &  0.55 & 0.12 &  0.46 & 0.13 &   0.26 & --   &  0.48 &  --   &  --   &   0.16 &  --   & --    & $s$-poor  \\
 N5286-1269G &  4988 & 2.24&  1.63 & $-$1.65 & 0.01 &  0.20 &    0.57 & 0.15 &   0.65 & 0.16 &  0.47 & 0.02 &  0.44 & 0.13 &$-$0.02 & --   &  0.23 &  0.08 &  --   &   1.14 &  1.03 & --    & $s$-rich  \\
  N5286-939G &  4972 & 2.27&  1.63 & $-$1.70 & 0.04 &  --   &    0.28 & --   &   0.36 & --   &  0.45 & 0.08 &  0.44 & 0.13 &   0.03 & --   &  0.30 &  --   &  --   &   0.61 &  0.56 & --    & $s$-rich  \\
 N5286-1197G &  5043 & 2.36&  1.63 & $-$1.80 & 0.03 &  0.76 &    --   & --   &   --   & --   &  0.49 & 0.22 &  0.50 & 0.14 &$-$0.04 & --   &  --   &  --   &  --   &   0.46 &  --   & --    & $s$-poor  \\
 N5286-1057G &  4974 & 2.10&  1.70 & $-$1.87 & 0.03 &  0.60 &    --   & --   &   --   & --   &  0.23 & --   &  0.36 & 0.16 &$-$0.04 & --   &  --   &  --   &  --   &$-$0.01 &  --   & --    & $s$-poor  \\
 N5286-1547G &  4838 & 2.00&  1.70 & $-$1.75 & 0.02 &  0.58 &    --   & --   &   --   & --   &  0.44 & 0.14 &  0.41 & 0.13 &   0.03 & --   &  0.33 &  0.10 &  0.06 &   0.05 &  --   & --    & $s$-poor  \\
 N5286-1077G &  5017 & 2.21&  1.66 & $-$1.74 & 0.03 &  --   &    0.21 & --   &   0.29 & --   &  0.35 & 0.23 &  0.34 & 0.12 &$-$0.09 & --   &  0.22 &  0.16 &  --   &   0.18 &  --   & --    & $s$-poor  \\
 N5286-5441G &  5104 & 2.08&  1.68 & $-$1.62 & 0.04 &  0.72 &    --   & --   &   --   & --   &  0.33 & 0.04 &  0.39 & 0.20 &   0.02 & --   &  --   &  0.12 &  0.03 &   0.28 &  --   & --    & $s$-poor  \\
 N5286-1607G &  5123 & 2.34&  1.62 & $-$1.62 & 0.03 &  --   &    0.57 & --   &   0.66 & --   &  0.53 & 0.16 &  0.60 & 0.26 &   0.06 & --   &  --   &  --   &  --   &   0.06 &  --   & --    & $s$-poor  \\
  N5286-969G &  5055 & 2.39&  1.60 & $-$1.63 & 0.08 &  --   &    0.45 & --   &   0.54 & --   &  0.42 & 0.07 &  0.41 & 0.21 &$-$0.01 & --   &  --   &  --   &  --   &   0.74 &  --   & --    & $s$-rich  \\
 N5286-1729G &  4881 & 2.08&  1.68 & $-$1.71 & 0.03 &  0.32 &    0.43 & 0.36 &   0.51 & 0.36 &  0.59 & 0.03 &  0.47 & 0.09 &$-$0.01 & --   &  0.47 &  --   &  --   &   0.88 &  1.00 & 0.05  & $s$-rich  \\
 N5286-1687G &  4441 & 1.14&  1.92 & $-$1.84 & 0.03 &  0.15 &    0.36 & 0.06 &   0.41 & 0.07 &  0.31 & 0.03 &  0.29 & 0.07 &   0.07 & 0.04 &  0.24 &  0.04 &  0.13 &   0.15 &  0.13 & 0.08  & $s$-poor  \\
 N5286-5541G &  4596 & 1.39&  1.85 & $-$1.81 & 0.03 &  0.48 & $-$0.12 & --   &$-$0.07 & --   &  0.29 & 0.04 &  0.35 & 0.12 &   0.00 & 0.03 &  0.33 &  0.11 &  0.05 &   0.33 &  0.11 & --    & $s$-poor  \\
 N5286-1737G &  4688 & 1.62&  1.80 & $-$1.88 & 0.03 &  0.18 &    0.43 & 0.12 &   0.50 & 0.12 &  0.39 & 0.02 &  0.37 & 0.11 &   0.03 & 0.04 &  0.33 &  0.13 &  0.06 &   0.02 &  0.25 & --    & $s$-poor  \\
 N5286-1537G &  4963 & 2.16&  1.68 & $-$1.90 & 0.04 &  --   &    --   & --   &   --   & --   &  0.49 & 0.07 &  0.43 & 0.13 &   0.10 & --   &  --   &  0.29 &  --   &   0.12 &  --   & --    & $s$-poor  \\
 N5286-5191G &  5196 & 2.12&  1.70 & $-$1.81 & 0.04 &  0.68 &    0.37 & --   &   0.46 & --   &  0.25 & --   &  0.64 & --   &   0.00 & --   &  0.29 &  --   &  --   &   0.58 &  --   & --    & $s$-poor  \\
 N5286-1567G &  4421 & 1.06&  1.94 & $-$1.83 & 0.04 &  0.41 &    0.11 & 0.11 &   0.15 & 0.13 &  0.32 & 0.07 &  0.28 & 0.08 &   0.02 & 0.06 &  0.29 &  0.10 &  0.14 &   0.07 &  0.14 & 0.01  & $s$-poor  \\
 N5286-1147G &  4660 & 1.54&  1.82 & $-$1.84 & 0.03 &  --   &    0.20 & 0.12 &   0.26 & 0.13 &  0.32 & 0.03 &  0.37 & 0.12 &   0.03 & 0.00 &  0.34 &  0.12 &  0.03 &$-$0.16 &  0.14 & 0.04  & $s$-poor  \\
 N5286-1659G &  5019 & 2.19&  1.65 & $-$1.65 & 0.02 &  0.29 &    0.36 & --   &   0.45 & --   &  0.35 & 0.08 &  0.47 & 0.15 &$-$0.05 & --   &  0.29 &  --   &  --   &   1.02 &  0.82 & --    & $s$-rich  \\
 N5286-1047G &  4840 & 1.81&  1.75 & $-$1.76 & 0.03 &  0.23 &    0.21 & --   &   0.28 & --   &  0.33 & 0.06 &  0.40 & 0.16 &   0.00 & --   &  0.24 &  0.04 &  0.11 &$-$0.02 &  --   & --    & $s$-poor  \\
 N5286-1557G &  5088 & 2.43&  1.61 & $-$1.77 & 0.03 &  --   &    --   & --   &   --   & --   &  0.48 & --   &  0.46 & 0.12 &$-$0.14 & --   &  0.22 &  --   &  --   &   0.54 &  --   & --    & $s$-poor  \\
 N5286-1227G &  5092 & 2.32&  1.64 & $-$1.78 & 0.06 &  --   &    --   & --   &   --   & --   &  0.45 & --   &  0.43 & 0.14 &$-$0.17 & --   &  0.41 &  --   &  --   &   0.37 &  --   & --    & $s$-poor  \\
 N5286-1599G &  5020 & 2.17&  1.65 & $-$1.61 & 0.03 &  --   &    0.50 & 0.06 &   0.58 & 0.05 &  0.52 & 0.04 &  0.48 & 0.15 &$-$0.06 & --   &  --   &  0.21 &  --   &   1.03 &  0.92 & --    & $s$-rich  \\
 N5286-1369G &  4813 & 1.77&  1.73 & $-$1.59 & 0.03 &  --   &    0.42 & 0.31 &   0.49 & 0.32 &  0.41 & 0.07 &  0.50 & 0.11 &$-$0.08 & --   &  0.34 &  0.14 &  0.00 &   1.20 &  1.13 & 0.15  & $s$-rich  \\
 N5286-1529G &  5140 & 2.46&  1.58 & $-$1.53 & 0.05 &  --   &    0.56 & --   &   0.65 & --   &  0.48 & 0.02 &  0.60 & 0.18 &$-$0.09 & --   &  0.42 &  --   &  --   &   1.35 &  1.07 & --    & $s$-rich  \\
  N5286-947G &  4921 & 2.15&  1.68 & $-$1.88 & 0.03 &  --   & $-$0.01 & --   &   0.06 & --   &  0.36 & --   &  0.56 & 0.35 &   0.22 & --   &  0.42 &  --   &  --   &   0.15 &  --   & --    & $s$-poor  \\
 N5286-1237G &  5042 & 2.27&  1.66 & $-$1.88 & 0.02 &  0.79 &    --   & --   &   --   & --   &  0.67 & --   &  0.54 & 0.15 &   0.11 & --   &  0.45 &  --   &  --   &   0.29 &  --   & --    & $s$-poor  \\
  N5286-827G &  5075 & 2.43&  1.60 & $-$1.65 & 0.03 &  --   &    --   & --   &   --   & --   &  0.59 & --   &  0.31 & 0.12 &$-$0.22 & --   &  0.13 &  --   &  --   &   0.03 &  --   & --    & $s$-poor  \\
 N5286-1017G &  5100 & 2.26&  1.62 & $-$1.56 & 0.04 &  --   &    0.32 & --   &   0.41 & --   &  0.48 & 0.13 &  0.28 & 0.33 &$-$0.06 & --   &  0.19 &  --   &  --   &$-$0.03 &  --   & --    & $s$-poor  \\
 N5286-1747G &  4897 & 1.89&  1.73 & $-$1.73 & 0.03 &  0.43 &    0.37 & 0.27 &   0.45 & 0.29 &  0.38 & 0.10 &  0.38 & 0.13 &   0.04 & --   &  0.37 &  0.12 &  0.08 &   0.09 &  --   & --    & $s$-poor  \\
  N5286-996G &  4959 & 2.08&  1.69 & $-$1.82 & 0.03 &  --   &    0.50 & --   &   0.59 & --   &  0.32 & 0.02 &  0.42 & 0.16 &$-$0.02 & --   &  0.28 &  --   &  --   &   0.25 &  --   & --    & $s$-poor  \\
 N5286-1649G &  4777 & 1.52&  1.81 & $-$1.62 & 0.03 &  0.06 &    0.34 & 0.00 &   0.41 & 0.00 &  0.34 & 0.04 &  0.39 & 0.12 &$-$0.09 & --   &  0.43 &  0.11 &  0.06 &   0.54 &  0.60 & 0.07  & $s$-rich  \\
  N5286-587G &  4714 & 1.51&  1.81 & $-$1.64 & 0.03 &  0.40 & $-$0.31 & --   &$-$0.26 & --   &  0.34 & 0.10 &  0.34 & 0.16 &$-$0.05 & --   &  0.25 &  0.02 &  0.01 &   0.60 &  0.43 & --    & $s$-rich  \\
   N5286-29G &  4963 & 2.22&  1.64 & $-$1.65 & 0.02 &  --   &    --   & --   &   --   & --   &  0.36 & 0.06 &  0.43 & 0.14 &$-$0.14 & --   &  0.41 &  --   &  --   &   1.09 &  --   & --    & $s$-rich  \\
  N5286-437G &  5063 & 2.26&  1.63 & $-$1.64 & 0.02 &  0.65 &    0.08 & --   &   0.16 & --   &  0.40 & 0.10 &  0.42 & 0.20 &$-$0.10 & --   &  0.22 &  --   &  --   &   0.27 &  --   & --    & $s$-poor  \\
   N5286-17G &  4966 & 2.16&  1.67 & $-$1.76 & 0.03 &  --   &    0.24 & --   &   0.32 & --   &  0.52 & --   &  0.41 & 0.18 &$-$0.13 & --   &  0.31 &  0.29 &  --   &$-$0.07 &  --   & --    & $s$-poor  \\
  N5286-719G &  4978 & 2.14&  1.67 & $-$1.71 & 0.03 &  0.80 &    0.21 & --   &   0.29 & --   &  0.41 & 0.23 &  0.41 & 0.14 &   0.05 & --   &  0.46 &  --   &  --   &   1.19 &  --   & --    & $s$-rich  \\
  N5286-169G &  4990 & 2.24&  1.64 & $-$1.72 & 0.03 &  --   &    0.19 & --   &   0.27 & --   &  0.42 & 0.06 &  0.45 & 0.13 &$-$0.05 & --   &  0.49 &  --   &  --   &   0.71 &  --   & --    & $s$-rich  \\
  N5286-07G &  4978 & 2.15&  1.68 & $-$1.83 & 0.04 &  0.33 &    --   & --   &   --   & --   &  0.47 & 0.34 &  0.28 & 0.12 &   0.03 & --   &  0.19 &  --   &  --   &   0.07 &  --   & --    & $s$-poor  \\
  N5286-289G &  5158 & 2.30&  1.60 & $-$1.47 & 0.04 &  0.18 &    0.21 & --   &   0.30 & --   &  0.45 & 0.08 &  0.48 & 0.25 &$-$0.01 & --   &  0.34 &  --   &  --   &   0.63 &  --   & --    & $s$-rich  \\
  N5286-379G &  4887 & 1.93&  1.70 & $-$1.58 & 0.03 &  0.05 &    0.31 & --   &   0.39 & --   &  0.36 & 0.07 &  0.42 & 0.12 &$-$0.08 & --   &  0.37 &  0.13 &  0.02 &   1.26 &  1.20 & 0.09  & $s$-rich  \\
  N5286-509G &  4881 & 1.94&  1.71 & $-$1.69 & 0.03 &  0.52 & $-$0.06 & --   &   0.01 & --   &  0.36 & 0.07 &  0.42 & 0.12 &$-$0.02 & --   &  0.33 &  0.18 &  0.01 &   0.85 &  0.87 & --    & $s$-rich  \\
  N5286-157G &  5118 & 2.42&  1.62 & $-$1.79 & 0.04 &  --   &    --   & --   &   --   & --   &  --   & --   &  0.47 & 0.15 &$-$0.14 & --   &  --   &  --   &  --   &   0.14 &  --   & --    & $s$-poor  \\
  N5286-707G &  5153 & 2.42&  1.61 & $-$1.71 & 0.04 &  --   &    0.34 & --   &   0.43 & --   &  --   & --   &  0.46 & 0.17 &$-$0.11 & --   &  0.34 &  --   &  --   &   0.10 &  --   & --    & $s$-poor  \\
  N5286-537G &  4838 & 1.85&  1.74 & $-$1.79 & 0.02 &  --   &    0.32 & --   &   0.40 & --   &  0.29 & 0.05 &  0.37 & 0.13 &   0.06 & --   &  0.22 &  0.17 &  --   &   0.33 &  --   & --    & $s$-poor  \\
  N5286-207G &  5177 & 2.45&  1.59 & $-$1.59 & 0.04 &  --   &    --   & --   &   --   & --   &  0.39 & --   &  0.42 & 0.30 &   --   & --   &  --   &  --   &  --   &   0.16 &  --   & --    & $s$-poor  \\
\hline
\end{tabular}
\end{table*}
\end{landscape}

%
\begin{table*}
\caption{Adopted temperatures (\teff\ [K]), surface gravities (\logg\ [cgs]), microturbolences (\vmicro\ [\kmsec]), metallicities (Fe\,{\sc i}) and chemical abundances ratios for the stars observed with UVES. \label{tab:abbUVES}}
\begin{tabular}{r ccc cc cc cc cc cc cc cc cc cc cc cc cc cc}
\hline\hline
  ID    & \teff&\logg&\vmicro&Fe\,{\sc i}     &$\pm$& Fe\,{\sc ii}     &$\pm$& O       & Na     &$\pm$& Mg   &$\pm$& Al   &$\pm$& Si   &$\pm$     \\ \hline
1219 & 4600 &1.20&1.90&$-$1.93 & 0.02 & $-$1.87 & 0.02 & $+$0.47 &   0.33 & 0.04 & 0.68 & 0.26 & 0.52 & 0.05 & 0.36 & 0.05\\ 
1439 & 4750 &1.70&1.71&$-$1.70 & 0.02 & $-$1.66 & 0.05 & $+$0.55 &   0.22 & 0.04 & 0.47 & 0.19 & 0.30 & 0.00 & 0.50 & 0.00\\ 
 859 & 4300 &0.85&1.98&$-$1.67 & 0.02 & $-$1.62 & 0.03 & $+$0.40 &   0.53 & 0.04 & 0.52 & 0.25 & 0.77 & 0.05 & 0.66 & 0.20\\ 
1309 & 4690 &1.50&1.71&$-$1.73 & 0.02 & $-$1.68 & 0.02 & $+$0.30 &   0.34 & 0.03 & 0.37 & 0.17 & 0.88 & 0.07 & 0.22 & 0.16\\ 
 579 & 4570 &1.05&1.85&$-$1.92 & 0.03 & $-$1.88 & 0.04 &      -- &   0.30 & 0.01 & 0.59 & 0.25 & 0.50 & 0.00 & 0.50 & 0.00\\ 
1339 & 4660 &1.50&1.90&$-$1.72 & 0.02 & $-$1.68 & 0.03 & $+$0.29 &   0.58 & 0.02 & 0.63 & 0.22 & 0.65 & 0.10 & 0.40 & 0.09\\ 
 177 & 4590 &1.25&2.00&$-$1.92 & 0.02 & $-$1.88 & 0.03 & $+$0.68 &$-$0.08 & 0.05 & 0.63 & 0.26 & 0.15 & 0.30 & 0.25 & 0.09\\ \hline\hline

  ID & Ca   &$\pm$& Sc\,{\sc ii}    &$\pm$& Ti\,{\sc i}  &$\pm$& Ti\,{\sc ii}  &$\pm$&  V  &$\pm$& Cr  &$\pm$  \\ \hline
1219 & 0.34 & 0.04 &   0.06 & 0.07 & 0.30 & 0.04 & 0.27 & 0.10 &   0.01 & 0.02 &$-$0.06 & 0.12\\       
1439 & 0.29 & 0.04 &   0.13 & 0.05 & 0.46 & 0.15 & 0.21 & 0.10 &   0.25 & 0.00 &$-$0.17 & 0.09\\       
 859 & 0.40 & 0.02 &   0.18 & 0.09 & 0.58 & 0.18 & 0.15 & 0.23 &   0.14 & 0.06 &   0.39 & 0.16\\       
1309 & 0.27 & 0.04 &   0.14 & 0.08 & 0.24 & 0.03 & 0.37 & 0.20 &$-$0.12 & --   &$-$0.12 & 0.06\\       
 579 & 0.29 & 0.05 &$-$0.01 & 0.06 & 0.34 & 0.06 & 0.13 & 0.18 &   --   & --   &$-$0.17 & 0.00\\       
1339 & 0.38 & 0.04 &   0.11 & 0.07 & 0.37 & 0.06 & 0.21 & 0.01 &   --   & --   &$-$0.01 & 0.08\\       
 177 & 0.24 & 0.04 &   0.06 & 0.04 & 0.22 & 0.07 & 0.20 & 0.11 &   0.17 & 0.27 &$-$0.20 & 0.01\\ \hline\hline

  ID & Mn     &$\pm$& Co     &$\pm$& Ni     &$\pm$& Cu     &$\pm$& Zn    &$\pm$& Y\,{\sc ii}     &$\pm$& Zr\,{\sc ii}    &$\pm$      \\ \hline
1219 &$-$0.40 & 0.04 &$-$0.08 & 0.15 & 	 0.00  & 0.04 &$-$0.57 & --   &  0.11 & -- & 	0.01 & 0.12 &	0.10 & -- \\
1439 &$-$0.48 & 0.05 & 0.25   & --   &$-$0.02  & 0.05 &$-$0.55 & --   &  0.13 & -- & 	0.70 & 0.06 &	0.55 & -- \\
 859 &$-$0.45 & 0.04 & 0.07   & 0.06 & 	 0.04  & 0.05 &$-$0.36 & 0.07 &  0.30 & -- & 	0.42 & 0.05 &	0.41 & -- \\
1309 &$-$0.43 & 0.04 & 0.21   & 0.18 &$-$0.05  & 0.04 &$-$0.67 & --   &  0.07 & -- & 	0.06 & 0.11 &	0.25 & -- \\
 579 &$-$0.45 & 0.04 & --     &  --  &$-$0.04  & 0.04 &$-$0.64 & --   &  0.37 & -- &$-$0.07 & 0.08 &	0.20 & -- \\
1339 &$-$0.40 & 0.05 & 0.06   & 0.07 & 	 0.02  & 0.04 &$-$0.35 & 0.11 &  0.56 & -- & 	0.58 & 0.07 &	0.73 & -- \\
 177 &$-$0.45 & 0.06 & 0.04   & --   &$-$0.05  & 0.04 &$-$0.58 & --   &  0.10 & -- &$-$0.07 & 0.06 &	0.20 & -- \\ \hline\hline

  ID &  Ba\,{\sc ii}  &$\pm$& La\,{\sc ii}  &$\pm$&Ce\,{\sc ii}   &$\pm$& Pr\,{\sc ii}  &$\pm$& Nd\,{\sc ii}  &$\pm$& Eu\,{\sc ii}  &$\pm$      \\ \hline
1219 & 	0.06 & 0.06 & 0.30 & 0.00 & 0.20 & --   & 0.40 & --    & 0.24 & --   & 0.42 & --   \\ 
1439 & 	0.74 & 0.06 & 0.72 & 0.04 & 0.54 & --   & 0.50 & --    & 0.54 & --   & --   & --   \\
 859 & 	0.73 & 0.09 & 0.82 & 0.07 & 0.65 & 0.01 & 0.59 & 0.02  & 0.72 & 0.02 & 0.25 & 0.10 \\ 
1309 & 	0.22 & 0.05 & 0.43 & 0.04 & 0.07 & --   & --   & --    & 0.30 & --   & 0.29 & --   \\ 
 579 & 	0.02 & 0.05 & 0.25 & 0.10 & --   & --   & --   & --    & 0.11 & --   & 0.36 & 0.07 \\ 
1339 & 	0.88 & 0.03 & 1.01 & 0.07 & 0.93 & 0.06 & 0.78 & 0.08  & 0.89 & 0.04 & 0.28 & --   \\ 
 177 & 	0.02 & 0.01 & 0.32 & 0.05 & 0.28 & --   & 0.35 & --    & 0.25 & --   & 0.30 & --   \\ \hline  
\end{tabular}
\end{table*}
%

%
\begin{table*}
\caption{Sensitivity of the derived UVES abundances to the uncertainties in atmospheric parameters and uncertainties due to the errors in the EWs measurements or in the $\chi$-square fitting procedure. The total internal uncertainty ($\sigma_{\rm total}$) has been obtained by considering the errors in atmospheric parameters, the covariance terms, and the EWs/fit terms.\label{tab:errUVE}}
\begin{tabular}{lcccccc}
\hline\hline
      &$\Delta$\teff &$\Delta$\logg&$\Delta$\vmicro&$\Delta$[A/H]& $\sigma_{\rm EWs/fit}$&$\sigma_{\rm total}$\\  
      & $\pm$50~K    & $\pm$0.16   & $\pm$0.11~\kmsec & $\pm$0.05~dex &     &                   \\
\hline
$\rm {[O/Fe]}$              & $\mp$0.03   & $\pm$0.07  & $\pm$0.04  & $\mp$0.00 & $\pm$0.01/$\pm$0.03 & $\pm$0.05/$\pm$0.06\\  
$\rm {[Na/Fe]}$             & $\mp$0.01   & $\mp$0.01  & $\pm$0.03  & $\mp$0.01 & $\pm$0.01/$\pm$0.02 & $\pm$0.04   \\  
$\rm {[Mg/Fe]}$             & $\mp$0.02   & $\mp$0.01  & $\pm$0.01  & $\mp$0.01 & $\pm$0.06           & $\pm$0.06   \\  
$\rm {[Al/Fe]}$             & $\mp$0.01   & $\pm$0.00  & $\pm$0.04  & $\mp$0.01 & $\pm$0.03/$\pm$0.07 & $\pm$0.05/$\pm$0.08\\
$\rm {[Si/Fe]}$             & $\mp$0.05   & $\pm$0.02  & $\pm$0.04  & $\pm$0.00 & $\pm$0.07           & $\pm$0.08   \\  
$\rm {[Ca/Fe]}$             & $\pm$0.00   & $\mp$0.01  & $\pm$0.00  & $\mp$0.01 & $\pm$0.02           & $\pm$0.02   \\  
$\rm {[Sc/Fe]}$\,{\sc ii}   & $\mp$0.06   & $\pm$0.07  & $\pm$0.02  & $\pm$0.01 & $\pm$0.04           & $\pm$0.03   \\  
$\rm {[Ti/Fe]}$\,{\sc i}    & $\pm$0.03   & $\mp$0.01  & $\pm$0.02  & $\mp$0.01 & $\pm$0.03           & $\pm$0.05   \\  
$\rm {[Ti/Fe]}$\,{\sc ii}   & $\mp$0.07   & $\pm$0.07  & $\pm$0.01  & $\pm$0.01 & $\pm$0.04           & $\pm$0.03   \\  
$\rm {[V/Fe]}$              & $\pm$0.00   & $\pm$0.00  & $\pm$0.04  & $\pm$0.00 & $\pm$0.08           & $\pm$0.09   \\
$\rm {[Cr/Fe]}$             & $\pm$0.03   & $\mp$0.01  & $\pm$0.00  & $\mp$0.01 & $\pm$0.06           & $\pm$0.06   \\
$\rm {[Mn/Fe]}$             & $\pm$0.05   & $\mp$0.01  & $\pm$0.04  & $\mp$0.02 & $\pm$0.01/$\pm$0.02 & $\pm$0.07   \\ 
$\rm {[Fe/H]}$\,{\sc i}     & $\pm$0.06   & $\mp$0.01  & $\mp$0.04  & $\mp$0.01 & $\pm$0.01           & $\pm$0.05   \\  
$\rm {[Fe/H]}$\,{\sc ii}    & $\mp$0.03   & $\pm$0.06  & $\mp$0.02  & $\pm$0.01 & $\pm$0.03           & $\pm$0.05   \\ 
$\rm {[Co/Fe]}$             & $\mp$0.01   & $\pm$0.01  & $\pm$0.04  & $\pm$0.00 & $\pm$0.11           & $\pm$0.12   \\
$\rm {[Ni/Fe]}$             & $\pm$0.00   & $\pm$0.01  & $\pm$0.03  & $\pm$0.00 & $\pm$0.03           & $\pm$0.04   \\
$\rm {[Cu/Fe]}$             & $\pm$0.02   & $\pm$0.00  & $\pm$0.03  & $\pm$0.00 & $\pm$0.01/$\pm$0.04 & $\pm$/0.04$\pm$0.06 \\
$\rm {[Zn/Fe]}$             & $\mp$0.07   & $\pm$0.04  & $\pm$0.01  & $\pm$0.01 & $\pm$0.08           & $\pm$0.08   \\  
$\rm {[Y/Fe]}$\,{\sc ii}    & $\mp$0.05   & $\pm$0.07  & $\pm$0.01  & $\pm$0.01 & $\pm$0.06           & $\pm$0.06   \\  
$\rm {[Zr/Fe]}$\,{\sc ii}   & $\mp$0.05   & $\pm$0.06  & $\pm$0.03  & $\pm$0.01 & $\pm$0.02/$\pm$0.08 & $\pm$0.02/$\pm$0.08\\
$\rm {[Ba/Fe]}$\,{\sc ii}   & $\mp$0.02   & $\pm$0.03  & $\mp$0.08  & $\pm$0.00 & $\pm$0.04           & $\pm$0.09   \\  
$\rm {[La/Fe]}$\,{\sc ii}   & $\mp$0.02   & $\pm$0.06  & $\pm$0.02  & $\pm$0.00 & $\pm$0.01           & $\pm$0.04   \\  
$\rm {[Ce/Fe]}$\,{\sc ii}   & $\mp$0.05   & $\pm$0.02  & $\pm$0.05  & $\pm$0.01 & $\pm$0.09           & $\pm$0.10   \\  
$\rm {[Pr/Fe]}$\,{\sc ii}   & $\mp$0.05   & $\pm$0.06  & $\pm$0.04  & $\pm$0.02 & $\pm$0.02/$\pm$0.07 & $\pm$0.03/$\pm$0.07\\ 
$\rm {[Nd/Fe]}$\,{\sc ii}   & $\mp$0.05   & $\pm$0.02  & $\pm$0.05  & $\pm$0.01 & $\pm$0.07           & $\pm$0.08   \\  
$\rm {[Eu/Fe]}$\,{\sc ii}   & $\mp$0.05   & $\pm$0.08  & $\pm$0.04  & $\pm$0.01 & $\pm$0.02/$\pm$0.08 & $\pm$0.04/$\pm$0.09\\  
\hline
\end{tabular}
\end{table*}
%

%
\begin{table*}
\caption{Sensitivity of the derived GIRAFFE abundances to the uncertainties in atmospheric parameters and uncertainties due to the errors in the $\chi$-square fitting procedure. The total internal uncertainty ($\sigma_{\rm total}$) has been obtained by considering the errors in atmospheric parameters, the covariance terms, and the fit terms.\label{tab:errGIR}}
\begin{tabular}{lcccccc}
\hline\hline
      &$\Delta$\teff &$\Delta$\logg&$\Delta$\vmicro&$\Delta$[A/H]& $\sigma_{\rm fit}$&$\sigma_{\rm total}$\\  
      & $\pm$50~K    & $\pm$0.20   & $\pm$0.20~\kmsec & $\pm$0.05~dex &     &                   \\
\hline
$\rm {[O/Fe]}$            & $\mp$0.03 & $\pm$0.09  & $\pm$0.04  & $\pm$0.00 & $\pm$0.03/$\pm$0.06 & $\pm$0.05/$\pm$0.08 \\  
$\rm {[Na/Fe]}$           & $\mp$0.02 & $\mp$0.02  & $\pm$0.04  & $\pm$0.03 & $\pm$0.05/$\pm$0.09 & $\pm$0.07/$\pm$0.10 \\  
$\rm {[Si/Fe]}$           & $\mp$0.02 & $\pm$0.02  & $\pm$0.06  & $\pm$0.01 & $\pm$0.02/$\pm$0.03 & $\pm$0.04/$\pm$0.05 \\ 
$\rm {[Ca/Fe]}$           & $\pm$0.00 & $\mp$0.04  & $\mp$0.03  & $\mp$0.01 & $\pm$0.02           & $\pm$0.02           \\ 
$\rm {[Sc/Fe]}$\,{\sc ii} & $\mp$0.03 & $\pm$0.07  & $\pm$0.05  & $\pm$0.03 & $\pm$0.02/$\pm$0.04 & $\pm$0.04/$\pm$0.05 \\  
$\rm {[Ti/Fe]}$           & $\pm$0.02 & $\mp$0.03  & $\pm$0.03  & $\mp$0.01 & $\pm$0.03/$\pm$0.04 & $\pm$0.04/$\pm$0.05 \\
$\rm {[Fe/H]}$            & $\pm$0.07 & $\pm$0.02  & $\mp$0.02  & $\pm$0.01 & $\pm$0.01/$\pm$0.02 & $\pm$0.07/$\pm$0.08 \\  
$\rm {[Ni/Fe]}$           & $\pm$0.00 & $\mp$0.01  & $\pm$0.03  & $\pm$0.00 & $\pm$0.03/$\pm$0.04 & $\pm$0.04/$\pm$0.05 \\ 
$\rm {[Ba/Fe]}$\,{\sc ii} & $\mp$0.04 & $\pm$0.03  & $\mp$0.10  & $\pm$0.01 & $\pm$0.07           & $\pm$0.15           \\  
$\rm {[La/Fe]}$\,{\sc ii} & $\mp$0.02 & $\pm$0.05  & $\pm$0.04  & $\pm$0.00 & $\pm$0.04/$\pm$0.07 & $\pm$0.06/$\pm$0.09 \\  
\hline
\end{tabular}
\end{table*}
%

%
\begin{table*}
\caption{
Mean GIRAFFE abundances for the total number of analysed stars, the $s$-rich, the $s$-poor groups. The averaged and rms values and the associated errors, have been computed by excluding values deviating more than 3$\times$68th percentile from the median. The total number of analysed stars for each group is also listed.\label{tab:meanGIRAFFE}}
\begin{tabular}{lcccc c cccc c cccc}
\hline\hline
Abundance &  Mean  & $\pm$ & $\sigma $& \#    &   &Mean  & $\pm$ & $\sigma $& \#&   &Mean  & $\pm$ & $\sigma$ & \#        \\
          &\multicolumn{4}{c}{All GIRAFFE stars} &   &\multicolumn{4}{c}{$s$-poor} &   &\multicolumn{4}{c}{$s$-rich}\\ \hline   
$\rm {[O/Fe]}$           &  $+$0.44 & 0.04 & 0.22 & 28       &   &$+$0.49 & 0.05 & 0.20 & 18   &   & $+$0.34 & 0.08 & 0.23 & 10  \\
$\rm {[Na/Fe]}$          &  $+$0.34 & 0.04 & 0.21 & 39       &   &$+$0.29 & 0.05 & 0.22 & 22   &   & $+$0.41 & 0.05 & 0.18 & 17  \\
$\rm {[Si/Fe]}$          &  $+$0.40 & 0.01 & 0.09 & 53       &   &$+$0.40 & 0.02 & 0.11 & 33   &   & $+$0.41 & 0.02 & 0.08 & 20  \\
$\rm {[Ca/Fe]}$          &  $+$0.41 & 0.01 & 0.09 & 55       &   &$+$0.40 & 0.02 & 0.10 & 35   &   & $+$0.42 & 0.01 & 0.05 & 20  \\
$\rm {[Sc/Fe]}$          &  $-$0.03 & 0.01 & 0.08 & 54       &   &$-$0.01 & 0.02 & 0.11 & 34   &   & $-$0.04 & 0.01 & 0.05 & 20  \\
$\rm {[Ti/Fe]}$\,{\sc i} &  $+$0.33 & 0.02 & 0.10 & 44       &   &$+$0.30 & 0.02 & 0.09 & 27   &   & $+$0.38 & 0.02 & 0.09 & 17  \\
$\rm {[Fe/H]  }$         &  $-$1.72 & 0.01 & 0.11 & 55       &   &$-$1.77 & 0.02 & 0.09 & 35   &   & $-$1.63 & 0.02 & 0.06 & 20  \\
$\rm {[Ni/Fe] }$         &  $+$0.13 & 0.01 & 0.07 & 27       &   &$+$0.11 & 0.01 & 0.05 & 17   &   & $+$0.13 & 0.02 & 0.07 & 10  \\
$\rm {[Ba/Fe]}$\,{\sc ii}&  $+$0.45 & 0.06 & 0.42 & 55       &   &$+$0.18 & 0.03 & 0.19 & 35   &   & $+$0.93 & 0.05 & 0.24 & 20  \\
$\rm {[La/Fe]}$\,{\sc ii}&  $+$0.61 & 0.09 & 0.40 & 21       &   &$+$0.14 & 0.01 & 0.03 & 8    &   & $+$0.88 & 0.07 & 0.23 & 13  \\
\hline
\end{tabular}
\end{table*}
%

%
\begin{table*}
\caption{
Mean UVES abundances for the total number of analysed stars, the $s$-rich and the $s$-poor group. \label{tab:meanUVES}}
\begin{tabular}{lcccc c cccc c cccc}
\hline\hline
Abundance &  Mean  & $\pm$ & $\sigma $& \#    &   &Mean  & $\pm$ & $\sigma $& \#&   &Mean  & $\pm$ & $\sigma $& \#\\
          &\multicolumn{4}{c}{All UVES stars} &   &\multicolumn{4}{c}{$s$-poor} &   &\multicolumn{4}{c}{$s$-rich}\\\hline
$\rm {[O/Fe]}$    &  $+$0.45 & 0.07 & 0.15 & 6        &   &$+$0.58 & 0.15 & 0.15 & 2    &   & $+$0.41 & 0.09 & 0.13 & 3\\
$\rm {[Na/Fe]}$   &  $+$0.32 & 0.09 & 0.22 & 7        &   &$+$0.18 & 0.16 & 0.23 & 3    &   & $+$0.44 & 0.14 & 0.20 & 3\\
$\rm {[Mg/Fe]}$   &  $+$0.55 & 0.04 & 0.11 & 7        &   &$+$0.63 & 0.03 & 0.05 & 3    &   & $+$0.54 & 0.06 & 0.08 & 3\\
$\rm {[Al/Fe]}$   &  $+$0.54 & 0.11 & 0.26 & 7        &   &$+$0.39 & 0.15 & 0.21 & 3    &   & $+$0.58 & 0.17 & 0.25 & 3\\
$\rm {[Si/Fe]}$   &  $+$0.41 & 0.06 & 0.16 & 7        &   &$+$0.37 & 0.09 & 0.13 & 3    &   & $+$0.52 & 0.09 & 0.13 & 3\\
$\rm {[Ca/Fe]}$   &  $+$0.31 & 0.02 & 0.06 & 7        &   &$+$0.29 & 0.04 & 0.05 & 3    &   & $+$0.36 & 0.04 & 0.05 & 3\\
$\rm {[Sc2/Fe]}$  &  $+$0.09 & 0.03 & 0.06 & 7        &   &$+$0.04 & 0.03 & 0.04 & 3    &   & $+$0.14 & 0.02 & 0.03 & 3\\
$\rm {[Ti1/Fe]}$  &  $+$0.36 & 0.05 & 0.13 & 7        &   &$+$0.29 & 0.05 & 0.06 & 3    &   & $+$0.47 & 0.07 & 0.10 & 3\\
$\rm {[Ti2/Fe]}$  &  $+$0.22 & 0.03 & 0.08 & 7        &   &$+$0.20 & 0.05 & 0.07 & 3    &   & $+$0.19 & 0.02 & 0.03 & 3\\
$\rm {[V/Fe]  }$  &  $+$0.09 & 0.07 & 0.15 & 5        &   &$+$0.09 & 0.11 & 0.11 & 2    &   & $+$0.19 & 0.08 & 0.08 & 2\\
$\rm {[Cr1/Fe]}$  &  $-$0.05 & 0.08 & 0.20 & 7        &   &$-$0.15 & 0.05 & 0.07 & 3    &   & $+$0.07 & 0.20 & 0.29 & 3\\
$\rm {[Mn/Fe] }$  &  $-$0.44 & 0.01 & 0.03 & 7        &   &$-$0.43 & 0.02 & 0.03 & 3    &   & $-$0.44 & 0.03 & 0.04 & 3\\
$\rm {[Fe/H]  }$  &  $-$1.80 & 0.05 & 0.12 & 7        &   &$-$1.92 & 0.00 & 0.01 & 3    &   & $-$1.70 & 0.02 & 0.03 & 3\\
$\rm {[Co/Fe] }$  &  $+$0.09 & 0.05 & 0.12 & 6        &   &$-$0.02 & 0.09 & 0.09 & 2    &   & $+$0.13 & 0.07 & 0.11 & 3\\
$\rm {[Ni/Fe] }$  &  $-$0.01 & 0.01 & 0.04 & 7        &   &$-$0.04 & 0.02 & 0.03 & 3    &   & $+$0.01 & 0.02 & 0.03 & 3\\
$\rm {[Cu/Fe] }$  &  $-$0.53 & 0.05 & 0.13 & 7        &   &$-$0.60 & 0.03 & 0.04 & 3    &   & $-$0.42 & 0.08 & 0.11 & 3\\
$\rm {[Zn/Fe] }$  &  $+$0.24 & 0.07 & 0.18 & 7        &   &$+$0.19 & 0.11 & 0.15 & 3    &   & $+$0.33 & 0.15 & 0.22 & 3\\
$\rm {[Y2/Fe] }$  &  $+$0.23 & 0.13 & 0.32 & 7        &   &$-$0.04 & 0.03 & 0.05 & 3    &   & $+$0.56 & 0.10 & 0.14 & 3\\
$\rm {[Zr2/Fe]}$  &  $+$0.35 & 0.09 & 0.23 & 7        &   &$+$0.17 & 0.04 & 0.06 & 3    &   & $+$0.56 & 0.11 & 0.16 & 3\\
$\rm {[Ba2/Fe]}$  &  $+$0.38 & 0.16 & 0.39 & 7        &   &$+$0.03 & 0.02 & 0.02 & 3    &   & $+$0.79 & 0.06 & 0.08 & 3\\
$\rm {[La2/Fe]}$  &  $+$0.55 & 0.12 & 0.30 & 7        &   &$+$0.29 & 0.03 & 0.04 & 3    &   & $+$0.85 & 0.10 & 0.15 & 3\\
$\rm {[Ce2/Fe]}$  &  $+$0.44 & 0.14 & 0.32 & 6        &   &$+$0.24 & 0.06 & 0.06 & 2    &   & $+$0.71 & 0.14 & 0.20 & 3\\
$\rm {[Pr2/Fe]}$  &  $+$0.52 & 0.09 & 0.17 & 5        &   &$+$0.38 & 0.04 & 0.04 & 2    &   & $+$0.62 & 0.10 & 0.14 & 3\\
$\rm {[Nd2/Fe]}$  &  $+$0.44 & 0.12 & 0.29 & 7        &   &$+$0.20 & 0.06 & 0.08 & 3    &   & $+$0.72 & 0.12 & 0.17 & 3\\
$\rm {[Eu2/Fe]}$  &  $+$0.32 & 0.03 & 0.06 & 6        &   &$+$0.36 & 0.04 & 0.06 & 3    &   & $+$0.27 & 0.02 & 0.02 & 2\\
\hline
\end{tabular}
\end{table*}
%

%
\begin{landscape}
\begin{table*}
\caption{
List of confirmed {\it anomalous} GCs, e.g. those with known metallicity variations, and their chemical properties. A sub-sample of these GCs have been classified as $s$-Fe {\it anomalous}, as they show also variations in $s$-process elements. For each object, and chemical property, we list the literature source.  \label{tab:anomaliclass}}
\begin{tabular}{l c l c c l c c l c c l c}
\hline\hline
GC                & \multicolumn{11}{c}{chemical abundance variations} & Proposed Class \\ 
\cline{2-12}
                  & metallicity   & Literature source        && $s$-elements &Literature source                &&$p$-capture elements & Literature source              && C+N+O & Literature source    &  \\
                  &               &                          &&              &                                 &&  in each Fe group   &                                &&       &                      & \\ \hline
$\omega$~Centauri & yes           & Norris et al. (1996);    && yes          & Norris \& Da Costa (1995);      && yes                 & Johnson \& Pilachowski (2010); &&yes    & Marino et al. (2012) &$s$-Fe-Anomalous   \\
                  &               & Suntzeff \& Kraft (1996) &&              & Smith et al. (2000);            &&                     & Marino et al.\,(2011)          &&       &                      &      \\
                  &               &                          &&              & Johnson \& Pilachowski (2010);  &&                     &                                &&       &                      &   \\
                  &               &                          &&              & Marino et al.\,(2011);          &&                     &                                &&       &                      & \\
                  &               &                          &&              & D'Orazi et al.\,(2011)          &&                     &                                &&       &                      & \\
                  &               &                          &&              &                                 &&                     &                                &&       &                      & \\
M\,22             & yes           & Marino et al. (2009);    && yes          & Marino et al. (2009, 2011, 2012)&& yes                 & Marino et al. (2009, 2011)     &&yes    & Marino et al. (2011) & $s$-Fe-Anomalous   \\
                  &               & Da Costa et al. (2009)   &&              &                                 &&                     &                                &&       & Alves Brito et al. (2012) &                  \\
                  &               &                          &&              &                                 &&                     &                                &&       &                      & \\
NGC\,1851         & possible small& Carretta et al. (2010)   && yes          & Yong \& Grundahl (2008);        && yes                 & Carretta et al. (2010);        &&yes    & Yong et al. (2014)   & $s$-Fe-Anomalous \\
                  &               & Gratton et al. (2013)    &&              & Villanova et al. (2010)         &&                     & Villanova et al. (2010)        &&       &                      & \\
                  &               & Marino et al. (2014)     &&              &                                 &&                     &                                &&       &                      & \\
                  &               &                          &&              &                                 &&                     &                                &&       &                      & \\
Terzan~5          & yes           & Ferraro et al. (2009)    && not studied  &                                 && not studied         &                                &&not-studied &                 & Anomalous \\
                  &               & Origlia et al. (2011)    &&              &                                 &&                     &                                &&       &                      & \\
                  &               & Massari et al. (2014)    &&              &                                 &&                     &                                &&       &                      & \\
                  &               &                          &&              &                                 &&                     &                                &&       &                      & \\
M\,54             & yes           & Carretta et al. (2010)   && not studied  &                                 && yes                 & Carretta et al. (2010)         &&not-studied &                 & Anomalous \\ 
                  &               &                          &&              &                                 &&                     &                                &&       &                      & \\
M\,2              & yes           & Yong et al. (2014)       && yes          & Lardo et al. (2013);            && yes                 & Yong et al. (2014)             &&not-studied &                 & $s$-Fe-Anomalous \\
                  &               &                          &&              & Yong et al. (2014)              &&                     &                                &&            &                 &                 \\
                  &               &                          &&              &                                 &&                     &                                &&       &                      & \\
NGC\,5824         & yes           & Da Costa et al.\, (2013) && not-studied  &                                 && not-studied         &                                &&not-studied &                 & Anomalous \\
                  &               &                          &&              &                                 &&                     &                                &&       &                      & \\
NGC\,5286         & yes           & This work                && yes          & This work                       && yes                 & This work                      &&not-studied &                 & $s$-Fe-Anomalous \\
\hline
\end{tabular}
\end{table*}
\end{landscape}
%

%
\begin{table*}
\caption{
Mean Fe and Ba abundance differences for the {\it anomalous} GCs M\,22, M\,2 and NGC\,5286. We list the differences among the $s$-poor and $s$-rich stars in M\,22 from Marino et al.\,2009, 2011, M\,2 from Yong et al.\,(2014).\label{tab:anomali}}
\begin{tabular}{l c c c }
\hline\hline
Abundance difference ($s$-rich$- s$-poor)&  M\,22           & M\,2             & NGC\,5286 \\   \hline
$\rm {\Delta [Fe/H]}$                & $+$0.15$\pm$0.02 & $+$0.17$\pm$0.04 & $+$0.17$\pm$0.01\\
$\rm {\Delta [Ba/Fe]}$               & $+$0.36$\pm$0.05 & $+$0.73$\pm$0.14 & $+$0.74$\pm$0.06\\
$\rm {\Delta [Ba/Fe]/\Delta[Fe/H]}$  & $+$2.4$\pm$0.5   & $+$4.3$\pm$1.3   & $+$4.4$\pm$0.5  \\
\hline
\end{tabular}
\end{table*}
%

\end{document}